\documentclass[%
rmp,
aps,
twocolumn,
nofootinbib,
longbibliography]{revtex4-2}

\usepackage{amsfonts,amsmath,amssymb,bbold}
\usepackage{mathtools}
\usepackage{color}
\usepackage{float}
\usepackage{graphicx}
\usepackage{dcolumn}
\usepackage{bm}
\usepackage{multirow}
\usepackage{nicefrac}
\usepackage{soul}
\usepackage{blindtext}
\usepackage{todonotes}
\usepackage{lineno}
\usepackage{comment}
\usepackage{physics}
\usepackage{tensor}
\usepackage[colorlinks=true,citecolor=blue,linkcolor=blue,urlcolor=blue]{hyperref}
\usepackage{tabularx}
\usepackage{printlen}
\usepackage{MnSymbol}

\uselengthunit{in} 
\setcitestyle{sort&compress}

\let\vec\mathbf

\begin{document}

\title{Massive quantum systems as interfaces of quantum mechanics and gravity}

\author{ Sougato Bose}
\affiliation{ Department of Physics and Astronomy, University College London, Gower Street, WC1E 6BT London, United Kingdom}
\author{ Ivette Fuentes}
\affiliation{ School of Physics and Astronomy, University of Southampton, Southampton SO17 1BJ, United Kingdom}
\affiliation{ Keble College, University of Oxford, Oxford OX1 3PG, United Kingdom}
\author{ Andrew A.~Geraci}
\affiliation{ Center for Fundamental Physics, Department of Physics and Astronomy, Northwestern University, Evanston, Illinois 60208, USA}
\affiliation{ Center for Interdisciplinary Exploration and Research in Astrophysics (CIERA)}
\author{Saba Mehsar Khan}
\affiliation{ Department of Physics, Lancaster University, Lancaster, LA1 4YB, United Kingdom}
\author{ Sofia Qvarfort}
\email{sofia.qvarfort@fysik.su.se}
\affiliation{ Nordita, KTH Royal Institute of Technology and Stockholm University, Hannes Alfv\'{e}ns v\"{a}g 12, SE-106 91 Stockholm, Sweden}
\affiliation{ Department of Physics, Stockholm University, AlbaNova University Center, SE-106 91 Stockholm, Sweden}
\author{ Markus Rademacher}
\affiliation{Department of Physics and Astronomy, University College London, Gower Street, WC1E 6BT London, United Kingdom}
\author{ Muddassar Rashid}
\affiliation{ Department of Physics, King’s College London, Strand, London, WC2R 2LS, United Kingdom}
\author{ Marko Toro\v{s}}
\affiliation{Faculty of Mathematics and Physics, University of Ljubljana, Jadranska 19, SI-1000 Ljubljana, Slovenia}
\author{ Hendrik Ulbricht}
\affiliation{ School of Physics and Astronomy, University of Southampton, Southampton SO17 1BJ, United Kingdom}
\author{\vspace{0.1cm} Clara C. Wanjura}
\affiliation{ Max Planck Institute for the Science of Light, Staudtstraße 2, 91058 Erlangen, Germany}
\affiliation{ Cavendish Laboratory, University of Cambridge, Cambridge CB3 0HE, United Kingdom}

\begin{abstract}
The traditional view from particle physics is that quantum gravity effects should only become detectable at extremely high energies and small length scales. Due to the significant technological challenges involved, there has been limited progress in identifying experimentally detectable effects that can be accessed in the foreseeable future. However, in recent decades, the size and mass of quantum systems that can be controlled in the laboratory have reached unprecedented scales, enabled by advances in ground-state cooling and quantum-control techniques. Preparations of massive systems in quantum states pave the way for the explorations of a low-energy regime in which gravity can be both sourced and probed by quantum systems. Such approaches constitute an increasingly viable alternative to accelerator-based, laser-interferometric, torsion-balance, and cosmological tests of gravity. 
In this review, we provide an overview of proposals where massive quantum systems act as interfaces between quantum mechanics and gravity. We discuss conceptual difficulties in the theoretical description of quantum systems in the presence of gravity, review tools for modeling massive quantum systems in the laboratory, and provide an overview of the current state-of-the-art experimental landscape. Proposals covered in this review include, among others, precision tests of gravity, tests of gravitationally-induced wavefunction collapse and decoherence, as well as gravity-mediated entanglement. We conclude the review with an outlook and summary of the key questions raised.
\end{abstract}
\maketitle

\tableofcontents

\newpage


%
\section{Introduction}
%
Inspired by particle physics, the central paradigm of gravitational physics indicates that quantum gravity effects should become detectable at extremely high energies and at extremely small length scales. Known as the Planck scale, this regime encompasses particle energies of $10^{19}$ GeV or length scales of $10^{-35}$ meters. However, accessing this parameter regime is extremely challenging, and there are few prospects for achieving the necessary technological progress in the next few decades. Tests of proposals such as string theory~\cite{dienes1997string, schwarz1999string}, loop quantum gravity~\cite{rovelli2008loop, ashtekar2021short}, and causal dynamical triangulation~\cite{loll2019quantum}, to mention just a few, therefore remain outstanding.  

At lower energies, however, quantum systems with masses several orders of magnitude higher than the atomic mass scale are starting to become accessible in the laboratory. At these scales, current theories predict that gravity should start affecting the dynamics of quantum states.  
A number of proposals and ideas have therefore been put forward. They encompass questions about superpositions of gravitational fields, gravity-induced wavefunction collapse via self-gravity or decoherence due to external gravitational fields, as well as the quantum nature of the gravitational field itself. The most encouraging aspect of these proposals is that many of them appear experimentally and technologically accessible in the near future.

Historically, the first tests of gravity (beyond drop tests performed by Galileo) were carried out by Cavendish in the 1790s. Here, a torsion balance was used to measure the gravitational constant $G$~\cite{newton1900experiments}. Since then, Torsion balance experiments have been a cornerstone of gravitational research, and the most accurate estimates of Newton's constant, namely $G = (6.67430 \pm 0.00015) \times 10^{-11}$\,m$^3$\,kg$^{-1}$s$^{-2}$ is based on number of torsion balance experiments~\cite{tiesinga2021codata}. Yet, $G$ remains one of the least precisely known fundamental constants, and experiments actually disagree on the value of $G$ more than they should based on the reported uncertainties~\cite{rothleitner2017invited}. The smallest detected gravitational coupling measured to date was observed between two millimeter-radius gold spheres~\cite{westphal2021measurement}, and the smallest separation which the gravitational potential has been measured at is 52\,$\mu$m~\cite{lee2020new}. 

More focused searches for quantum gravity have been considered in the context of particle accelerators and tests of the Standard Model. The current energy scale of the LHC is 6.8 TeV per beam and 13.6 TeV during collisions, which is 16 orders of magnitude away from the energies of the Planck scale ($10^{19}$ GeV). Nevertheless, several proposals predict that gravity might interact more strongly at energies below the Planck scale due to the existence of additional dimensions~\cite{arkani1998hierarchy,dimopoulos2001black}. Concretely, any signs of gravity should appear mainly as missing energy signatures due to direct graviton production. As of yet, no such evidence has been conclusively found, and the technological challenges involved in reaching higher energy with particle accelerator scales are substantial. 

Another way to test gravity is to turn to cosmological observations. Indications of quantum gravity could potentially be found in the signatures of $\gamma$-ray bursts~\cite{amelino1998tests}. 
Another key question is how quantum gravity effects influenced the early formation of the universe. While many other effects have likely been washed out during the latter stages of the universe's expansion, the detection of primordial gravitational waves could shed light on quantum gravity effects present shortly after the Big Bang~\cite{kamionkowski2016quest}. In addition, non-Gaussian signatures in the cosmic microwave background (CMB) might provide additional insights into this period~\cite{komatsu2010hunting}.  
Other astrophysical tests of quantum gravity have also been proposed, such as signatures in the light of distant quasars~\cite{lieu2003phase,ragazzoni2003lack}. However, no such signals have yet been found.

\begin{figure}[t]
    \centering
    \includegraphics[width=.49\textwidth]{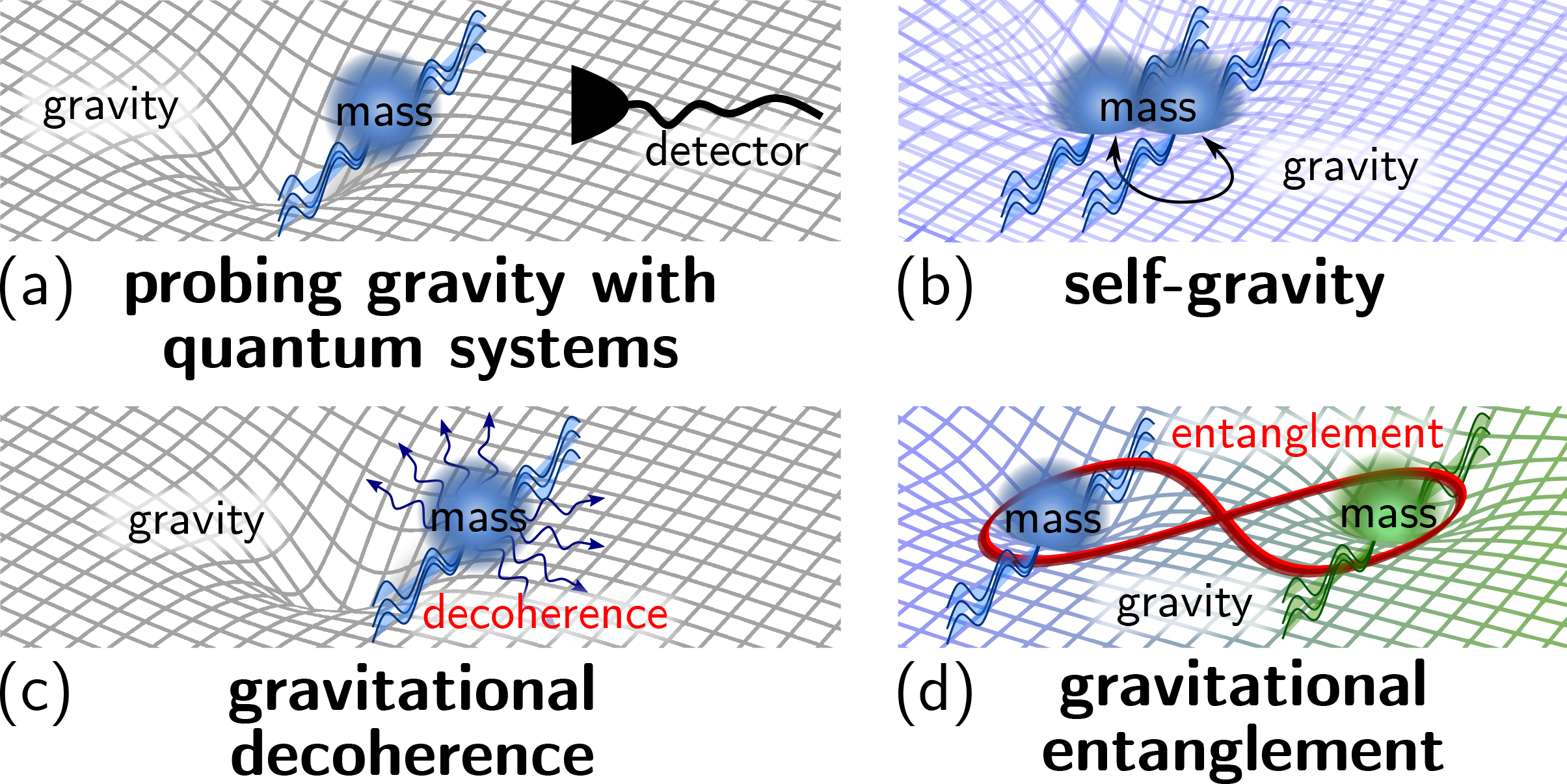}
    \caption{\textbf{Intersection of quantum mechanics and gravity.} The figure shows the possible tests of gravity that can be performed with quantum systems. 
    (a)~Quantum-enhanced measurements of gravitational effects. (b)~Quantum superpositions are unstable due to gravitational self-energy. (c)~An external gravitational field that acts as the environment causes the quantum system to decohere. (d)  Two quantum systems in spatial superpositions become entangled through gravity.
    }
    \label{fig:QuantGravity}
\end{figure}

The detection of gravitational waves by the Laser Interferometer Gravitational-Wave Observatory (LIGO) collaboration~\cite{abbott2016observation} has opened yet another avenue for tests of quantum gravity. Many theories of quantum gravity (such as loop quantum gravity~\cite{rovelli2008loop, ashtekar2021short} or string theory~\cite{dienes1997string}) require modification to the classical Einstein-Hilbert action, which in turn affects the propagation of gravitational waves~\cite{alexander2008gravitational}. The detection of primordial gravitational waves could potentially also shed light on which effective theories of gravity are valid at lower scales, which is otherwise known as the so-called Swampland problem of string theory~\cite{dias2019primordial}. In addition, the development of LISA provides prospects for tests of the equivalence principle and Lorentz invariance~\cite{barausse2020prospects}.

In this review, we aim to provide an alternative viewpoint to the paradigm of accelerator-based, cosmological, and laser-interferometric tests of gravity. 
The core question that this review thus seeks to address is: \textit{What aspects of gravity can be tested with massive quantum systems, and what can we learn from the outcome of these tests? } Here, we define \textit{massive quantum systems} as systems with masses far beyond the single-atom mass scale, such as micromechanical resonators, levitated nano-beads, as well as  Bose-Einstein condensates (BECs).  We are primarily concerned with the non-relativistic regime, where velocities are lower compared with the speed of light. Our goal is to gather together the tools and ideas necessary for testing gravity at low energies with massive quantum systems. Some of these notions are sketched in Figure~\ref{fig:QuantGravity}. We also hope that this review article will serve as a useful introduction and overview of the field for those who are just setting out to explore these questions. We also refer to the following works, which summarize tests that can be performed with tabletop experiments~\cite{carney2019tabletop} and superconductors~\cite{gallerati2022interaction}. 

The review is structured as follows. 
In Section~\ref{sec:theory:background}, we provide an overview of tensions between quantum mechanics and gravity, as well as a conceptual overview of how gravity can be incorporated into quantum mechanics at low energies. 
In Section~\ref{sec:theory:tools}, we provide an overview of theory tools for modeling massive quantum systems in the laboratory. Many of these tools are directly applicable to the proposed tests that follow in the next section.    
In Section~\ref{sec:gravity:tests}, we review proposals for tests of gravity with massive quantum systems. They include precision tests of gravity, searches for gravitational decoherence and wavefunction collapse, schemes for entangling massive quantum systems through gravity, and more. 
In Section~\ref{sec:experiments}, we provide an overview of state-of-the-art gravity tests and  experimental platforms which appear promising for tests of gravity. 
The review is concluded with some final remarks in Section~\ref{sec:discussion}. 

Before proceeding, we note that it would be impossible to give a completely balanced overview of such a broad and diverse field. We hope that this review provides a snapshot of the field today and that it ultimately helps focus efforts toward realizing some of the proposals that have been put forward for testing the interplay between quantum mechanics and gravity.


%
\section{Consolidating quantum mechanics and gravity} \label{sec:theory:background}
%

\begin{figure}[t]
   \centering
   \includegraphics[width=.5\textwidth]{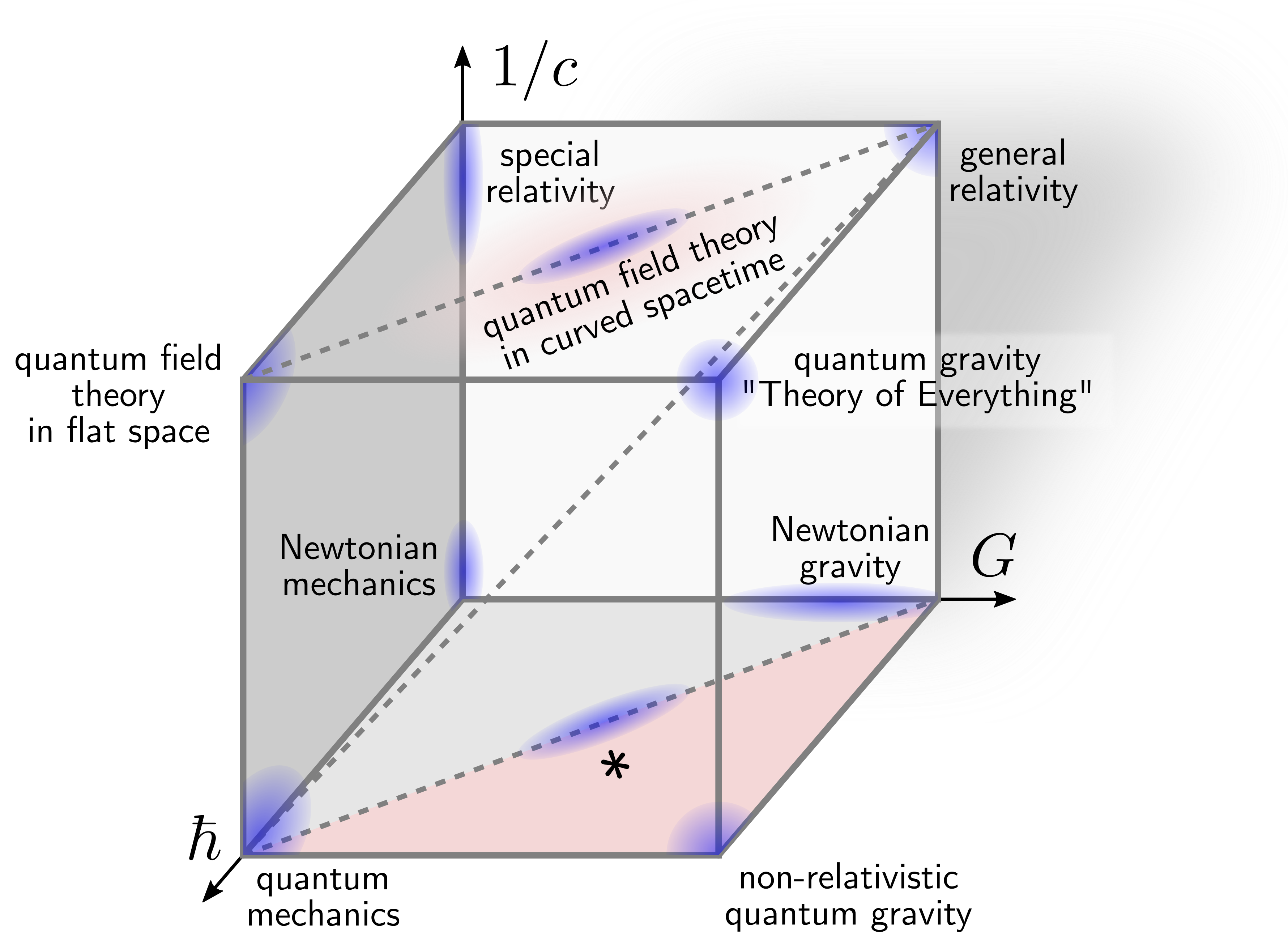}
   \caption{\textbf{Cube of theories.}
   Current theories are placed in a cube where the axes are small expansion parameters: the speed of light $1/c$, Plank's constant $\hbar$, and the gravitational constant $G$.
   The axes are the speed of light $c$, Plank's constant $h$, and the gravitational constant $G$. While some physicists aim at building a ``Theory of Everything''  that includes quantum physics (depending on $\hbar$) and general relativity (depending on both $c$ and $G$), others suggest that gravity should not be quantized.
   This review mainly focuses on the regime of the lower, light-red triangle, and the upper diagonal (quantum field theory).
   *Interplay of quantum mechanics and gravity at low energies (this could include Schrödinger-Newton-like equations, collapse models, gravitational phase shifts, etc., see Sec.~\ref{sec:incorporating:gravity} and Sec. \ref{sec:gravity:tests}).
   Figure adapted from~\cite{Bronstein1933K}, slightly modified for our purposes.
   }
   \label{fig:cube}
\end{figure}

A limitation in developing a theory of quantum gravity has been the inability to resolve the persisting tensions between the fundamental principles of quantum physics and general relativity. Current theories are good approximations in certain regimes. The relations between current theories can be found in the cube of theories in Fig.~\ref{fig:cube}, see Ref.~\cite{Bronstein1933K}. A ``Theory of Everything'' that combines quantum physics and general relativity is expected to be a theory in which the speed of light $c$, Planck's constant $\hbar$, and the gravitational constant $G$ all play significant roles. Interestingly, scientists disagree on the need to quantize gravity. In this section, we discuss how gravity is different from other forces, and why it has been so difficult to construct a unified theory.

We start by exploring different attempts to incorporate gravity into quantum mechanics at low energies. We cover modifications of quantum dynamics that include gravity as a phase term or driving term. We then motivate moving to quantum field theory in curved spacetime and perturbative quantum gravity (Sec.~\ref{sec:incorporating:gravity}). Then, we summarize the challenges that arise when incorporating general relativity into the way we generally do non-relativistic quantum mechanics in first quantization (Sec.~\ref{sec:postulates}). Many of these challenges have been discussed throughout the literature. While we here attempt to provide an outline of the main ideas and research directions, it is by no means a complete account of the history or challenges that arise in the context of developing a fully-fledged theory of quantum gravity.

\subsection{Incorporating gravitational effects into quantum mechanics at low energies} \label{sec:incorporating:gravity}

The goal of this section is to provide a high-level outline of the main theoretical ideas that enable a limited consolidation of quantum mechanics and gravity. For each case, we detail the underlying assumptions that enable the treatment and discuss the validity and limits of the theory. Every time we introduce a new tool (such as quantum field theory), we motivate the leap beyond the current framework. However, we note that current tools are often not enough and that theory must ultimately be guided by experiments.

\subsubsection{Newtonian potential in the Schrödinger equation} \label{sec:Newtonian:Schrödinger}

The second postulate of quantum mechanics dictates that the evolution of a single or composite quantum wavefunction $\Psi(t,x)$ in time is described by the Schrödinger equation. A natural starting point when attempting to incorporate gravity into the dynamics of a quantum system is by including it as a potential term in the Schrödinger equation. However, time in the Schrödinger equation is absolute, in contrast to general relativity, where time is an observer-dependent quantity. To use the Schrödinger equation, we must therefore make the following assumptions: (i) we consider a single inertial frame where time is well-defined, (ii) the gravitational field is weak, and (iii) the quantum particles do not travel at relativistic speeds. With these assumptions, it is possible to include the Newtonian potential from a source mass into the Schrödinger equation for a quantum particle with mass $m$ as follows:
\begin{align} \label{eq:Schrodinger:equation}
	\mathrm{i}\hbar \frac{\partial}{\partial t} \Psi(t,\vec{x})
	& = \left[\frac{\vec{p}^2}{2m} +  V_N(\vec{x})\right] \Psi(\vec{x}), 
\end{align}
where $\Psi(\vec{x})$ is the wavefunction in the position basis, for which we have denoted the position vector $\vec{x}$, and the momentum operator $ p_i\equiv-i \hbar \partial/\partial x_i$ for the direction $x_i$. The gravitational constant $G$ explicitly appears in the Newtonian potential $V_N(\vec{x}) =- G m m_S /|\vec{x} - \vec{x}_S|$, where $m_S$ is the source mass and $\vec{x}_S$ is the position of the source mass. 
For small radial displacements $\delta x \gg R$ away from the source located at distance $R$, such as for a particle moving in Earth's gravitational field, we may approximate the Newtonian potential as $G m m_S /|\vec{x} - \vec{x}_S| \approx mg \delta x $ where we have introduced the gravitational acceleration $g = m_S G /R^2$ (omitting the constant part of the potential and higher order corrections). For such a linear potential, the solutions to the Eq. \eqref{eq:Schrodinger:equation} are given by  Airy functions \cite{griffiths2018introduction}, and the Feynman path-integral propagator is given by \cite{sakurai1995modern}:
\begin{align}
    &\langle \vec{x}_n , t_n| \vec{x}_{n-1}, t_{n-1} \rangle \nonumber \\
    &\qquad  = \sqrt{\frac{m}{2\pi i \hbar \Delta t}} \mathrm{exp} \left[ i \int^{t_n}_{t_{n-1}} dt \frac{ \frac{1}{2} m \dot{\vec{x}}^2 - m g \delta x}{\hbar} \right], 
\end{align}
where $\Delta t=t_{n} - t_{n-1}$ denotes the time-increment.

\begin{figure}[t]
   \centering
   \includegraphics[width=.5\textwidth]{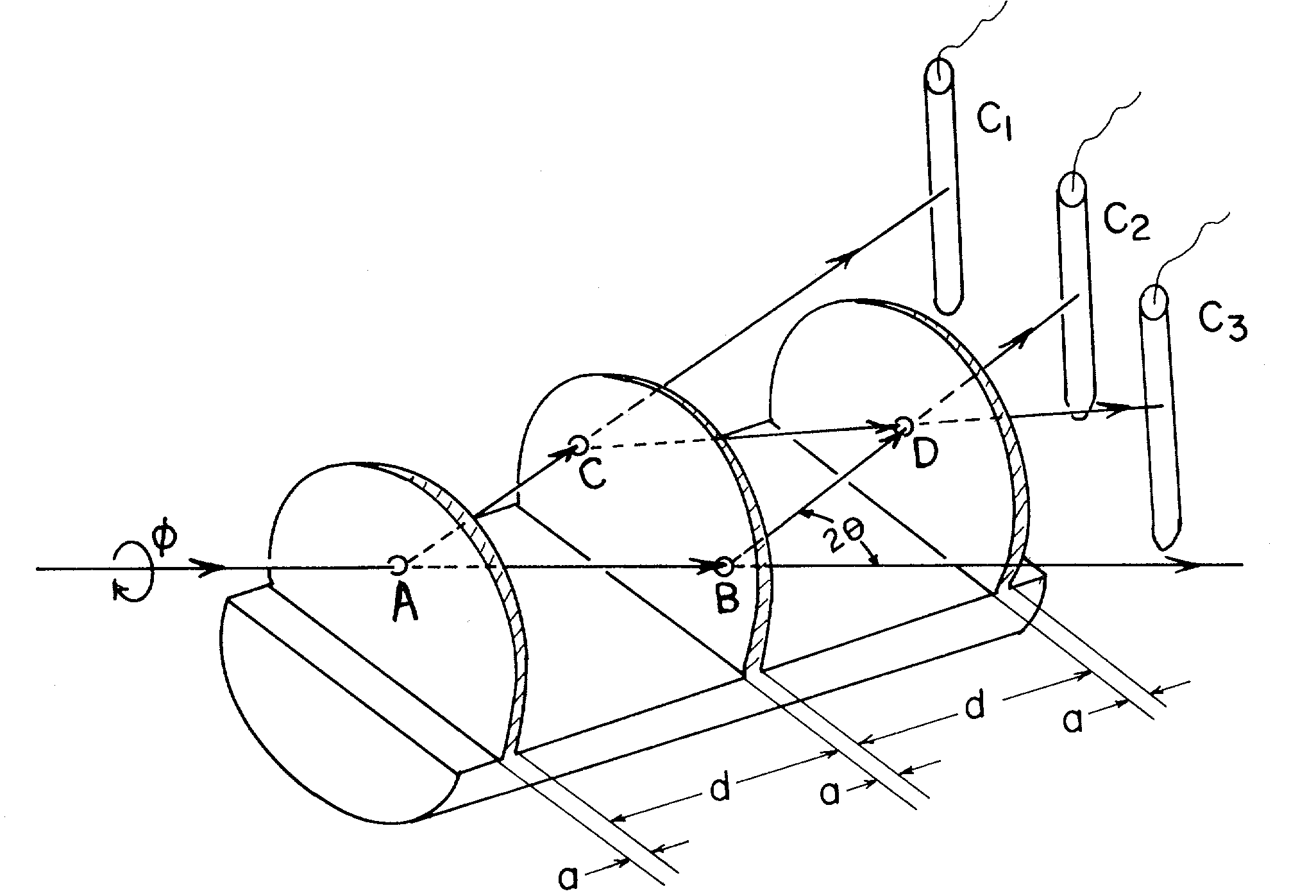}
   \caption{\textbf{Superpositions of neutrons along different paths in a gravitational field}. Reprinted with permission from the original work Collela R, Overhauser A W and Werner S A 1975 Phys. Rev. Lett. 34 1472~\cite{colella1975observation}. Copyright 1975 by the American Physical Society. }
\label{fig:COW:experiment}
\end{figure}

Indeed, the fact that gravity can be included into the Schrödinger equation as in Eq.~\eqref{eq:Schrodinger:equation} has been experimentally verified. 
In their pioneering experiment, Colella, Overhauser, and Werner (COW) demonstrated that neutrons passing through the Earth's gravitational field acquire a phase-shift that can be observed through interference~\cite{colella1975observation} (with the initial theoretical proposal outlined in~\cite{overhauser1974experimental}). This effect is sometimes referred to as \textit{gravity induced quantum interference}~\cite{rauch2015neutron}.

For the description of the COW experiment, we follow~\cite{abele2012gravitation}. The neutrons are placed into a superposition of two spatial locations. Each branch then traverses a path at two different heights above the Earth. The two paths enclose a parallelogram (see the original sketch of the experiment in Fig.~\ref{fig:COW:experiment}). 
One branch of the superposition takes the upper path $A \rightarrow C \rightarrow D$, and the lower branch takes the path $A \rightarrow B \rightarrow D$. 
The momentum of each neutron is determined through energy conservation, which dictates that the sum of the  kinetic and potential energy must remain the same:
\begin{align}
    E_0 = \frac{\hbar^2 k_0^2}{2m _n} = \frac{\hbar^2 k^2 }{2m_n} + m_n g H(\phi), 
\end{align}
where $m_n$ is the neutron mass, $g$ is the gravitational acceleration, and $H(\phi)=H_0 \sin(\phi)$ is the height which depends on the orientation of the setup $\phi$ (i.e., $\phi=\pi/2$ corresponds to the maximum height difference $H_0$, i.e., the height difference between the segments AB and CD for vertical orientation). The difference in height between the two paths means that the momentum $p = \hbar k$ on the higher path CD must be less than the momentum $p_0 = \hbar k_0$ on the lower path AB. 
The corresponding accumulated phase difference $\Delta \Phi_{\mathrm{COW}}$ between the two different paths is given by
\begin{align}\label{eq:cowphase}
    \Delta \Phi_\mathrm{COW} = \Delta k S \approx - q_\mathrm{COW} \sin(\phi) , 
\end{align}
where $\Delta k = k - k_0$, and $S$ is the path length of the segments AB and CD.  It is instructive to write the final phase using the geometric factor $q_\mathrm{COW} = \lambda m_n^2 g A_0/\hbar^2 $, where $A_0 = H_0 S$ is the area of the parallelogram (i.e., interferometric area), and $\lambda$ is the neutron wavelength. Depending on the value of the phase difference $\Delta \Phi_\mathrm{COW}$ we then observe either destructive or constructive interference of each individual neutron with itself as it is recombined at the end of the interferometric paths. 

Gravitational phase shifts, such as the one in Eq.~\eqref{eq:cowphase}, have sparked numerous investigations, not only with neutrons but also with cold atoms, where experiments span from tests of the equivalence principle to searches of dark matter~\cite{tino2021testing}. However, subtleties arise when additional corrections from general relativity are similarly included as phase shifts. We return to this question in Sec.~\ref{sec:relativistic:Schrödinger}.

\subsubsection{Gravity and the quantum harmonic oscillator} \label{sec:gravity:QHO}
Before going beyond Newtonian gravity, we briefly mention another route for including gravity as a perturbation into the dynamics of a quantum system. 
One of the few known analytic solutions to the Schrödinger equation apart from the hydrogen atom is the quantum harmonic oscillator (QHO). Here, the potential $V(\vec{x})$ in Eq.~\eqref{eq:Schrodinger:equation} is quadratic in $\vec{x}$, such that in a single spatial dimension $x$, $V(x) \propto x^2$. The resulting solutions describe a harmonic oscillator with quantized energy levels. The QHO is important in the context of tests of gravity because gravitational effects can be described as effective interaction terms as modifications to the quadratic trapping term. 

In the language of second quantization, the Hamiltonian of the QHO reads
\begin{align}
    \hat H_\mathrm{QHO} = \frac{1}{2}  m \omega_m^2 \hat x^2 + \frac{1}{2m} \hat p^2,
    \label{eq:harmonicOscillator}
\end{align}
where $m$ is the mass of the system, $\omega_m$ is the angular frequency, and where  $\hat x$ and $\hat p$ are position and momentum operators. In second quantization, $\hat x$ and $\hat p$ are given in terms of the annihilation and creation operators $\hat a$ and $\hat a^\dag$ as 
\begin{align} \label{eq:zpm}
   &\hat x = \sqrt{\frac{\hbar}{2\omega_m m}} \left( \hat a^\dag + \hat a \right),
    &&\hat{p} = i \sqrt{\frac{\hbar \omega_m m}{2}} \left( \hat a^\dag - \hat a \right), 
\end{align}
where $[\hat a, \hat a^\dag] = 1$.

Gravitational effects can be included in the description of the center-of-mass dynamics of these quantum systems.
We consider a point-like gravitational source of mass $m_S$ situated at a distance $r_0$ away from the quantum systems~\cite{scala2013matter, qvarfort2018gravimetry, raetzel2018dynamical}. The Newtonian potential is given by $G m m_S /r_0$. Assuming that the quantum system that probes the gravitational field is perturbed by a small distance $\delta x$, we expand the Newtonian potential in terms of $\delta x$ 
\begin{equation} \label{eq:newt_pot_exp}
    V(r - \delta x) \approx \frac{G m m_S}{r_0} \left( 1 + \frac{\delta x}{r_0} + \frac{(\delta x)^2}{2 r_0^2}  + \mathcal{O}[(\delta x)^2]\right). 
\end{equation}
Inserting this potential in the Hamiltonian of the QHO and replacing the perturbation $\delta x$ with the quantum operator $\hat x$ we obtained a modified Hamiltonian of the QHO:
\begin{align} \label{eq:QHO:Hamiltonian:gravity}
    \hat H_{QHO} = \frac{1}{2} m \omega_m^2 \hat x^2 + \frac{1}{2m} \hat p^2 + \mathcal{G}_1 \hat x + \mathcal{G}_2 \hat x^2 + \mathcal{O}( \hat x^3), 
\end{align}
where we have defined
\begin{align}
    &\mathcal{G}_1 = \frac{G m m_S}{r_0^2},  
    &&\mathcal{G}_2 =  \frac{G m m_S }{2r_0^3}, 
\end{align}
and where higher orders of the perturbation can be similarly defined, although the resulting nonlinear equations of motion are generally challenging to solve. 
We provide an overview of quantum sensing of gravitational fields with quantum optomechanical systems and with Bose-Einstein condensates, where the force enters as described here in Sec.~\ref{sec:quantum:enhanced:sensing}.

\subsubsection{Gravity beyond the Schrödinger equation} \label{sec:relativistic:Schrödinger}

Thus far in our presentation, incorporating gravity into quantum mechanics has been straight-forward since both the Newtonian potential and the Schrödinger equation are non-relativistic and share a joint notion of absolute time. However, problems start to arise as we wish to go further and include additional effects from general relativity, e.g., time-dilation.

Consider, for example, a quantum particle in a spatial superposition where each branch of the superposition follows a different spacetime trajectory, not unlike the COW experiments discussed in Section \ref{sec:Newtonian:Schrödinger}. In that case, we assumed that a weak gravitational effect introduces a potential difference. However, if the gravitational effect is strong, such that the background spacetime can no longer be considered flat across the relevant length scale, the two branches of the superposition experience different proper times throughout their trajectories and should, therefore, evolve at different rates.  
There is no prescription for how to perform calculations in this scenario in the absence of an external observer. However, some studies have considered using the Schrödinger equation to describe the system only in the reference frame in which the system is measured.  That is any evolution of the quantum states, as seen in the laboratory frame, can be described as a result of some effective dynamics that arise from gravity. In~\cite{zych2011quantum}, for example, it was proposed that internal degrees-of-freedom of particles can act as clocks that record the elapsed proper time. The addition of internal clock states solves the challenge of interpreting a phase shift as either a potential shift or redshift due to differences in proper time. Similarly, in~\cite{pikovski2015universal}, it was shown that the effects of time-dilation, as seen from an external observer, result in decoherence in composite particles. That is, by defining a Hamiltonian for the center-of-mass and internal degrees of freedom, general relativistic corrections are incorporated into the full dynamics of the particle. When the superposition branches are brought back together, the effect manifests as decoherence. Several other mechanisms that cause decoherence have been derived using similar semi-classical arguments. We cover these in Sec.~\ref{sec:grav:dec:section}. It has been pointed out in the literature that there are inconsistencies that arise when introducing proper times in quantum superpositions~\cite{marzlin1995dipole,sonnleitner2018mass,schwartz2019post2}. The argument is that classical systems couple to gravity via the {\it minimum coupling} principle, which is diffeomorphism invariant, and that the coupling is not consistent with Galilean invariant equations such as the Schrödinger equation~\cite{schwartz2019post1}. 
However, the argument of ~\cite{pikovski2015universal} in response is that there is no inconsistency if no non-relativistic physics is imposed and the correct Schrödinger equation canin fact be derived with relativistic extensions. 

Work on describing post-Newtonian phases in atom interferometry considers the free propagation of the atoms along spacetime geodesics. The atom-light interaction is described in a covariant manner to calculate the leading order general relativistic effects~\cite{dimopoulos2008general,werner2023atom}. In addition, these considerations offer a possible reinterpretation of the COW experiment (see Sec.~\ref{sec:Newtonian:Schrödinger}). The question becomes: Can the phase shift that is detected by an atom interferometer be interpreted as a gravitational redshift? This interpretation was first suggested in~\cite{muller2010precision}, and was followed by a vigorous debate in the community (see Refs [236–244] in~\cite{tino2021testing}). The ambiguity arises because the phase shift in the atom interferometer can either be interpreted as an effective potential shift or due to redshift, which has resulted from the differences in proper times. 

Many of the works listed above consider interferometry of quantum states in the presence of gravity. Recombining the two states at the end of the interferometry process naturally involves taking a notion of an appropriate inner product. It is important to note that the inner product between quantum states in non-relativistic quantum mechanics, where there is an absolute time, is not Lorentz-invariant. In the position representation, the inner product is given by 
\begin{align}\label{eq:innerProduct}
 \int_{\mathbb{R}^3} \mathrm{d}\mathbf{x} \, \psi_j^*(t,x) \psi_\ell(t,x) = \delta_{j,\ell}, 
\end{align}
where $\psi_j(t,x)$ are wavefunctions and where $\delta_{j,\ell}$ is the Kronecker delta-function. A consequence is that two quantum states in different inertial frames or different spacetime locations cannot be consistently compared. Since the description of measurements and averages requires the inner product, the experimental observations cannot be described appropriately with this inner product. This poses a problem often overlooked when describing quantum interferometry in curved spacetime. However, as long as a single laboratory frame is considered, the inner product is well-defined.  In Sec. \ref{sec:QFTCS}, we introduce the Klein-Gordon inner product, which is appropriate for (scalar) relativistic quantum fields.

If we truly wish to describe quantum systems in a manner that is consistent with general relativity, we must use a covariant formalism where the equations and the inner products are Lorentz invariant. Quantum field theory in curved spacetime enables such a description in the low energy regime.

\subsubsection{Quantum field theory in curved spacetime} \label{sec:QFTCS}
Thus far, we discussed approaches for describing the effects of weak gravitational effects using the Schrödinger equation. Such schemes consider a single inertial frame or study the differences between two inertial frames as an effective Hamiltonian. That is, relativistic corrections are treated as dynamical perturbations in the Schrödinger equation, where time remains absolute. However, in relativity, measurements of well-defined quantities must coincide in different frames, and a consistent description in non-inertial frames is also required. This is only possible through a covariant formalism. The question becomes: can quantum systems be described using equations that are Lorentz invariant? 

The answer is affirmative within some restrictions. It is possible to describe some aspects of the interplay of quantum physics and general relativity using quantum field theory (QFT) in curved spacetime (CS). QFT in CS is a semi-classical approach that considers the behavior of quantum fields on a classical spacetime background. Crucially, spacetime is not quantum, rather, it is a solution of Einstein field equations. The formalism describes multi-particle effects, and, interestingly, it turns out that considering single particles such as single atoms is non-trivial. Quantum field theory in curved spacetime has enabled the study of some effects in quantum physics and quantum information in relativistic settings, including entanglement and its applications in non-inertial frames and curved spacetime. Section~\ref{sec:non-inertial} includes a discussion on the degradation of entanglement as seen by observers in uniform acceleration. 

Here, we provide a brief overview of QFT in CS. A more comprehensive account of this research field can be found in textbooks by \cite{birrell1982quantum,fulling1989aspects,schweber2005introduction,parker2009quantum}. We limit our discussion to a scalar field, which is the simplest case.  See~\cite{hollands2015quantum} for a full review of QFT in CS, including other cases, such as the Dirac equation, which describes fermionic fields. In the case of a single scalar field, the Schrödinger equation is replaced by the Klein-Gordon equation, which reads (having set $\hbar = c = 1$)
\begin{equation}
    ( \square_g - m^2 ) \phi = 0,   
\end{equation}
where $\square_g = g^{\mu \nu} \nabla_\mu \nabla_\nu$ is the D'Alembertian operator associated with the metric $g$ and $\phi$ is the scalar field with mass $m$. 

Under superficial inspection, the Klein-Gordon equation in flat spacetime looks very similar to the Schrödinger equation.
The main difference is that it has a second derivative in both the spatial and temporal coordinates, making it invariant under Lorentz transformations as required by relativity.
Historically, the Klein-Gordon equation was derived from a relativistic (classical) Hamiltonian of a particle and then interpreting momentum and position as operators~\cite{schweber2005introduction}
\begin{align}
    H_\mathrm{r} = \sqrt{(\bm p c)^2 + (m c^2)^2}, 
\end{align}
with momentum vector $\mathbf{p}$. 
To avoid the problem of the square root, which would appear if we inserted $H_{\mathrm{r}}$ into the Schrödinger equation, the formalism considers instead the squared operator equation
\begin{align}
    -\hbar^2 \frac{\partial^2}{\partial t^2} \phi(t,\bm r) = [\bm{\hat p}^2 c^2 + (mc^2)^2]\phi(t,\bm r).
\end{align}
In contrast to the Schrödinger equation, the resulting wave equation is Lorentz invariant
\begin{align}\label{eq:KleinGordonEquation}
    \left[\left(\frac{1}{c^2}\frac{\partial^2}{\partial t^2}-\nabla^2\right)+\left(\frac{mc}{\hbar}\right)^2\right] \phi(t,\bm r) = 0.
\end{align}
However, interpreting $\phi(t,\bm r)$ as a wave function is generally problematic. This is because the probability density $\rho\equiv\frac{\mathrm{i}\hbar}{2mc^2}(\phi^*[\partial_t\phi]-[\partial_t\phi^*]\phi)$ and probability current $j_\ell\equiv\frac{\hbar}{2m\mathrm{i}}(\phi^*[\partial_\ell\phi]-[\partial_\ell\phi^*]\phi)$ defined such that they satisfy the continuity equation $\nabla\cdot\bm j+\partial_\mu\rho = 0$ can take negative values due to the second derivative in time in Eq.~\eqref{eq:KleinGordonEquation}~\cite{schweber2005introduction}. 
Specifically, the Klein-Gordon inner product for $\phi(t,\mathbf{r})$ is derived from the continuity equation and is given by
\[(\phi(t,\mathbf{r}),\psi(t,\mathbf{r}))=-i\int_{\Sigma}(\psi^*\partial{\mu}\phi-(\partial_{\mu}\psi^*)\phi)d\Sigma^{\mu},\] 
where $\Sigma$ is a spacelike hypersurface. While the inner product is Lorentz invariant by construction, it yields negative probabilities in some cases. Therefore, interpreting $\phi(t,\bm r)$ as the wave function of a single particle is inconsistent with quantum mechanics. 

The problem can be solved in special cases when the spacetime has specific symmetries. In these cases, one can construct an operator-valued function by associating creation and annihilation operators $\hat a_k$ and $\hat a_k^{\dagger}$ for a mode $k$ to the positive and negative mode solutions $ u_k$ and $ u_k^{\ast}$ of the Klein-Gordon equation. 
\begin{align}\label{eq:quantum field}
    \hat{\phi}(t,\bm r)=\int dk ( u_k \hat a_k+u_k^{\ast} \hat a_k^{\dagger}). 
\end{align}
This operator-valued function obeys the Klein-Gordon equation, and particle states, with positive norms, are defined by the action of creation operators on the vacuum state. The vacuum state is defined by $a_k\ket{0}=0$. The operators act on the Fock space $\bigoplus_{n=0}^{\infty}\mathcal{H}^{\otimes n}$, where $\ket{0}\in\mathbb{C}$, $\mathcal{H}$ is the single particle Hilbert space, $\mathcal{H}^{\otimes n}$ the n-particle sector  and $\bigoplus$, $\otimes$ the direct sum and tensor product, respectively. This construction is only possible when the solutions of the Klein-Gordon equation can be classified in positive and negative frequency mode solutions. Crucially, such a classification is only possible when the spacetime admits a time-like Killing vector field. A Killing vector field is the tangent vector space of transformations that leave the metric invariant. Spacetimes that admit a global time-like Killing vector field are stationary, such as Minkowski or the Schwarzschild spacetime. A consistent theory can be constructed for spacetimes with these symmetries having a well-defined probability distribution.  Well-known examples are quantum fields in non-inertial frames and eternal black holes~\cite{schweber2005introduction,birrell1982quantum}. 

A key problem in QFT in CS is that particles are not well defined. Only observers flowing along time-like Killing vector fields can describe particle states in a meaningful way. In general, curved spacetimes do not admit time-like Killing vector fields globally. Moreover, in the case that the spacetime does have a global time-like Killing vector field, the vector field is not necessarily unique. A consequence of this is that the field can be equivalently quantized in several different bases corresponding to different Killing observers. Using the Klein-Gordon inner product, it is possible to find a unitary transformation, called a Bogoliubov transformation, that relates the solutions to the equation in the different basis. This induces a transformation between the creation and annihilation operators in the old basis $\hat a_j$ and $\hat a_j^\dag$ and new operators $\hat b_k$ and $\hat b_k^\dag$ associated to the solutions in a different basis. In the new frame, the operators are given by
\begin{align}
    \hat b_k  = \int dk \left( \alpha_{kj } \hat a _j  + \beta_{jk} \hat a_j^\dag \right), 
\end{align}
where $\alpha_{kj}$ and $\beta_{jk}$ are called Bogoliubov coefficients. A consequence is that the vacua are not equivalent, and the particle content of the field is observer-dependent. 
The vacuum that was annihilated by the mode operator $\hat a_j \ket{0}$ in the first frame no longer appears empty in the second frame, since 
\begin{align}
    \bra{0}\hat b_k^\dag \hat b_k \ket{0} = \int dk |\beta_{kj}|^2 \neq 0. 
\end{align}
This has important consequences in the study of entanglement in relativistic quantum fields since the notion of subsystems are indispensable to store information (see Sec. \ref{sec:non-inertial}).

Some spacetimes do not admit global time-like Killing vector fields but do have spacetime regions where particles can be well defined. An example is the metric that describes a toy model for the expansion of the universe, known as the Friedmann–Lemaître–Robertson–Walker (FLRW) metric. The spacetime is not stationary, and time-like Killing vector fields are defined only in the past and future infinity regions (see \cite{birrell1982quantum} and references within). It is possible to show, using Bogoliubov transformations, that the vacuum state at past infinity has entangled particles in the future infinity region \cite{ball2006entanglement}.

Some particularly interesting consequences of QFT in non-inertial frames and CS are the Unruh-Davies-Fulling effect~\cite{fulling1973nonuniqueness, davies1975scalar, unruh1976notes} and the closely related Hawking radiation effect~\cite{hawking1975particle}. The inertial vacuum appears populated by particles in a thermal state for uniformly accelerated observers. A region of spacetime becomes inaccessible to non-inertial observers due to their acceleration. Tracing over the field modes in the casually discounted region leads to mixed states.  In the case of uniformly accelerated observers in flat spacetime, the Minkowski vacuum corresponds to a thermal state with Unruh temperature
\begin{align}
    T_{\mathrm{Unruh}} = \frac{\hbar a}{2 \pi c k_B} , 
\end{align}
where $k_B$ is the Boltzmann constant. A similar situation occurs in black hole spacetimes, where observers hovering outside the horizon loose access to the region inside the black hole. The inertial vacuum state corresponds to a thermal state for observers at a fixed distance from an eternal black hole. This spacetime is stationary and the black hole mass is constant. In the case of a collapsing star, the spacetime is not stationary and there is energy flux known as Hawking radiation, where
\begin{align}
    T_{\mathrm{Hawking}} = \frac{\hbar c^3}{8 \pi GM k_B} , 
\end{align}
is the Hawking temperature for a black hole with mass $M$.

These results further emphasize the notion that the vacuum in a curved spacetime is not unique, which has implications for the coherence and entanglement of quantum systems. We explore the consequences for entanglement in curved spacetime in Sec.~\ref{sec:non-inertial}.

 A main lesson that we learn from the development of QFT in CS is that fields, and not particles, are fundamental. Particles are derived notions that do not always have a viable interpretation. QFT is a multi-particle theory and single particles can be described using this formalism only when energies are not sufficient to create new particles. In this low-energy case, it is possible to restrict the system to the single particle sector because the energies present are not high enough to create new particles. In the next section, we will discuss attempts to construct a covariant description of the quantum harmonic oscillator in the presence of curved spacetime using the Klein-Gordon equation and the restriction to the single particle sector mentioned above.  

 A full reconciliation between quantum mechanics and special relativity requires QFT (in flat spacetime), which has been demonstrated numerous times in particle accelerators. However, QFT in curved space still awaits experimental corroboration. In Sec. \ref{sec:QFT:CS}, we discuss proposals to test its key predictions using Bose-Einstein condensates.  Although QFT in curved spacetime enables the study of some effects at the interface of quantum physics and GR, including entanglement and decoherence (see Sec. \ref{sec:grav:dec:section}), a full reconciliation between the theories must include the effects of quantum matter on the background metric itself. These effects, known as back-reaction, are out of the scope of QFT in CS. That is, QFT in CS is limited by a semi-classical description where the spacetime is assumed to be a classical background given by Einstein's equations, and only fields are quantized. Ultimately, the difficulty in including back-reaction in a covariant theory of quantum fields is the main difficulty in developing a theory of quantum gravity.

\subsubsection{Harmonic oscillator in the presence of gravity using the Klein-Gordon equation} 

An alternative approach to describe a harmonic oscillator in the presence of a gravitational field beyond the Newtonian approximation is to use a Klein-Gordon equation and the Klein-Gordon inner product, which, as introduced in the section above, are compatible with both general relativity and quantum physics. The Klein-Gordon equation describes a scalar field in curved spacetime. However, a single particle (such as an atom) in the presence of the gravitational field of a spherical mass can be described by restricting the solutions to the single particle sector and using the Schwarzschild metric~\cite{marzlin1995dipole, sonnleitner2018mass, schwartz2019post2, huimann2020quantum}. The problem with this approach is that the Klein-Gordon equation does not have a trapping potential term. 

A solution to this was proposed in~\cite{huimann2020quantum} by designing an effective spacetime metric that included not only the external gravitational field but also mimicked the relevant features of an oscillating trapping potential. The effective metric reduces to the Newtonian potential in the non-relativistic approximation, and the equation reduces to the Schrödinger equation of a harmonic oscillator in the presence of Newtonian gravity. However, solving the equation beyond this approximation is very challenging, and only some solutions are possible in special cases.

An alternative approach which also uses a restriction of the dynamics to the single particle sector,
considers a classical system coupling to gravity via minimum coupling and then quantizes the system via canonical quantization~\cite{schwartz2019post1}. This approach was inspired on work computing relativistic corrections of an atom interacting with the electromagnetic field~\cite{sonnleitner2018mass} and on studies of the dipole coupling between a system of N particles with total charge zero and the electromagnetic field in the presence of a weak gravitational field~\cite{marzlin1995dipole}. More recently, a full first-order post-Newtonian expansion has been performed in~\cite{schwartz2019post2}.

\subsubsection{Entanglement and decoherence in non-inertial frames and black holes} \label{sec:non-inertial}
Entanglement is a key notion in quantum mechanics and is often regarded as the true herald of quantumness. In a single inertial frame, the Schrödinger equation readily describes how two subsystems interacting via a potential gradually become entangled. However, describing entanglement in relativistic settings is more complicated. Entanglement strongly depends on the notion of subsystem and bipartition. In the quantum theory, subsystems can always be defined independently of the observer. As a consequence, entanglement is conserved for moving observers. 

In the relativistic case, entanglement is only invariant in flat spacetime and if observers move with constant velocity. Interestingly, it was shown that the Minkowski vacuum contains spatial correlations that can produce entanglement between initially uncorrelated atoms interacting with the vacuum state~\cite{valentini1991non, reznik2003entanglement,wang2021coherently}. 

Consider two inertial observers in flat spacetime who are performing an experiment to determine the degree of entanglement between two particles, such as two photons or two fermions. We assume that they find that the systems are maximally entangled. If two uniformly accelerated observers try to determine the degree of entanglement in the same system, they find that there are many particles isntead of just two. The notion of the system's bipartition is lost due to the Fulling–Davies–Unruh effect, which was introduced in Sec. \ref{sec:QFTCS}. The inertial vacuum state corresponds to a thermal state for uniformly accelerated observers.  In QFT, a well-defined notion of the subsystem is only possible when global bosonic or fermionic modes with sharp frequency are considered. This is because the frequency is invariant, although the number of particles in the mode varies with acceleration. For uniformly accelerated observers, a region of spacetime becomes inaccessible, and global states become more mixed at higher accelerations, decreasing entanglement. Non-uniform motion and, thus, gravity produce decoherence~\cite{fuentesschuller2005alice}. For localized systems, such as moving cavities \cite{bruschi2012voyage} or propagating wave packets in curved spacetime \cite{bruschi2014spacetime}, motion and gravity can either degrade states or create entanglement \cite{friis2013relativistic}.

In curved spacetime, the situation is even more complex because inertial observers disagree with the particle content of the field. As a consequence of this, there is no well-defined notion of entanglement in curved spacetime. Entanglement between global modes can only be studied in spacetimes that have asymptotically flat spacetimes such as black holes \cite{adesso2009correlation,jing2023bosonic,wu2023classifying} and cosmological toy models \cite{ball2006entanglement}.

The study of the observer-dependent nature of entanglement~\cite{alsing2012observer} in relativistic settings has been a topic of interest in the field of {\it relativistic quantum information}. For an overview of the field, see the special issue~\cite{mann2012relativistic}. The field is concerned with studying relativistic effects on quantum technologies, including quantum communications ~\cite{bruschi2014spacetime}, and on addressing fundamental questions in quantum field theory ~\cite{lopp2018light,wu2023quantum,barman2023spontaneous}, black holes~\cite{Ng2022little}, cosmology~\cite{bubuianu2021kaluzaklein} and high-energy physics~\cite{bertlmann2001bell,naikoo2020quantum} with an information-theoretical perspective.   

A good example where notions of quantum information are applicable to fundamental questions is the well-known \textit{information loss paradox} in black holes.  Information stored in pure states in the spacetime of a black hole is lost due to states becoming completely mixed after the black hole evaporates via Hawking radiation. Here, the interplay of quantum field theory and general relativity leads to a paradox, the resolution of which seems to require giving up fundamental principles such as unitarity, locality, or the equivalence principle. Quantum fields in black hole spacetimes give rise to one of the starkest indications of the incompatibility of quantum theory and general relativity. A large amount of work has focused on addressing this problem using quantum information, see for example the recent papers~\cite{yoshida2019firewalls,penington2020entanglement}. The question becomes: could entanglement carry the lost information out of the black hole? The distribution of entanglement, via the monogamy of entanglement, between modes inside and outside of the black hole~\cite{adesso2009correlation} could play a role in the potential resolution to the paradox~\cite{merali2013astrophysics}. However, this resolution requires entanglement to be somehow broken at the horizon. It was conjectured that observers falling into a black hole encounter a firewall made of high-energy quanta at (or near) the event horizon, which breaks the entanglement~\cite{almheiri2021entropy}. However, there is still an ongoing discussion in the community on whether this resolves the matter or not.

\subsubsection{Perturbative quantum gravity} \label{sec:QLG:and:PQG}

In the preceding sections, we have assumed that gravity is a background gravitational field obtained by solving Einstein's equations with classical sources (for example, the background gravitational field created by the Earth). Such analysis does, however, not take into account that the quantum matter (i.e., the quantum system in the laboratory) can also be a source of gravity. This effect is known as \textit{gravitational backreaction} and is one of the many challenging problems that a fully-fledged quantum theory of gravity should address. The backreaction from a quantum system could be naively included in Einstein field equations with both the spacetime metric and the stress-energy tensors promoted to quantum operators.  However, when we try to perturbatively quantize gravity, we are faced with the problem of an infinite number of free parameters coming from the high-energy regime that need to be fixed using experimental data, i.e., we get a theory that is non-renormalizable and thus does not have predictive power. The full quantization of gravity is an open problem~\cite{weinberg1980ultraviolet,niedermaier2007asymptotic,reuter2007Functional,shomer2007pedagogical}.  

Instead, we here limit the discussion of the gravitational backreaction to the perturbative regime of gravity at low energies. In this regime, general relativity can be quantized by following analogous steps, as with any other field theory. This was done in the seminal paper~\cite{gupta1952quantizationA} using the Gupta–Bleuler formalism~\cite{Gupta1950Theory,Bleuler1950Eine} applied to the approximate linear form of Einstein’s gravitational field and later generalized beyond the linear case~\cite{gupta1952quantizationB}. Quantum general relativity can be treated as an effective field theory (EFT) at low energies using the covariant Feynman path integral approach, which allows making predictions without the full knowledge of the theory at high energies~\cite{donoghue1994general}. Within this framework, we first expand the metric $g_{\mu \nu}$ as
\begin{align}  \label{eq:perturbative}
    \hat{g}_{\mu \nu} = \eta_{\mu \nu} + \hat{h}_{\mu \nu}, 
\end{align}
where $\eta_{\mu \nu}$ is the Minkowski spacetime metric (or, in general, some other background gravitational field $\bar{g}_{\mu \nu}$), and $\hat h_{\mu \nu}$ contains the fluctuations of the metric which we quantize. Specifically, we can then obtain the graviton propagator:
\begin{equation} \label{eq:graviton}
    \frac{i P_{\mu\nu,\alpha\beta}}{k^2+i \epsilon},
\end{equation}
where $k^\mu$ is the four momenta ($k^2=k_\mu k^\mu$), and the projection operator is given by (in the harmonic gauge)
\begin{equation}
    P_{\mu\nu,\alpha \beta}=\frac{1}{2}\left( \eta_{\mu\alpha}\eta_{\nu\beta}+\eta_{\mu\beta}\eta_{\nu\alpha}-\eta_{\mu\nu}\eta_{\alpha\beta}\right).
\end{equation}
The interaction Lagrangian is given by:
\begin{align} \label{eq:grav_lagrang}
   \mathcal{L}_\text{int}=\frac{1}{2} \hat{h}^{\mu \nu}   \hat{T}_{\mu\nu},
\end{align}
where $\hat{T}_{\mu \nu}$ is now the stress-energy tensor produced by quantum systems. Starting from Eq.~\eqref{eq:grav_lagrang}, we can obtain matter-graviton vertices. In addition, we also have graviton-graviton vertices as the graviton couples to all energetic particles, including to itself.  Once the Feynman rules are obtained, we can then perform calculations similarly as done in other quantum field theories (see, for example, the book~\cite{scadron2006advanced}).

Let us suppose we have two non-relativistic massive quantum systems. We can write the corresponding stress-energy tensor as: 
\begin{equation}
    \hat{T}_{\mu\nu} =\hat{T}_{\mu\nu}^{(m)} +\hat{T}_{\mu\nu}^{(M)},
\end{equation}
where  $\hat{T}_{\mu\nu}^{(m)}$ ($\hat{T}_{\mu\nu}^{(M)}$) is the contribution from system of mass $m$ ($M$). Using perturbation theory in the EFT context discussed in Eqs.~\eqref{eq:perturbative}-\eqref{eq:grav_lagrang}, we can then find the corrections to the Newtonian potential:
\begin{align} \label{eq:corrected:Newt:pot}
    \hat{V} &= - \frac{G M m}{\hat{r}} \biggl[ 1 + 3\frac{G(M+m)}{\hat{r} c^2} + \frac{41}{10\pi} \frac{G \hbar}{\hat{r}^2 c^3} \biggr],  
\end{align}
where $\hat{r}$ denotes the distance between the two systems. The first term in Eq.~\eqref{eq:corrected:Newt:pot} is the tree-level contribution, while the second and third terms come from one loop Feynman diagrams. These latter terms have been calculated with three techniques: Feynman diagrams~\cite{bjerrumbohr2003quantum, kirilin2002quantum}, unitarity-based methods~\cite{bjerrumbohr2014shell, holstein2016analytical}, and dispersion relations~\cite{bjerrumbohr2003quantum}. In addition, this result in Eq.~\eqref{eq:corrected:Newt:pot}  applies to particles of any spin and is thus universal~\cite{holstein2008spin}. 

To conclude this section, we note that there are many ways in which gravitational effects can be incorporated into the dynamics of quantum systems. To establish which ones are accurate, we must ultimately be guided by experiments.

\subsection{Summary of challenges} \label{sec:postulates}

We have outlined ways in which gravity can be consolidated with quantum mechanics in a limited way, although many conceptual and mathematical challenges remain.  Here we summarize the challenges by examining the postulates of quantum mechanics one by one. For each challenge, we mention the resolution when one exists (for example, quantum field theory successfully combines quantum mechanics with special relativity). The remaining challenges must ultimately be determined by experiments.

\subsubsection{Quantum states and the superposition principle}
        
 The first postulate states that non-relativistic quantum mechanics (NRQM) in first quantization associates a Hilbert space with every quantum system by representing the states of a system with vectors in a Hilbert space. To preserve probabilities, physical states $\ket{\psi_j(t)}$ must be normalized with respect to the inner product  $\braket{\psi_j(t)}{\psi_\ell(t)}=\delta_{j,l}$, shown in Eq.~\eqref{eq:innerProduct}.
Physical quantities are given in terms of expectation values, which are evaluated using this inner product.
A quantum superposition corresponds to a state $\ket{\Psi(t)}$ that is a linear combination of basis states $\ket{\psi_j(t)}$ and amplitudes $c_j$,
        \begin{align}\label{eq:superpositionState}
            \ket{\Psi(t)} = \sum_jc_j\ket{\psi_j(t)}, 
        \end{align}
in which $\sum_j\lvert c_j\rvert^2=1$ ensures that the superposition state is normalized, hence allowing for a probabilistic interpretation of the theory.
Any such superposition remains a valid quantum state.
        
Several conflicts between this postulate and relativity can be identified:

        \begin{enumerate}
        \item[(i)] To satisfy Lorentz invariance, space and time must enter on an equal footing.
        As mentioned in Sec.~\ref{sec:relativistic:Schrödinger}, the inner product in Eq.~\eqref{eq:innerProduct} is not Lorentz invariant~\cite{birrell1982quantum}, which implies that physical quantities in NRQM are not compatible with physical quantities in relativity. 
        
        \item[(ii)] The wave functions $\ket{\psi_j(t)}$, $\ket{\psi_\ell(t)}$ in Eq.~\eqref{eq:innerProduct} and \eqref{eq:superpositionState} are evaluated at equal times $t$. Time enters as a (global) parameter, while in special and general relativity, it is a relative concept that depends on the given world line. 
        Furthermore, whereas quantum states can be in a superposition of several spatial locations, in curved spacetime, time can pass at different rates in different locations.

        \item[(iii)] In both special and general relativity, the times at which events occur are observer-dependent. For space-like events, the order in which they occur may change. Such explicit notions of causality are not part of the framework of NRQM but must instead be added by hand. 
        \item[(iv)] It has been argued that the superposition principle is in conflict with the principle of covariance \cite{penrose1986gravity,penrose1996gravity} and with the equivalence principle \cite{howl2019exploring}. An argument challenging this view has been recently put forward~\cite{giacomini2022quantum}.
        \end{enumerate}
As we saw in Sec.~\ref{sec:QFTCS}, some of these points can be addressed by moving to quantum field theory and considering fields rather than particles.

\subsubsection{Quantum state evolution} \label{sec:postulate:II}
        
The time evolution of quantum states in NRQM is given by the Schrödinger equation in Eq.~\eqref{eq:Schrodinger:equation}. There are a number of conflicts with general relativity:

\begin{enumerate}
    \item[(i)]
    As we saw in Sec.~\ref{sec:QFTCS}, Lorentz invariance requires that derivatives with respect to time and space are of the same order, which is not the case for the Schrödinger equation in Eq.~\eqref{eq:Schrodinger:equation}. However, relativistically invariant versions of the Schördinger equation, such as the Klein-Gordon equation (Eq.~\eqref{eq:KleinGordonEquation}) or the Dirac equation, in conjunction with relinquishing the notion of single-particle states are needed to overcome this inconsistency, as illustrated by quantum field theory. 

    \item[(ii)] In quantum mechanics, energies are quantized; while they are not in general relativity, they are closely related to mass and the metric through Einstein's field equations. Two possible approaches to this apparent conflict are to either (i)~'quantize gravity', i.e.~develop a theory of quantum gravity in which the gravitational field is quantized, or (ii)~to 'gravitize quantum mechanics', i.e. to preserve the principles of general relativity, such as the equivalence principle, to modify quantum mechanics. The question of how to resolve these issues remains very much open. 

    \item[(iii)] As detailed in Sec.~\ref{sec:non-inertial}, the black hole information paradox (see~\cite{Raju2022Lessons} for a review) poses another challenge as it seems to require giving up unitarity. 
    This question similarly remains open. 
\end{enumerate}

\subsubsection{Quantum measurements}
The process of performing measurements in general relativity is straightforward, and up to limitations due to the measurement apparatus, we assume that we can measure with arbitrary precision.
However, in quantum mechanics, (projective) measurements are performed according to the Born rule: possible measurement outcomes are the eigenvalues $\lambda_j$ of Hermitian operators $\hat A$ (observables) and the associated probability to observe this measurement result is the projection of the system's state $\ket{\psi}$ onto the associated eigenstate $\ket{\lambda_j}$ of the observable, $\lvert\braket{\lambda_j}{\psi}\rvert^2$.
The fact that we measure observables that do not necessarily commute imposes limits to the precision with which we can measure different observables at the same time, most famously captured in the Heisenberg uncertainty for position and momentum
\begin{align}\label{eq:HeisenbergUncertainty}
    \mathrm{var}(\hat x(t))\, \mathrm{var}(\hat p(t)) \geq \frac{\hbar^2}{4}.
\end{align}
Several conflicts with general relativity arise from this statement: 
\begin{itemize}
    \item[(i)]~In the context of relativity, we do not encounter the same limitations on measurement precision. Classical variables can be measured to arbitrary precision without state-update resulting from the measurement. 
    \item[(ii)] Without the measurement postulate (that is, external observers), there are no events in quantum mechanics. On the other hand, both special and general relativity are fundamentally based on the notion of events. Quantum superpositions are not compatible with the notion of a single event, such as a measurement, in spacetime. There have, however, been proposals for an event-based formulation of quantum mechanics, which fundamentally modifies the Born rule~\cite{giovannetti2023geometric}. 
    \item[(iii)] Problems also arise in QFT in CS. On one hand, the theory inherits the measurement problem from quantum theory, and on the other hand, new problems arise due to causality. Here, it has been shown that projective measurements on quantum fields lead to faster-than-light signaling~\cite{sorkin1993impossible}. Finding ways to give a resolution to this problem is an active research field~(see, for example, ~\cite{fewster2020quantum}). 
\end{itemize}

Finally, the measurement problem in quantum mechanics, which states that there is no consistent dynamic description of the measurement process, also applies in the context of gravity. The issue is partially addressed by collapse theories, which, while they have not yet been experimentally verified, propose a dynamical mechanism (see Section~\ref{sec:grav:collapse}). 

\subsubsection{Composite quantum systems and entanglement}
In quantum mechanics, we use the tensor product to compose a system out of multiple subsystems, e.g. $\ket{\psi}=\ket{\psi}_A \otimes \ket{\psi}_B$.
We saw in Sec.~\ref{sec:QFTCS} that in quantum field theory in curved spacetime, the definition of sub-systems is problematic since the notion of particle number is observer-dependent~\cite{fuentesschuller2005alice,alsing2012observer}. We have already identified the crucial issues, which are:

\begin{itemize}
    \item[(i)]  A consequence of the tensor product structure for composite systems in quantum theory is that multi-partite systems can be entangled, which means that entanglement becomes an observer-dependent quantity. In the famous EPR paradox, this leads to a violation of causality of locality, i.e.~we need to allow for faster-than-light effects if the theory is to remain local. 
    \item[(ii)] The notion of entanglement requires the Hilbert space partition to be well-defined. This is commonly done in terms of particles or modes. However, in curved spacetime, the notion of particles is ill-defined. Generally, different inertial observers in curved space see different particle content in the field. Particles can only be well defined in rare spacetimes in which the metric is globally invariant under spatial translations or in which spacetime has regions where the metric has this symmetry. In most cases, it is not possible to define a Hilbert space partition and study entanglement in composite systems.
\end{itemize}

The challenges listed here are all difficult to address, and in many cases their resolution should ultimately be determined by experiments. To enable these experiments, we now proceed to review tools and methods used to model such systems in the laboratory.


%
\section{Theoretical frameworks for modeling massive quantum systems in the laboratory} \label{sec:theory:tools}

%
To test the effects of gravity, which are often extremely small, with massive quantum systems, it is crucial to model the proposed experiment accurately.  
Here, we briefly account for common theoretical tools used to describe mechanical resonators in the laboratory. Firstly, we discuss ways in which a probe can interact with the massive system (Sec.~\ref{sec:QOMS:Hamiltonian}). We then cover models of open-system dynamics (Sec.~\ref{sec:open:system:dynamics}), which are needed to model the experiments. We cover measurements and control schemes necessary for readout (Sec.~\ref{sec:control}), as well as quantum metrology tools (Sec.~\ref{sec:sensing}). A snapshot of experiments with massive quantum systems is given in Sec.~\ref{sec:State-of-the-art} for testing gravity and for generating quantum states in Sec.~\ref{sec:controlling:massive:systems}.

%
\subsection{Coupling a mechanical mode to a probe} \label{sec:QOMS:Hamiltonian}
%
A key challenge in controlling massive systems in the laboratory is the fact that they often cannot be measured directly. In order to manipulate and control these massive systems, we must first couple them to a probe. We review two such models here.

\subsubsection{Optomechanical interaction} \label{sec:optomechanical:Hamiltonian}
We first consider the case where a mechanical mode couples to a cavity mode (which can be optical, microwave, electrical, or magnetic). This brief exposition largely follows the review~\cite{aspelmeyer2014cavity}. See also~\cite{barzanjeh2022optomechanics} for further reading. In many systems, the frequency of the cavity mode depends on the center-of-mass position of the mechanical resonator. When consider the frequency shift to first-order in the position of the mechanical oscillator, we acquire a coupling between the occupation number of the cavity mode and the position of the oscillator. The derivation often depends on platform-specific details. For example, in a Fabry-P\'{e}rot moving-end mirror cavity, the cavity deforms due to photon pressure~\cite{law1995interaction}. For levitated nano-particles, the light-matter interaction can instead be derived by assuming the trapped sphere to be smaller than the laser waist of the beam~\cite{romeroisart2011optically}. Similarly, in some electro-mechanical systems, the motion of the resonator couples to capacitance which in turn induces a frequency shift~\cite{regal2011cavity}. 

In all of the cases above, we arrive at the following \textit{cavity optomechanical Hamiltonian} (we denote it in this way even though the cavity field might not be an optical mode) 
\begin{align} \label{eq:optomechanical:hamiltonian}
    \hat H = \hbar \omega_{\mathrm{c}} \hat a^\dag \hat a + \hbar \omega_{\mathrm{m}} \hat b^\dag \hat b - \hbar g_0 \hat a^\dag \hat a \bigl( \hat b^\dag + \hat b \bigr), 
\end{align}
in which $\hat a$ and $\hat a ^\dag$ are the annihilation and creation operators for the radiation mode with free angular frequency $\omega_{\mathrm{c}}$, and where $\hat b$ and $\hat b^\dag$ are the annihilation and creation operator for the mechanical mode with free angular frequency $\omega_{\mathrm{m}}$. The operators obey the canonical commutator relation $[\hat a, \hat a^\dag] = [\hat b, \hat b^\dag] = 1$. The optomechanical coupling $g_0$ has units of angular frequency and encodes the strength of the interaction between the cavity and mechanical modes. In most experimental realizations, the coupling is defined as the optical frequency shift per displacement $g_0 \equiv - x_{ZPF}\partial\omega_c /\partial x$, where $x_{ZPF} = \sqrt{\hbar/2m \omega_m}$ is the zero-point fluctuation. The probe fields act as the means for both readout and control. Crucially,  the interaction between the bosonic mode and mechanical resonator enables the detection of extremely small displacements, e.g.~due to gravitational effects. 
The nonlinear quantum dynamics generated by the Hamiltonian in Eq.~\eqref{eq:optomechanical:hamiltonian} were first solved in~\cite{bose1997preparation,mancini1997ponderomotive}, where it was shown that both the cavity mode and the mechanical mode evolve into highly non-classical superpositions of coherent states. Particularly, this is a classic way to generate Schr\"{o}dinger cat states of the macroscopic mechanical mode~\cite{bose1999scheme,marshall2003towards,qvarfort2018gravimetry}. The solutions were later generalized to time-dependent couplings~\cite{qvarfort2019enhanced}. As detailed in Section~\ref{sec:gravitational:decoherence}, the system dynamics of this Hamiltonian have been used for a number of proposals related to the detection of gravitational decoherence.   

The Hamiltonian in Eq.~\eqref{eq:optomechanical:hamiltonian} describes an idealized system isolated from its environment. In a realistic setting, both the bosonic mode and mechanical modes undergo dissipation, thermalization, and decoherence (see Section~\ref{sec:open:system:dynamics} for details). 
To replenish the lost quanta from the radiation mode, an external source is used to pump the system. Such a pump is modeled with a bosonic pump term $\hat H_d = \alpha(t) \hat a + \alpha^*(t) \hat a^\dag$, where $\alpha(t)$ is a complex drive amplitude. However, with the inclusion of such a term, the dynamics induced by the Hamiltonian in Eq.~\eqref{eq:optomechanical:hamiltonian} can no longer be solved exactly~\cite{Qvarfort2022Solving}. A common method to proceed is to solve the system dynamics perturbatively, or by examining the steady-state for weak driving, see e.g.~\cite{rabl2011photon,nunnenkamp2011single}. 

For a strong enough pump, the system dynamics can be approximated as linear. Here, the term `linear' refers to the fact that the resulting Heisenberg equations of motion contain only linear operator terms. The inclusion of a pump term (strongly) driving mode $\hat a$ lets us separate $\hat a$ into the classical amplitude of the drive $\alpha$ and the fluctuations $\delta \hat a$, such that $\hat a = \alpha + \delta \hat a$\footnote{Depending on the application, we may also consider $\hat a = \langle \hat a\rangle + \delta \hat a$ and $\langle\delta \hat a\rangle=0$. This leads to a Hamiltonian of a similar form, see~\cite{aspelmeyer2014cavity}.}.
The interaction term in Eq.~\eqref{eq:optomechanical:hamiltonian} becomes
\begin{align}
    \hat H_I = - \hbar g_0 ( \alpha + \delta \hat a )^\dag ( \alpha + \delta \hat a )( \hat b^\dag + \hat b). 
\end{align}
 When $|\alpha| \gg \langle\delta\hat a \rangle$, the cubic term $- \hbar g_0 \delta \hat a^\dag \delta \hat a$ can be removed because it is smaller by a factor of $|\alpha|$ than the other terms. The remaining linear Hamiltonian is
\begin{align} \label{eq:linearized:optomechanical:Hamiltonian}
    \hat H_I \approx - \hbar g_0 ( \alpha^* \delta \hat a  + \alpha \delta \hat a^\dag )( \hat b^\dag + \hat b). 
\end{align}
This Hamiltonian is a common starting point for a number of investigations and accurately describes a range of experiments (see references in~\cite{aspelmeyer2014cavity}). For example, within the resolved sideband regime where $\omega_m \gg \kappa$, it is possible to engineer either a beam-splitter interaction $\hat a^\dagger \hat b+\mathrm{h.c.}$ or a two-mode squeezing term $\hat a^\dagger \hat b^\dagger+\mathrm{h.c.}$ by optically pumping on either the red ($\omega_c - \omega_m$) or blue ($\omega_c + \omega_m$) sideband. The red-sideband interaction is necessary for implementing e.g.~side-band cooling~\cite{liu2013review}. Another way to couple an optical and mechanical mode is through a dissipative coupling rather than a dispersive one, where the displacement of the mechanical resonator directly modulates the decay rate of the cavity~\cite{elste2009quantum}. 

The dynamics of the nonlinear Hamiltonian in Eq.~\eqref{eq:optomechanical:hamiltonian} cannot be solved exactly in the presence of a pump term and optical dissipation. However, by engineering the system such that the optical mode dissipates from the cavity on a time scale much faster than that of the mechanical element, the interaction between the optical and mechanical modes can instaed be described as an instantaneous interaction. This is also known as the unresolved sideband regime, where $\omega_m \ll \kappa$. The framework, often referred to as \textit{pulsed optomechanics} was developed in~\cite{vanner2011pulsed}. 
By considering the Langevin equations in the unresolved sideband regime, the light-matter interaction can be modeled as an instantaneous unitary operator of the form $
\hat U =  e^{i \mu \hat n_l \hat X_m}$ \cite{pikovski2012probing},
where $\mu$ is a dimensionless coupling which depends on the pulse shape, $\hat n_l$ is the number of photons in the pulse entering or leaving the cavity, and $\hat X_m = (\hat b^\dag+ \hat b )/\sqrt{2}$ is the mechanical quadrature. For an adiabatic cavity with $\kappa \gg \tau_{in}^{-1}$, where $\tau_{in}$ is the characteristic timescale of the input pulse, the value of $\mu$ becomes $\mu = \sqrt{8} g_0/\kappa$, where $g_0$ is the optomechanical coupling and $\kappa$ is the optical dissipation rate~\cite{clarke2023cavity}. 
It is also possible to start from the linearized optomechanical Hamiltonian in Eq.~\eqref{eq:linearized:optomechanical:Hamiltonian} and derive a pulsed interaction that couples the position quadratures of the mechanical mode and the probe field~\cite{khosla2013quantum, bennett2016quantum}. The resulting unitary operator is $
    \hat U_{\mathrm{Lin}} = e^{i \chi  \hat X_c \hat X_m }$,
where $\chi$ again depends on the pulse shape, and where $\hat X_c$ is the amplitude quadrature of the input pulse (as opposed to the cavity quadrature). In the adiabatic regime and for an input coherent pulse,  $\chi = 4 g_0 \sqrt{N}/ \kappa$, where $N$ is the average number of photons in the input pulse~\cite{vanner2011pulsed}.  A closely related idea to  pulsed optomechanics is that of stroboscopic optomechanics~\cite{brunelli2020stroboscopic}, where a train of short pulses of light is injected into the cavity. 
Proposals using nonlinear pulsed optomechanics include the generation of cat-states~\cite{clarke2018growing,ringbauer2018generation},  entangled states~\cite{clarke2020generating,neveu2021preparation}, and entangled cat-states~\cite{kanarinaish2022two}. State-preparation using linearized pulsed optomechanics to generate cat-like states has also been put forward using the addition and subtraction of phonons from the mechanical state \cite{milburn2016nonclassical} and by swapping the mechanical state with a photon-subtracted state of light \cite{hoff2016measurement}. 
Pulsed optomechanics has given rise to a number of protocols intended to test fundamental physics, such as tests of modified commutator relations detailed in Section~\ref{sec:GUP}.

\subsubsection{Coupling to a two-level system } \label{sec:two:level:coupling} 
Instead of a probe field, it is also possible to couple the mechanical resonator to a two-level system. The advantage of such a coupling is that the high level of control that has been achieved for two-level systems can now be indirectly applied to the mechanical resonator.  Collectively, these systems are sometimes known as \textit{hybrid optomechanical systems} since they couple bosonic continuous degrees-of-freedom to a two-level qubit system.  Examples include nitrogen-vacancy centers (NV) embedded into a nanodiamond~\cite{neukirch2015multi, hoang2016electron}, or superconducting resonator qubits coupled to mechanical modes~\cite{oconnell2010quantum, chu2018creation, satzinger2018quantum, arrangoizarriola2019resolving}. See e.g.~\cite{rogers2014hybrid, chu2020perspective} for dedicated reviews.  

In the case of a spin coupled to mechanical motion, the same notions apply to the coupling between the spin of an ion and its center-of-mass position~\cite{cirac1995quantum}. Applying a magnetic field gradient to a trapped ion couples the internal and motional states of the system through the Zeeman effect. The spin-mechanical Hamiltonian reads, to first order in the position operator, 
\begin{align}
    \hat H =\hbar \omega_{\rm{m}} \hat b^\dag \hat b + \frac{1}{2}\hbar \omega_0 \hat \sigma_z  +  \frac{\hbar  \lambda}{2}  ( \hat b^\dag + \hat b) \hat \sigma_z, 
\end{align}
where $\omega_0$ is the angular frequency of the qubit system, $\sigma_z$ is the Pauli operator denoting the free energy,  $\hat b, \hat b^\dag$ denote the annihilation and creation operators of the mechanical mode, and $\lambda$ is a coupling constant that depends on the platform in question. 

Superconducting qubits coupled to a mechanical mode are more commonly modeled using the Jaynes-Cummings Hamiltonian:
\begin{align}
    \hat H_{JC} = \hbar \omega_{\rm{m}} \hat b^\dag \hat b + \hbar \Omega \frac{\sigma_z }{2} + \frac{\hbar \lambda}{2} \left( \hat b \sigma_+ + \hat b^\dag \sigma_- \right), 
\end{align}
where $\Omega$ is the oscillation frequency of the superconducting qubit, $\sigma_z$ is the Puali matrix, while $\sigma_- = \sigma_x - i \sigma_y $ and $\sigma_+ = \sigma_x + i \sigma_y$ are the raising and lowering operators in terms of Pauli matrices. Here $\lambda$ again denotes the strength of the coupling between the mechanical mode and the qubit. 
A number of theoretical proposals have utilized the two-level system coupling for e.g.~enhancing the optomechanical coupling strength~\cite{heikkila2014enhancing, pirkkalainen2015cavity}, state preparation~\cite{yin2013large, kounalakis2020flux}, as well as cooling~\cite{martin2004ground, hauss2008dissipation, jaehne2008ground, nongthombam2021ground}.

\subsection{Open-system dynamics for massive quantum systems} \label{sec:open:system:dynamics}

To detect the extremely weak effects of gravity, we must be able to distinguish them against any underlying noise floor. Additionally, proposals such as gravitational decoherence and gravity-induced state reduction stipulate that gravity itself manifests as a noise signature (see Sec.~\ref{sec:gravitational:decoherence}). Both of these considerations necessitate the use of accurate noise models for massive quantum systems. Here we review the main tools for modeling decoherence in massive quantum systems. For a dedicated review on noise models for mechanical resonators in the quantum regime, see~\cite{bachtold2022mesoscopic}.

\subsubsection{Quantum master equations} \label{sec:master:equations}

Quantum master equations describe the quantum state evolution in a situation in which a system, e.g., a mechanical oscillator or other probes, is coupled to a larger environment that we cannot control or measure, such as a thermal bath. Apart from modeling the interaction of a probe with its environment, this description is relevant for describing gravitational decoherence, gravitational collapse, see Sec.~\ref{sec:grav:dec:section}, and gravitational entanglement, see Sec.~\ref{sec:entanglement:mediated:gravity}. 

The dynamics of a quantum system coupled to the environment, or bath, can be described with the Hamiltonian $\hat H$, which contains a term for the system dynamics $\hat H_\mathrm{s}$ (e.g., a quantum system in the laboratory, such as a cavity, a mechanical oscillator or atoms), the environment or bath $\hat H_\mathrm{b}$ (e.g., a thermal bath) and a term describing the coupling between system and bath $\hat H_\mathrm{sb}$. The full Hamiltonian is
\begin{align}\label{eq:systemBathCoupling}
    \hat H & = \hat H_\mathrm{s} + \hat H_\mathrm{b} + \hat H_\mathrm{sb}.
\end{align}
The fully evolved state of the system given an initial state $\ket{\Psi_0}$ is  $\ket{\Psi(t)} = \hat U(t) \ket{\Psi_0}$, where $\hat U(t) = e^{- i \hat H t/\hbar}$. In principle, any type of system-bath coupling $\hat H_\mathrm{sb}$ is possible, but they might not always lead to analytically solvable dynamics~\cite{Gardiner2000Quantum}. 

In the laboratory, we often do not have access to the bath degrees-of-freedom which are therefore traced out from the quantum state. The result is a mixed state, where the degree of mixedness is captured by the \emph{purity} $\tr\left( \hat \rho^2 \right) \leq1$, saturating the bound $\tr \left( \hat \rho^2\right)=1$ only when the state is pure.
	
We now consider a dynamical equation for the evolution of the reduced system density matrix $\hat \rho_{\mathrm{s}}$~\cite{breuer2002theory}. Such an equation is known as a \textit{master equation}. To derive the simplest possible master equation, we assume (i) that the system and bath are initially in a product state $\hat \rho_{\mathrm{s,b}} = \hat \rho_{\mathrm{s}} \otimes \hat \rho_{\rm{b}}$, (ii) that the coupling between the system and the bath is weak (also known as the Born approximation), (iii) that the environment does not retain a memory of the interactions (the Markov approximation), and, (iv), that fast-rotating terms can be discarded (the secular approximation). By tracing out the bath modes $\hat b_\ell$, we obtain the Gorini-Kossakowski-Sudarshan-Lindblad master equation for the evolution of $\hat \rho_{\mathrm{s}} (t)$~\cite{gorini1976completely,lindblad1976generators} (commonly just referred to as the Lindblad equation): 
	\begin{align}\label{eq:Lindblad:equation}
		\dot{\hat{\rho}}_{\mathrm{s}}(t) & = -\frac{\mathrm{i}}{\hbar}[\hat H_\mathrm{s}, \hat \rho_{\mathrm{s}}(t)] \nonumber \\
  &\qquad + \sum_\ell \left(\hat  L_\ell \hat \rho_{\mathrm{s}}(t) \hat L_\ell^\dagger-\frac{1}{2}\{\hat \rho_{\mathrm{s}}(t), \hat L_\ell^\dagger \hat L_\ell\} \right). 
	\end{align}
Here, $\hat H_\mathrm{s}$ is the system Hamiltonian and $\hat L_\ell$ denote the Lindblad \emph{jump operators}. The Lindblad equation can also be written in shorthand as  $\dot{\hat{\rho}}_{\mathrm{s}} = -i [\hat H_{\mathrm{s}}, \hat \rho_{\mathrm{s}}]/\hbar + \sum_\ell \hat{\mathcal{D}}[\hat L_\ell ] \rho_\mathrm{s}(t)$, 
where $
    \hat{\mathcal{D}} [\hat L_\ell] \hat \rho(t) =  \hat L_\ell \hat \rho(t) \hat L_\ell^\dag - \hat L_\ell^\dag \hat L_\ell \hat \rho(t)/2  - \hat \rho(t) \hat L_\ell^\dag \hat L_\ell/2$
is called the standard dissipator. 

For mechanical resonators coupled to probe fields or two-level systems, there are commonly two types of noise that affect the system: dissipation and scattering processes in the probe field, and thermalization processes in the phononic modes~\cite{aspelmeyer2014cavity}.
Mechanical dissipation and thermalization arise due to processes specific to the system. In clamped systems, for example, unwanted thermal excitations are transferred via the physical point of attachment. Both optical and mechanical noise can be modeled with the Langevin equation in the linear optomechanical regime (see Sec.~\ref{sec:langevin:equation}). A common way to model Markovian dissipation and thermalization is to assume Lindblad operators of the form $\hat L_1 = \sqrt{(1 + n_{\mathrm{th}}) \Gamma} \hat b$ and $\hat L_2 = \sqrt{\Gamma n_{\mathrm{th}}} \hat b^\dag$, where $\Gamma$ is the mechanical linewidth and $n_{\mathrm{th}} = [\mathrm{exp}(\hbar \omega /k_\mathrm{B} T) - 1]^{-1}$ is the thermal occupation number of the bath, for which $\omega$ is the bath frequency, $k_\mathrm{B}$ is Boltzmann's constant, and $T$ is the temperature of the bath. 

In levitated systems, environmental noise arises in part due to collisions between the system and surrounding gas particles. A number of such processes correspond to position localization, which can be described with the following master equation~\cite{romeroisart2011quantum}:
\begin{align}
\bra{x}\dot{\hat{\rho}}(t) \ket{x'} &= \frac{i}{\hbar} \bra{x} [\hat \rho(t), \hat H] \ket{x'} \nonumber \\
&\quad - \Gamma(x-x') \bra{x}\hat \rho(t) \ket{x'}.
\end{align}
Here, the form of $\Gamma(x-x')$ depends on the nature of the noise. 
In the limit where the decoherence decay depends quadratically on $|x -x'|$, the equation simplifies to 
\begin{align}
\dot{\hat{\rho}}(t)  = \frac{i}{\hbar} [\hat \rho(t), \hat H]  - \Lambda [\hat x, [\hat x, \hat  \rho(t)]], 
\end{align}
where $\Lambda$ is the dissipation rate. The dynamics has been solved for a coupling o a probe field in~\cite{bassi2005towards} using a stochastic unraveling method (see~\cite{adler2005towards} for details).

We now briefly discuss generalizations to the strong coupling and the non-linear regime as well as to non-Markovian dynamics. 
Commonly, dissipation of the probe field and mechanical mode are treated separately. However, such a treatment is not always valid when the probe field and mechanics are strongly coupled. Instead, we must consider a dressed Lindblad equation~\cite{hu2015quantum}, which was solved in~\cite{torres2019optomechanical} using a damping basis approach~\cite{briegel1993quantum}.  In the nonlinear regime, a solution to the Lindblad master equation has been found for weak dissipation~\cite{mancini1997ponderomotive} and was later generalized to arbitrary $\kappa$~\cite{qvarfort2021master}, although a closed-form expression for the evolved state cannot be obtained. 
The Lindblad equation for an optomechanical system in the nonlinear regime was solved for dissipation of the mechanical mode~\cite{bose1997preparation, mancini1997ponderomotive}, with Lindblad operators $\hat L = \sqrt{\gamma} \hat b^\dag \hat b$, where $\gamma$ is the mechanical dissipation rate.

Models assuming a Markovian environment are often sufficient to capture the open dynamics of the system accurately. However, there are certain cases where a non-Markovian description is necessary. Non-Markovianity arises when the bath retains a memory of the interaction with the system, and information can flow back into the system~\cite{breuer2002theory}. Generally, to model such non-Markovian noise, we must either consider the Caldeira-Leggett master equation for Brownian motion~\cite{caldeira1983path} or solve a general non-Markovian master equation~\cite{hu1992quantum}, although some care must be taken since the dynamics predicted by these approaches do not automatically guarantee physical states as solution~\cite{Kohen1997Phase}. Non-Markovian dynamics can also be modeled using Lindblad-type equations with time-dependent noise rates~\cite{zhang2012general}. Some studies of mechanical resonators indicate the requirement for non-Markovian dynamics. For example, it was shown that in clamped systems, the resulting noise spectrum was consistent with a non-Markovian spectrum~\cite{groblacher2009observation}. In addition, theoretical works indicate that modeling the effects of damped tunneling two-level systems on a nanomechanical flexing beam resonator gives rise to non-Markovian noise~\cite{remus2009damping}.  
Optomechanical systems in non-Markovian environments have been modeled, although generally without the use of a master equation. A Feynman-Vernon influence functional method was used to study sideband cooling in non-Markovian environments~\cite{triana2016ultrafast}, and the influence of non-Markovian noise on the optomechanical nonlinearity was considered in~\cite{qvarfort2023enhanced}.

Beyond analytic solutions, master equations are often solved numerically. Useful tools include, for example, the \texttt{QuTiP} Python package~\cite{johansson2012qutip}. See also~\cite{campaioli2023tutorial} for a tutorial.

\subsubsection{Langevin equations and input-output formalism} \label{sec:langevin:equation}

The quantum Langevin equations model the non-unitary evolution of the quantum modes in the Heisenberg picture. The interaction between an input mode and the system is imprinted on output fields, which are detected in experiments. The Langevin equations also are a useful tool for modeling quantum metrology, see Sec.~\ref{sec:sensing}, and in particular experimental tests of gravity, e.g., weak-force detection, see Sec.~\ref{sec:quantum:enhanced:sensing}.

The Langevin equations for the bosonic field $\hat a_j$ read (see~\cite{caldeira1983quantum, Gardiner2000Quantum} for a derivation)
\begin{align}\label{eq:LangevinEquations}
	\dot{\hat{a}}_j & = -\frac{\gamma_j+\gamma_{j,\mathrm{in}}}{2} \hat a_j -\frac{\mathrm{i}}{\hbar} [\hat H_\mathrm{s}, \hat a_j] - \sqrt{\gamma_{j,\mathrm{in}}} \hat a_{j,\mathrm{in}}.
\end{align}
Note that following standard convention, $\hat a_{j,\mathrm{in}}$ has units of $\mathrm{s}^{-1/2}$. This input field could be noise from a (heat) bath or a coherent probe field. In a concrete setting, $\gamma_{j,\mathrm{in}}$ could be the coupling rate between the system and a waveguide used to couple the probe field into the system. If the system also experiences loss into other channels at rate $\gamma_j$, this rate is added to the overall decay rate. %

The input fields are connected the the output fields via input-output boundary conditions~\cite{Caves1982Quantum,Gardiner1985Input,clerk2010introduction}
\begin{align} \label{eq:input:output}
     \hat a_{j,\mathrm{out}} = \hat a_{j,\mathrm{in}} + \sqrt{\gamma_{j,\mathrm{in}}} \hat a_j.
\end{align}
By solving the internal system dynamics as a function of the input fields, it becomes possible to model the output fields purely as a function of the input fields. 

The Langevin equations for a cavity optomechanical system with Hamiltonian given by Eq.~\eqref{eq:optomechanical:hamiltonian} are~\cite{bowen2015quantum}
\begin{equation} \label{eq:langevin:equation:optomechanics}
\begin{split} 
    \dot{\hat{a}} &= - \frac{\kappa}{2} \hat a + i \Delta \hat a + i g_0 \hat a (\hat b^\dag + \hat b ) - \sqrt{\kappa_\mathrm{in}} \hat a_\mathrm{in} ,  \\
    \dot{\hat{b}} &= - \frac{\Gamma}{2} \hat b  - i \omega_\mathrm{m} \hat b + i g_0 \hat a^\dag \hat a - \sqrt{\Gamma_\mathrm{in}} \hat b_\mathrm{in}, 
\end{split}
\end{equation}
in which $\Delta=\omega-\omega_\mathrm{c}$ is the detuning of the light frequency $\omega_\mathrm{l}$ from the cavity frequency $\omega_\mathrm{c}$, Eq.~\eqref{eq:optomechanical:hamiltonian}, $\kappa$ is the optical linewidth, $\kappa_\mathrm{in}$ is the rate by the probe field dissipates away from the cavity, $\Gamma$ is the mechanical linewidth, and $\Gamma_\mathrm{in}$ is the coupling rate for thermal heating. These equations are challenging to solve in the nonlinear regime but can be linearized by considering a strong optical pump field (see Section~\ref{sec:optomechanical:Hamiltonian}). Such a treatment lies at the basis of many models of optomechanical systems~\cite{aspelmeyer2014cavity}.

In a typical experimental setting, we are interested in the frequency-dependent response to an input $\hat a_{j,\mathrm{in}}$, which we obtain by means of the Fourier transform from Eqs.~\eqref{eq:LangevinEquations}, $\hat a(\omega)\equiv\frac{1}{\sqrt{2\pi}} \int_{-\infty}^\infty \mathrm{d}t \, e^{i\omega t} \hat a(t)$. Experimentally, the Fourier transform is calculated over a finite time-window $[-\tau,\tau]$, which converges to the Fourier integral in the limit $\tau\to\infty$. The Fourier transform is also the main analytic method by which the Langevin equations can be solved, provided that they are linear in terms of the operators they contain. In that case, we can apply the Fourier transform to Eq.~\eqref{eq:LangevinEquations} to derive a scattering matrix 
\begin{align}
    	S(\omega) = \mathbb{1} + \sqrt{\gamma_\mathrm{in}} (\mathrm{i}\omega\mathbb{1}+M)^{-1} \sqrt{\gamma_\mathrm{in}}, 
\end{align}
where $\gamma_\mathrm{in}=\mathrm{diag}(\gamma_{1,\mathrm{in}},\dots,\gamma_{N,\mathrm{in}})$ contains the input noise terms and where the elements of  $M$ are defined from the Langevin equations in Eq.~\eqref{eq:LangevinEquations} as $\dot{\hat{a}}_j = \sum_\ell M_{j,\ell} \hat a_\ell - \sqrt{\gamma_j}\hat a_{j,\mathrm{in}}$. The scattering matrix allows us to relate the input and output fields as
\begin{align}
       \hat a_{j,\mathrm{out}}(\omega)=\sum_\ell S_{j,\ell}(\omega) \hat a_{\ell,\mathrm{in}}(\omega), 
\end{align}
where $S_{j, \ell}$ are the matrix elements of $S$.
The transmission between the $j$th input and the $\ell$th output port is given by $T_{\ell,j}(\omega)=\lvert S_{\ell,j}(\omega)\rvert^2$.
Interactions such as single- or two-mode squeezing can give rise to $\lvert S_{\ell,j}(\omega)\rvert^2\geq 1$, which is referred to as gain $\mathcal{G}=\lvert S_{\ell,j}(\omega)\rvert^2$.
These quantities are relevant for characterizing sensors, devices such as isolators, circulators, and (directional) amplifiers, and other scattering experiments such as optomechanical induced transparency (OMIT) experiments~\cite{xiong2018fundamentals}.

We now turn our attention to (quantum) noise in the context of the Langevin equations.
For a general treatment of noise, we refer to the designated review~\cite{clerk2010introduction} and specifically in the context of cavity-optomechanics to~\cite{aspelmeyer2014cavity}.

Quantum systems are typically hard to isolate and are susceptible to noise and dissipation. For instance, the real-time motion of a mechanical oscillator subjected to fluctuating thermal Langevin force was measured in~\cite{Hadjar1999High}.
These forces can be included straightforwardly in the Langevin equations via the input fields $\hat a_{j,\mathrm{in}}$, Eq.~\eqref{eq:LangevinEquations}.
Instead of recording real-time trajectories, it is typically more convenient to record the noise power spectral density $\mathcal{S}_{\hat O \hat O }(\omega)$ defined for some system operator $\hat O$. The spectral density describes the intensity of the noise at a given frequency.
In practice, we obtain $\mathcal{S}_{\hat O \hat O}(\omega)$ by averaging over many experimental runs. According to the Wiener-Khinchin theorem~\cite{wiener1930generalized,khintchine1934korrelationstheorie} this is equivalent to calculating the Fourier-transform of the auto-correlation:  
\begin{align}\label{eq:noiseSpectrum}
    \mathcal{S}_{\hat O^\dagger \hat O}(\omega) & \equiv \int_{-\infty}^\infty \mathrm{d}t \, e^{i\omega t} \langle \hat O^\dagger(t) \hat O(0)\rangle.
\end{align}
We obtain the noise spectral density from the Langevin equations in Eq.~\eqref{eq:LangevinEquations} by calculating correlators of, e.g., the fields $\langle \hat a_j^\dagger(\omega) \hat a_j(\omega)\rangle$, or position $\langle \hat x_j(\omega) \hat x_j(\omega)\rangle$, taking into account the input fluctuations $\hat a_{j,\mathrm{in}}(\omega)$.
The assumption we made of Markovian noise translates to the fact that $\hat a_\mathrm{in}$ are uncorrelated in time. This is also known as Gaussian white noise, and the vacuum fluctuations are given by 
\begin{align}
    \langle \hat a_\mathrm{in}(t) \hat a^\dagger_\mathrm{in}(t')\rangle & = (n_\mathrm{th} + 1)\delta(t-t'),  \\
    \langle \hat a^\dagger_\mathrm{in}(t) \hat a_\mathrm{in}(t')\rangle & = n_\mathrm{th}\delta(t-t'),
\end{align}
with the number of thermal bosonic excitations. 
Note that, for the case of an optical probe field, the environment corresponds to the vacuum, meaning that $n_\mathrm{th} =0 $.

As an example, let us consider the position noise of a single harmonic oscillator with frequency $\omega_\mathrm{m}$ and damping rate $\Gamma$. The spectral density is given by~\cite{clerk2010introduction}
\begin{align}\label{eq:spectralDensityOscillator}
    \mathcal{S}_{xx} = 2\pi x_\mathrm{xpf}^2 \bigg(n_\mathrm{th}(\hbar\omega_\mathrm{m}) & \frac{\Gamma}{(\omega_\mathrm{m}+\omega)^2+(\Gamma/2)^2} \notag \\
    + [n_\mathrm{th}(\hbar\omega_\mathrm{m}) + 1] & \frac{\Gamma}{(\omega_\mathrm{m}-\omega)^2+(\Gamma/2)^2}\bigg), 
\end{align}
where $n_\mathrm{th}(\hbar\omega_\mathrm{m})$ is the expected number of particles according to the Bose-Einstein statistic, and where $x_\mathrm{xpf} = \sqrt{\hbar /(2 \omega_\mathrm{m} m)}$ is the zero-point fluctuation, for which $m$ is the mass of the oscillator.
The area under the spectral density $\mathcal{S}_{\hat x \hat x}(\omega)$ with $\hat x$ the position operator is proportional to $\langle \hat x^2\rangle$.  In thermal equilibrium at large temperatures, $k_\mathrm{B}T\gg\hbar \Omega$, $\langle \hat x^2 \rangle $ is proportional to the temperature according to the fluctuation-dissipation theorem.

In general, we note that the spectral density in Eq.~\eqref{eq:spectralDensityOscillator} is not symmetric in $\omega$ due to spontaneous emission, which, classically, it would be. 
As important application to optomechanics, the asymmetry of the noise spectral  density of the radiation pressure when driving on the red-detuned sideband allows to cool the mechanical oscillator.
The optical noise spectral density then enters in the net optical cooling rate of the mechanical oscillator~\cite{Marquardt2007Quantum}.
If, in addition, the light used to drive the mechanical oscillator is squeezed, side-band cooling allows the cooling of the mechanical mode to the ground state~\cite{clark2017sideband}.
The spectral density is also relevant to sensing applications as it determines the signal-to-noise ratio~\cite{clerk2010introduction,Lau2018Fundamental}, Sec.~\ref{sec:sensing}.

\subsection{Measurement and control of massive quantum systems} \label{sec:control}
To control the massive quantum systems in the laboratory, we need to be able to manipulate their motion and perform accurate measurements. This is particularly important when measuring weak gravitational effects. Here we summarize the key ideas behind different measurement and control schemes that are used in the various proposals covered in Section~\ref{sec:gravity:tests}. 
For an in-depth discussion of control and measurement of quantum systems, we refer to the designated reviews~\cite{jacobs2006straightforward,clerk2010introduction} as well as the following textbooks~\cite{wiseman2009quantum,jacobs2014quantum}.

\subsubsection{Quantum measurements} \label{sec:quantum:measurements}
    To extract information about how gravity affects quantum systems, we must perform a measurement. There are a number of different measurement types and schemes. Here we briefly review the most common ones. 
    
    Projective measurements, {also known as von Neumann measurements}, model the measurement apparatus as a macroscopic pointer that can be read out classically. Strong correlations between the system and the pointer let us determine the state of the system unambiguously by measuring the pointer. 
    However, the measurement destroys the coherence of the wave function, subsequently destroying information about the conjugate observable and leading to back-action quantum noise~\cite{jacobs2006straightforward,clerk2010introduction}.

    Quantum non-demolition measurements~\cite{braginsky1980quantum,braginsky1995quantum,braginsky1996quantum,peres1993quantum}, on the other hand,  present a special case in which the eigenstates of the observable we are measuring are also eigenstates of the system, or equivalently the measured observable $\hat A$ commutes with the Hamiltonian $\hat H$, $[\hat H,\hat A]=0$, and thus $\hat H$ and $\hat A$ are simultaneously diagonalizable. Measuring multiple times yields the same result and allows for improved measurement accuracy, which is crucial for certain force-sensing schemes. 
    We discuss such back-action evasion schemes in more detail in Section~\ref{sec:BAE}.

    Weak measurements only extract partial information about an observable and thus do not fully destroy the information about the conjugate observable. For a detailed discussion, we refer to the pedagogical review~\cite{jacobs2006straightforward}.
    The main idea is to construct operators $\hat P_m$ such that $\sum_m \hat P_m^\dagger \hat P_m = \mathbb{1}$. The state after the measurement expressed in terms of the projectors $\hat P_n$ and the state $\hat\rho$ before the measurement is then given by
    \begin{align}
        \hat \rho_\mathrm{f}=\frac{\hat P_n^\dagger \hat \rho \hat P_n}{\tr\,(\hat P_n^\dagger \hat \rho \hat P_n)}, 
    \end{align}
    with the probability $\tr\,(\hat P_n^\dagger \hat \rho \hat P_n)$ to obtain this outcome\footnote{An operator of the form $\hat M=\sum_{n=a}^b \hat P_n^\dagger \hat P_n$ is a positive operator and $\tr(\hat M \hat \rho)$ gives the probability that $n\in[a,b]$. Therefore, this operator defines a positive operator-valued measure (POVM).}.
    Note that we recover a von Neumann measurement when measuring in the eigenbasis, i.e. setting $P_n=\ketbra{n}{n}$.
    Rather than measuring in the eigenbasis, we define $P_n$ as a weighted sum over different eigenstates that peaks at a specific eigenstate but has a certain width~\cite{jacobs2006straightforward}. A small width corresponds to a strong measurement, with the limit of zero width corresponding to a von Neumann measurement. A large width performs a weak measurement. The measurement strength $k$ is typically defined as the inverse of this width.
    It was suggested that weak measurements can lead to more accurate measurements of gravitational forces~\cite{Kawana2019Amplification}.

    Rather than measuring a system once, it can be interesting to continuously extract information from the system. Together with feedback, such measurement strategies can, for instance, be employed to squeeze or cool levitated mechanical systems~\cite{Genoni2015Quantum} that can be used for gravity tests, see Sec.~\ref{sec:tests:levitated}.
    A theory for such continuous measurements can be constructed from a sequence of time intervals $\Delta t$ during which weak measurements are performed with a measurement strength proportional to $\Delta t$.
    In the limit of infinitesimally short time intervals, we obtain a stochastic equation of motion due to the random nature of the measurements (see~\cite{jacobs2006straightforward} for a derivation).
    The measurement current of the continuously observed observable $A$ with a weak measurement is then given by
    \begin{align}\label{eq:measurementRecord}
        \mathrm{d} I(t) = \sqrt{k} \langle A(t) \rangle \mathrm{d} t + \mathrm{d} W(t), 
    \end{align}
    in which $k$ is the measurement strength, $\mathrm{d}W(t)$ is the standard Wiener increment describing the white imprecision noise in the measurement current and fulfills\footnote{$\llangle .\rrangle$ denotes the average over all possible measurement outcomes. For a pedagogic introduction to stochastic calculus, we refer to~\cite{jacobs2006straightforward}.} $\llangle \mathrm{d} W\rrangle=0$ and $\llangle \mathrm{d} W^2\rrangle=\mathrm{d} t$~\cite{clerk2010introduction}.
    The density matrix $\hat\rho_\mathrm{c}$ determining the expectation value $\langle A(t) \rangle$ is now conditional on the measurement current and evolves according to the stochastic master equation
    \begin{align}\label{eq:densityMatrixContinuousMeasurement}
        \mathrm{d}\hat \rho_\mathrm{c} =
        & -\frac{i}{\hbar}[\hat H_\mathrm{s},\hat \rho_\mathrm{c}] \mathrm{d}t + \frac{k}{4} \mathcal{D}[\hat A]\hat \rho_\mathrm{c} \mathrm{d} t \notag \\
        & + \frac{\sqrt{k}}{2} [\hat A \hat \rho_\mathrm{c} + \hat \rho_\mathrm{c} \hat A - 2\langle \hat A\rangle \hat \rho_\mathrm{c}] \mathrm{d} W, 
    \end{align}
    with $\mathcal{D}[.]\hat\rho_\mathrm{c}$ defined as below Eq.~\eqref{eq:Lindblad:equation}.
    The evolution of this conditioned quantum state is referred to as quantum trajectory.
    This equation can only be solved analytically in a special case~\cite{jacobs2006straightforward,Wiseman1996Quantum} and, in most cases, has to be solved numerically.

    Averaging over all possible measurement results, i.e., the observer does not retain the measurement current, Eq.~\eqref{eq:densityMatrixContinuousMeasurement} simplifies to
    \begin{align}
        \mathrm{d}\llangle\hat \rho_\mathrm{c}\rrangle & = -\frac{i}{\hbar}[\hat H_\mathrm{s}, \llangle\hat \rho_\mathrm{c}\rrangle] \mathrm{d}t + \frac{k}{4} \mathcal{D}[\hat A] \llangle\hat \rho_\mathrm{c}\rrangle \mathrm{d} t,
    \end{align}
    since $\llangle  \hat \rho_\mathrm{c}\mathrm{d} W\rrangle=0$ as $\hat \rho_\mathrm{c}$ and $\mathrm{d} W$ are statistically independent~\cite{jacobs2006straightforward}.

Continuous measurements can also be described with Langevin equations. Here, the recorded measurement current is then determined by the output fields.
This is particularly convenient for describing homodyne and heterodyne detection.
For a homodyne detection, the output at frequency $\omega_0$ is combined with a signal of a local oscillator at the same frequency $\omega_0$ in an interferometer. The detected current (in a rotating frame with frequency $\omega_0$) is then given by\footnote{Note that we encounter different conventions for the definition of $I(t)$ which is sometimes defined as $I = (e^{-i\phi} b_\mathrm{out} + e^{i\phi} b_\mathrm{out}^\dagger)/2$ or simply $I = e^{-i\phi} b_\mathrm{out} + e^{i\phi} b_\mathrm{out}^\dagger$.}~\cite{Barchielli2015Quantum}
\begin{align}\label{eq:homodyneCurrent}
    \hat I(t) \equiv (e^{-i\phi} \hat b_\mathrm{out}(t) + e^{i\phi} \hat b_\mathrm{out}^\dagger(t))/\sqrt{2}.
\end{align}
Here, $\phi$ is a phase difference depending on the optical path that determines the observed quadrature. 
For $\phi=0$, we obtain $\hat I(t)=(\hat b_\mathrm{out}(t) + \hat b_\mathrm{out}^\dagger(t))/\sqrt{2}=q_\mathrm{out}(t)$ while $\phi=\pi/2$ yields $\hat I(t) = -i(\hat b_\mathrm{out}(t) - \hat b_\mathrm{out}^\dagger(t))/\sqrt{2} = p_\mathrm{out}(t)$.

In a heterodyne detection scheme, the output at frequency $\omega_0$ is combined with a signal of a local oscillator at a different frequency $\omega_1$ detuned from the output frequency by $\Delta\equiv\omega_1-\omega_0$, thereby giving rise to the heterodyne current
\begin{align}\label{eq:heterodyneCurrent}
    \hat I(t) \equiv (& e^{-i(\phi-\Delta t)} \hat b_\mathrm{out}(t)
    + e^{i(\phi-\Delta t)} \hat b_\mathrm{out}^\dagger(t))/\sqrt{2}. 
\end{align}
As a result, the measured quadrature oscillates in time, providing information about both amplitude and phase. However, this comes at the cost of an added half-quantum of noise~\cite{bowen2015quantum}. Homodyne and heterodyne noise spectra are discussed in detail in~\cite{Barchielli2015Quantum,bowen2015quantum}.

\subsubsection{Feedback and feedforward}

    Along with continuous measurements, we can continuously apply operations on the system to steer it toward a desired state.
    Creating or stabilizing certain quantum states 
    can be advantageous in the context of metrology. In particular, superposition states can be used to test theories about quantum mechanics and gravity.
    Furthermore, it has been shown that certain entangled states can improve the sensitivity and signal-to-noise ratio~\cite{leibfried2004heisenberg,roos2006designer}.
    Feedback is also employed for squeezing or cooling in many of the experimental tests of gravity in Sec.~\ref{sec:State-of-the-art}, and, in general, for the control of quantum systems, see Sec.~\ref{sec:controlling:massive:systems}.
    The feedback can either explicitly depend on the measurement current $I(t)$ obtained in Eq.~\eqref{eq:measurementRecord} (closed loop) or not (open loop feedback). Apart from applying feedback, we can also use the continuously measured quantum system to control a second quantum system which is then referred to as feedforward.

    The optimal control of quantum systems has been investigated over a long time~\cite{peirce1988optimal,dahleh1990optimal,judson1992teaching,warren1993coherent,wiseman1993quantum, wiseman1994quantum1,magrini2021real}.
    Feedback can either be applied semi-classically---here, the measurement current that is used to provide feedback is obtained with a classical sensor---or fully quantum---here, the detectors and sensors are all quantum systems.
    The main idea behind classical, continuous feedback is to apply a semi-classical potential that steers the system coherently towards the desired quantum state~\cite{caves1987quantum,wiseman1994quantum1,Doherty2000Quantum,Lloyd2000Coherent}.
    For instance, we may measure the position of a mechanical oscillator and then perform a displacement operation to shift its position. 

    In general, the quantum state evolution now explicitly depends on the stochastic measurement current $I(t)$, which will be different in each experimental run, resulting in conditional dynamics described by a stochastic master equation. We obtain the stochastic measurement current, Eq.~\eqref{eq:measurementRecord}, by measuring an observable $\hat A$, which is determined according to the stochastic master equation for continuous feedback, Eq.~\eqref{eq:densityMatrixContinuousMeasurement}. This current~\eqref{eq:measurementRecord} is then fed back to drive the system system via the Hamiltonian
    \begin{align}
        \hat H_\mathrm{fb} = \sqrt{\kappa_\mathrm{fb}} I(t-\tau) \hat B,
    \end{align}
    with $\kappa_\mathrm{fb}$ the feedback strength, $\tau$ some time delay, and in which $\hat B$ encodes the operation that is chosen to be applied based on the measurement outcome. Note that the feedback operation $\hat B$ may also involve the measured observable $\hat A$.
    Furthermore, the choice of $\hat B$ explicitly depends on the measurement current in the case of closed-loop feedback, while it does not for open-loop feedback.
    
    Instead, we can look at the unconditional dynamics by averaging over all measurement outcomes
    \begin{align}\label{eq:densityMatrixFeedback}
        \frac{\mathrm{d}\hat \rho}{\mathrm{d} t} =
        & -\frac{i}{\hbar}[\hat H_\mathrm{s},\hat \rho] + \frac{k}{4} \mathcal{D}[\hat A]\hat \rho + \kappa_\mathrm{fb} \mathcal{D}[\hat B]\hat\rho \notag \\
        & - i \frac{\sqrt{k \kappa_\mathrm{fb}}}{2} [\hat B, \hat A \hat\rho + \hat\rho \hat A].
    \end{align}
    Here,
    $- i \frac{\sqrt{k \kappa_\mathrm{fb}}}{2} [\hat B, \hat A \hat\rho + \hat\rho \hat A]$ encodes a linear restoring term and dissipation, e.g., in an optomechanical system, this could be a restoring force. 
    The term $\kappa_\mathrm{fb} \mathcal{D}[\hat B]\hat\rho$ describes additional fluctuations as a consequence of the feedback.

    Analogous to the scenario described above, the measurement current of system A can force a second system B in a feedforward scheme.
    The main difference is that $\hat B$ now denotes an operator of the other system B.
    For instance, reservoir-engineered non-reciprocity can be thought of as an autonomous feed-forward scheme~\cite{Metelmann2017Nonreciprocal} in which the measurement results of one system are used to drive another system but not vice versa.
    
    Rather than controlling the quantum system based on the classical measurement record, it is also possible to replace the sensors and controllers with quantum systems that coherently interact with the system to be controlled~\cite{Lloyd2000Coherent,Nurdin2009Coherent,Nelson2000Experimental}. 
    Coherent feedback protocols can outperform measurement-based schemes~\cite{Hamerly2012Advantages,Hamerly2013Coherent} because they can exploit a geodesic path 
    in Hilbert space that is forbidden to measurement-based schemes~\cite{Jacobs2014Coherent}.
    A convenient way to describe coherent feedback is with a Langevin equation formalism~\cite{Gardiner2000Quantum}, Eq.~\eqref{eq:LangevinEquations}. Here, the output field $\hat a_\mathrm{out}$ is fed to the input field $\hat b_\mathrm{in}$ of the mode that is to be controlled with some time delay $\tau$:
    \begin{align}
        \hat b_\mathrm{in}(t) & = \sqrt{\kappa_\mathrm{fb}} \hat a_\mathrm{out}(t-\tau). 
    \end{align}
    A number of schemes have been proposed and implemented to control mechanical oscillators via feedback, such as feedback cooling~\cite{chang2010cavity,Li2011Millikelvin,Hamerly2012Advantages,Hamerly2013Coherent,Genoni2015Quantum,Jain2016Direct,vovrosh2017parametric,setter2018real,rademacher2022nonequilibrium,Guo2022Coherent,Mansouri2022Cavity,Harwood2021Cavity}, schemes to control squeezing, entanglement and state transfer~\cite{Harwood2021Cavity}, or to control the motional state of the mechanical oscillator, its resonance frequency and damping rate~\cite{ernzer2022optical}.
    Feedback cooling allowed the cooling of a 10 kg mass in the LIGO detector close to its ground state~\cite{Abbott2009Observation,whittle2021approaching-1} and a millimeter-sized membrane resonator was cooled to the ground state with measurement-based feedback~\cite{rossi2018measurement}. Measurement-based feedback cooling has also been demonstrated in electromechanical systems~\cite{Wang2022Fast} where it was also demonstrated that feedback can lead to dynamically stability in situations that would be unstable without feedback.
    Feedback schemes are also employed to equalize mechanical loss rates in experiments with multiple mechanical oscillators~\cite{Poggio2007Feedback,delPino2022Non,Wanjura2022Quadrature}.

\subsection{Quantum metrology with massive quantum systems}
\label{sec:sensing}
Since gravity is extremely weak, we may sometimes wish to quantify the sensitivity of a quantum sensor to ensure it is powerful enough. The main tool used for this is quantum estimation theory, also referred to as quantum metrology. We here outline the key concepts and refer to the following reviews for more detailed reading~\cite{paris2009quantum, clerk2010introduction, toth2014quantum}. Detection methods for optomechanical systems are reviewed in~\cite{poot2012mechanical}. Many of these tools are used for precision tests of gravity (see Sec.~\ref{sec:quantum:enhanced:sensing}).

\subsubsection{Langevin description of a quantum sensor} \label{sec:langevin:sensor}

This exposition of a sensing scheme follows~\cite{clerk2010introduction,Lau2018Fundamental}.
In a typical metrology setting, we would like to infer an infinitesimal change in a small parameter $\epsilon$ that the Hamiltonian $\hat H(\epsilon)$ depends on. Expanding $\hat H(\epsilon)$ to first order in $\epsilon$, we have 
\begin{align}
    \hat H & = \hat H_0 + \epsilon \hat V + O(\epsilon^2), 
\end{align}
where $\hat H_0$ is the free Hamiltonian and where $\hat V$ is the operator that encodes  $\epsilon$. To extract changes in $\epsilon$, we probe the system governed by $\hat H$ with a probe field $\hat a_\mathrm{in}$ and currents the response $\hat a_\mathrm{out}$ which then depends on $\epsilon$. 
For small $\epsilon$, we can write $\hat a_\mathrm{out} \cong \hat a_\mathrm{out}^{(0)} + \lambda\epsilon$ where $\lambda$ is a linear response coefficient.

To characterize the resolving power of a quantum sensor, we can calculate the signal-to-noise ratio (SNR) by comparing the integrated signal power to the noise power.
The measurement current can, for instance, be obtained via homodyne detection (see Sec.~\ref{sec:quantum:measurements}). The power associated with the signal is then given by the expectation value of the time-integrated measurement current $\hat m(t) \equiv \int_0^t \mathrm{d}\tau \, \hat I(t)$, where $\hat I(t)$ is defined in Eq.~\eqref{eq:homodyneCurrent}. The measurement current should be compared to the power without the perturbation $\epsilon$, so we define the power difference $\mathcal{P}$ of the signal with and without the perturbation $\epsilon$ as
\begin{align}
    \mathcal{P} = \left[\langle \hat m(t)\rangle - \langle \hat m(t)\rangle\rvert_{\epsilon=0}\right]^2.
\end{align}
The associated noise power is then given by
\begin{align} \label{eq:noise:ratio}
    \mathcal{N} \equiv \langle \delta \hat m(t) \delta \hat m(t)\rangle = t \mathcal{S}_{II}(0), 
\end{align}
where $\delta \hat m(t) = \hat m(t) - \langle \hat m(t)\rangle$. Here, $\mathcal{S}_{II}(0) =\frac{1}{2}\int \mathrm{d}t e^{\mathrm{i}\omega t} \langle \left\{\delta I(t), \delta I(0)\right\}\rangle$ is the noise spectral density defined in Eq.~\eqref{eq:noiseSpectrum} of the measurement current at $\omega=0$. Note that Eq.~\eqref{eq:noise:ratio} is linear in time $t$ because we consider the integrated measurement current.  
The signal-to-noise ratio (SNR) is then given by the ratio of $\mathcal{P}$ and $\mathcal{N}$, $\rho_\mathrm{SNR}\equiv\mathcal{P}/\mathcal{N}$.
For applications such as gravitational wave detection~\cite{Caves1979Microwave}, force sensing~\cite{caves1980measurement}, and force gradient sensing~\cite{Rudolph2022Force}, it is vital to ensure that the signal is stronger than the noise, such that $\rho_{SNR} \geq 1$. Since the noise increases with $t$, we require the measurement current to also accumulate information about $\epsilon$ at the same rate. Therefore, it is crucial to retain long coherence times in the system so that a strong signal can be retained throughout. 

Quantum mechanics puts a limit~\cite{caves2012quantum} on the added noise $\mathcal{A}$ of an amplifier when referred to the amplification gain $\mathcal{G}$. In particular, we find for the variance $\langle\Delta \hat a_\mathrm{out}\rangle \equiv \langle \hat a_\mathrm{out}^\dagger \hat a_\mathrm{out}\rangle - \lvert\langle \hat a_\mathrm{out}\rangle\rvert^2$ the expression
$\langle\Delta \hat a_\mathrm{out}\rangle = \mathcal{G}(\langle\Delta \hat a_\mathrm{in}\rangle + \mathcal{A})$ in which quantum mechanics restricts $\mathcal{A}\geq\frac{1}{2}$.

It was proposed that non-reciprocity and non-Hermitian topology are promising resources for quantum sensors; the first allows to overcome fundamental constraints on the signal-to-noise ratio of conventional sensors~\cite{Lau2018Fundamental,Kononchuk2022Exceptional,slim2023optomechanical}, and, in addition, the second can lead to an exponentially-enhanced sensitivity~\cite{McDonald2020Exponentially,Koch2022Quantum}. Both non-reciprocity~\cite{Metelmann2015Nonreciprocal} and non-Hermitian topological chains~\cite{Wanjura2020Topological,McDonald2018Phase} can be engineered in driven-dissipative quantum systems, e.g., based on optomechanics~\cite{Mercier2020Nonreciprocal,delPino2022Non,Youssefi2022Topological}.

\subsubsection{Standard quantum limit} \label{sec:SQL} 

To perform tests of fundamental physics, see Sec.~\ref{sec:State-of-the-art}, it is often crucial to measure the oscillator position accurately. However, Heisenberg's uncertainty principle states that it is impossible to simultaneously know the position and momentum of a single quantum system with high accuracy. A measurement of the mass's position necessarily introduces back action on its momentum. In this context, we often speak of a standard quantum limit (SQL) that limits the accuracy of position measurements as the system evolves in time. 

The SQL is straightforward to derive for an effective free mass $m_{\rm{eff}}$ that is harmonically trapped with frequency $\omega_m$. Its position quadrature $\hat X_m(t)$ evolves in time as 
{
\begin{align}
\hat X_m (t) = \hat X_m(0) \cos(\omega_m t ) + \frac{\hat P_m(0)}{m_{\rm{eff}}\omega_m} \sin(\omega_m t), 
\end{align}}
where $\hat X_m(0)$ and $\hat P_m(0)$ are the {position and momentum} operators at $t = 0$, which remain unchanged during intervals smaller than the damping time. Then, considering the Heisenberg uncertainty principle, we find 
\begin{align}
\Delta X_m(t) \Delta Y_m(t) &\geq \frac{1}{2} |\langle[\hat X_m(0), \hat Y_m(0)]\rangle| = x_{\rm{zpf}}^2 , 
\end{align}
where {we have introduced the quadrature $\hat{Y}_m(t)=\frac{\hat P_m(t)}{m_{\rm{eff}}\omega_m}$ and} $x_{\mathrm{zpf}} = \sqrt{\frac{\hbar}{2 m_{\rm{eff}} \omega_m}}$ is the zero-point fluctuation {($\hat{X}_m(t)$ and $\hat{Y}_m(t)$ are both expressed in units of length for easier comparison).} Thus, any measurement that tries to measure both quadratures with equal precision is limited to $\Delta X_m(t) = \Delta Y_m(t) = x_{\rm{zpf}}$.  See~\cite{caves1980measurement} for a derivation of the SQL for a single quantum oscillator.

When the system is coupled to an external probe field, we consider an \textit{optomechanical SQL} where the position of a mechanical resonator is detected through phase measurements~\cite{bowen2015quantum}. There are small fluctuations in the probe field itself, which is known as shot-noise. Shot-noise can be decreased by increasing the number of quanta in the probe field, which improves the signal-to-noise ratio and makes detection easier. The increase in the field quanta does, however, lead to stronger recoil in the mechanical system, known as (quantum) back-action noise or radiation pressure noise. The result is a fluctuation force on the mechanical resonator. By balancing these two sources of noise, we arrive at the optomechanical SQL, which sets the limit on the achievable accuracy of position measurements. The SQL can be calculated by deriving the output spectrum of the measured probe field and balancing the resulting shot-noise and radiation-pressure noise. We refer to~\cite{clerk2010introduction, bowen2015quantum} for the full derivation.
With the Langevin equations in Eq.~\eqref{eq:langevin:equation:optomechanics} to model the dynamics, as well as the input-output relations shown in Eq.~\eqref{eq:input:output}, we find that the measured quadrature of the mechanical mode in Fourier space results in the following symmetrized power spectral density~\cite{bowen2015quantum}
\begin{align} \label{eq:SQL:spectrum}
    S_{\mathrm{det}}(\omega) = \frac{1}{8 \eta \Gamma |C_{\mathrm{eff}}|} + 2 \Gamma |\chi(\omega)|^2 |C_{\mathrm{eff}}|, 
\end{align}
where $C_{\mathrm{eff}}$ is the effective optomechanical cooperativity, defined as $C_{\mathrm{eff}} = C/(1 - 2 i \omega /\kappa)^2$, where $C = 4 g_0^2/\kappa \Gamma$ is the optomechanical cooperativity, for which $g_0$ is the optomechanical coupling, and $\kappa$ and $\Gamma$ are the optical and  mechanical linewidths, respectively. $\eta$ is the detection efficiency. The expression $\chi(\omega)$ in Eq.~\eqref{eq:SQL:spectrum} is the mechanical susceptibility, which is given by 
\begin{align}
    \chi(\omega) = \frac{\omega_m}{\omega_m^2 - \omega^2 - i \omega \Gamma}. 
\end{align}
To balance the two terms in Eq.~\eqref{eq:SQL:spectrum}, we require the optimal effective cooperativity to be 
\begin{align}
    |C_{\mathrm{eff}}^{\mathrm{opt}} | = \frac{1}{4 \eta^{1/2} \Gamma |\chi(\omega)|}, 
\end{align}
the symmetrized spectrum at the SQL is then given by 
\begin{align}
    S^{\mathrm{SQL}}_{\mathrm{det}}(\omega) = |\chi(\omega)|, 
\end{align}
where we have assumed an optimal detection efficiency with $\eta = 1$. That is, in the optimum case, the spectrum is given by the susceptibility $\chi(\omega)$ of the resonator.

In practice, there are a number of additional noise sources that can be included in the measured noise spectrum, such as measurement imprecision and amplifier noise, see Sec.~\ref{sec:langevin:sensor}.
For example,~\cite{magrini2021real} provides a detailed analysis of noise budgeting in an optomechanical experiment for the purpose of quantum-limited measurements,
and~\cite{Martynov2016Sensitivity} lists and characterizes a number of relevant noise sources, such as thermal noise, laser noise and electronic noise in the LIGO detector which are also relevant to many other experiments. See also~\cite{danilishin2012quantum} for a review of how quantum noise can be calculated in a gravitational-wave detector.

Note that the SQL is by no means a fundamental limit, as opposed to the Heisenberg limit (see Sec.~\ref{sec:quantum:fisher:information}). It can be evaded by using squeezed states, which reduce the noise in the measured quadrature~\cite{bowen2015quantum}.

\subsubsection{Classical and quantum Fisher information} \label{sec:quantum:fisher:information}
Previously, we focused on how well a detector can probe a signal against a noisy environment. However, we can also ask how much information a quantum system can fundamentally accumulate about a specific effect. This notion is captured by the Fisher information, which is a valuable metrology tool relevant for many of the weak-force detection schemes described in Sec.~\ref{sec:quantum:enhanced:sensing}. See~\cite{paris2009quantum,giovannetti2011advances} for comprehensive introductions. 

Consider a specific measurement with outcomes $\{x\}$ performed on the quantum state $\hat \rho(\theta)$, where $\theta$ is the parameter that we wish to estimate. The distribution of the measurement outcomes is given by $p(x|\theta) = \mathrm{tr}( \hat \Pi_x \hat \rho_\theta)$, where $\hat \Pi_x$ is a POVM element which models the measurements. The classical Fisher information (CFI) corresponds to the amount of information about $\theta$ gained from this measurement series. It is given by~
	\begin{align}
		I_\mathrm{F}(\theta) & = \int \mathrm{d}x \, p(x\vert\theta) \left(\frac{\partial\ln p(x\vert\theta)}{\partial\theta}\right)^2.
	\end{align}
The CFI can be generalized in the quantum case by optimizing over all possible measurements of the quantum state. This is known as the quantum Fisher information (QFI). The QFI can also be viewed as a distance measure that quantifies the change of the state due to the parameter $\theta$. That is, given the two quantum states $\hat \rho_\theta$ and $\hat \rho$, the most general form of the QFI is 
	\begin{align}  \label{eq:Bures:distance:QFI}
		I_\mathrm{F}(\theta) & = 4 \left(\left.\frac{\partial d_\mathrm{B}(\hat \rho_{\theta},\hat \rho)}{\partial \theta}\right\rvert_{\epsilon=0}\right), 
	\end{align}
where $d_\mathrm{B}$ is the Bures distance~\cite{bures1969extension, helstrom1967minimum}
\begin{align}
d_\mathrm{B}(\hat \rho_1,\hat \rho_2)=\sqrt{2(1-\sqrt{\mathcal{F}(\hat \rho_1,\hat \rho_2)})},
\end{align}
for which $\mathcal{F}$ is  the fidelity $\mathcal{F}(\hat \rho_1, \hat \rho_2)=\left(\tr[ \sqrt{\sqrt{\hat \rho_1} \hat \rho_2\sqrt{\hat \rho_1}}] \right)^2$. For pure states $\ket{\Psi(\theta)}$ which encode the parameter $\theta$, the QFI becomes
 \begin{align}
 I_{\mathrm{F}}(\theta) = 4 \left( \braket{\partial_\theta \Psi(\theta)} - |\langle\Psi(\theta) | \partial_\theta \Psi(\theta)\rangle|^2 \right), 
 \end{align}
 where $\partial_\theta$ denotes the partial derivative with respect to $\theta$. 
The QFI can also be computed for initially mixed states $\hat \rho$ that evolve unitarily, such that  $\hat\rho(\theta)=\hat U_\theta \hat\rho(0) \hat U_\theta^\dagger$. When the initial state can be decomposed in terms of an orthonormal basis $\hat \rho(0) = \sum_n \lambda_n \ket{\lambda_n}\bra{\lambda_n}$  the QFI can be written as~\cite{pang2014quantum,liu2014quantum}
    \begin{align} \label{eq:mixed:state:QFI}
    I_{\mathrm{F}}(\theta)
    =& \;4\sum_n \lambda_n\,\left(\bra{\lambda_n}\mathcal{\hat H}_\theta^2\ket{\lambda_n} - \bra{\lambda_n}\mathcal{\hat H}_\theta\ket{\lambda_n}^2\right)\nonumber\\
    &-8\sum_{n\neq m}
    \frac{\lambda_n \lambda_m}{\lambda_n+\lambda_m}
    \bigl| \bra{\lambda_n}\mathcal{\hat H}_\theta \ket{\lambda_m}\bigr|^2,
    \end{align}
    where the second sum is over all terms with $\lambda_n+\lambda_m\ne0$, $\lambda_n$ is the eigenvalue of the eigenstate $\ket{\lambda_n}$, and where the Hermitian operator $\mathcal{\hat H}_\theta$ is defined as
    $\mathcal{\hat H}_\theta=-i\hat U^\dagger_\theta \partial_\theta{\hat
      U}_\theta$. The expression in Eq.~\eqref{eq:mixed:state:QFI} can also be extended to the multi-parameter case~\cite{liu2019quantum}, where it sometimes is possible to extract more information than in the single-parameter case~\cite{paris2009quantum}. 

When the quantum system interacts with an environment, the QFI can be challenging to compute since Eq.~\eqref{eq:mixed:state:QFI} no longer holds. In addition to the general expression in Eq.~\eqref{eq:Bures:distance:QFI}, the QFI can be defined in terms of the symmetric logarithmic derivative $\hat L_\theta$~\cite{helstrom1969quantum,holevo2011probabilistic} $
I_{\mathrm{F}} = \mathrm{tr}(\hat \rho \hat L_\theta^2)$,
where $\hat L_\theta$ is given by 
\begin{align}
\partial_\theta \hat \rho = \frac{1}{2} ( \hat \rho \, \hat L_\theta  + \hat L_\theta \hat \rho ).
\end{align}
If an expression for $\hat L_\theta$ is found, the QFI can be immediately calculated. 
One way to solve this equation for $\hat L_\theta$ by treating it as a Lyapunov matrix equation, which has a general solution~\cite{paris2009quantum}
\begin{align}
    \hat L_\theta = 2 \int^\infty_0 dt \, \mathrm{exp}[- \hat \rho_\theta t] \partial_\theta \, \hat \rho_\theta \, \mathrm{exp}[- \hat \rho_\theta t]. 
\end{align}
When the channel that encodes the parameter $\theta$ can be represented with Kraus operators, a general upper bound to the QFI can be derived~\cite{escher2011general}.

A key feature of the QFI and CFI is that they fundamentally relate to the variance of the parameter $\theta$ through the Cram\'er–Rao bound~\cite{helstrom1969quantum,Braunstein1994Statistical}
	\begin{align}
		\mathrm{var}(\theta) & \geq \frac{1}{\mathcal{M}I_\mathrm{F}(\theta)}, 
	\end{align}
 where $\mathcal{M}$ is the number of measurements performed, and where for the QFI, the inequality is saturated. Since $\mathcal{M}$ is always finite, it is generally desirable to maximize the Fisher information to reduce the variance of the estimated parameter.  The Cram\'er–Rao bound is applicable to any quantum system and provides a generalized uncertainty relation~\cite{Braunstein1996Generalized} even when no Hermitian operator can be associated with the parameter of interest, e.g., as in the case of phase estimation~\cite{helstrom1969quantum,Braunstein1994Statistical,Braunstein1996Generalized}.
The Cramer-Rao bound also relates to the so-called Heisenberg limit, which is defined as the scaling of the variance $\mathrm{Var}(\theta)$ of a parameter $\theta$. For classical systems, the scaling of $\mathrm{Var}(\theta)$ is at most $1/N$, where $N$ is the number of probes used, in accordance with the central limit theorem~\cite{giovannetti2011advances}. As opposed to the standard quantum limit (see Sec.~\ref{sec:SQL}), the Heisenberg limit is a hard limit that depends on the number of resources in the system~\cite{zwierz2010general}. These resources can either refer to a number of subsystems or the translational power of the Hamiltonian, which is higher for nonlinear dynamics (that is, Hamiltonian terms with products of more than quadratic operators). For example, a self-Kerr Hamiltonian with a term proportional to $(\hat a^\dag \hat a)^2$ has more translational power than that with just $\hat a^\dag \hat a$.
The Heisenberg scaling does, however, go beyond $1/N$ for certain initial quantum states. For example, the QFI for phase estimation with NOON states scales as $1/N^2$~\cite{dowling1998correlated, bollinger1996optimal}. The QFI has also been used to investigate certain relativistic settings, see e.g.~\cite{pinel2013quantum,ahmadi2014quantum,Tian2015Relativistic,Hao2016Quantum}  and has been proposed as a probe of spacetime structure~\cite{Du2021Fisher}. 


%
\section{Proposed tests of gravity with massive quantum systems} \label{sec:gravity:tests} 
%

Equipped with tools to model massive systems in the laboratory, we now ask the question: What are the possible ways in which gravity influences quantum systems, and how can these effects be detected? A number of diverse and creative proposals have been put forward that probe the properties of gravity, quantum mechanics, and their interfaces. The goal of this section is to outline the main directions of research and key proposals that allow tests to be performed with massive quantum systems. 
We begin by considering gravity from a classical source, such as the Earth's gravitational fields, and its detection by quantum systems (Sec.~\ref{sec:quantum:enhanced:sensing}). Specifically, we focus on proposals for how quantum properties can enhance the sensitivity of the probe. We proceed to consider proposals where gravity causes decoherence of a quantum probe (Sec.~\ref{sec:grav:dec:section}), including different types of decoherence proposals as well as nonlinear modifications of quantum theory. We then review recent proposals for detecting gravitationally induced entanglement (Sec.~\ref{sec:entanglement:mediated:gravity}). The final part (Sec.~\ref{sec:other:tests}) outlines additional proposals that do not strictly fit into the other sections but are still relevant to the topic of this review.

%
\subsection{Precision tests of gravity} \label{sec:quantum:enhanced:sensing}
%

One approach for precision tests of gravity relies on the use of sensitive mechanical resonators.  This section provides a brief review of weak-force sensing with massive systems in the quantum regime. 
We here restrict ourselves to proposals where the sensor is in the quantum regime, but where gravity originates from a classical source. This should be contrasted with the tests discussed in Sec.~\ref{sec:entanglement:mediated:gravity}, where the quantum system itself is considered as a source of the gravitational field. Many precision tests of gravity have already been performed with classical mechanical resonators and atom interferometers. See Sec.~\ref{sec:State-of-the-art} for an overview of these experiments and Fig.~\ref{fig:sensitivity:vs:mass} for a summary of force sensitivities that have been achieved to date.

\subsubsection{Weak-force detection with back-action evading measurements} \label{sec:back:action:evasion}
\label{sec:BAE}
The large mass of massive quantum systems (compared with the mass scale of single atoms) means that they couple more strongly to gravity. A common goal of precision gravimetry with massive quantum systems is to resolve the force that affects the center of mass of the mechanical resonator. Usually, a probe field or two-level system is used for the control and readout of the sensor. For a probe field, back-action noise and the inherent uncertainty of field fluctuations give rise to the standard quantum limit (SQL), which we reviewed in Sec.~\ref{sec:SQL}, beyond which displacements cannot be resolved. The limits for a moving-end mirror were first discussed in~\cite{arcizet2006beating}, and the first experimental observation of radiation pressure due to shot noise was performed in~\cite{purdy2013observation}. 

Back-action evading (BAE) schemes constitute a key resource for weak-force sensing since they allow for an increase in measurement precision without adding additional noise during readout. 
The effect of measurement backaction can be circumvented if, rather than attempting to measure both of the mechanical quadratures, one only couples the light field to one of the quadratures such that it becomes a conserved quantity~\cite{thorne1978quantum,braginsky1980quantum}.
Concretely, this means that if we couple only the $\hat X_{\rm{m}} = ( \hat b^\dag + \hat b)/\sqrt{2}$ quadrature of the mechanical oscillator to the radiation pressure of the photon field $\propto \hat n$, the interaction Hamiltonian is $\hat H_\mathrm{int}\propto  \hat n \hat X_{\rm{m}}$ which implies that $[\hat H_\mathrm{int},\hat X_{\rm{m}}]=0$. An observable that commutes with the Hamiltonian is also referred to as a quantum non-demolition (QND) variable since it can be measured repeatedly without destroying the quantum state.

Coupling only one quadrature to the light field is experimentally challenging since it would require a time-dependent coupling between the quadratures and the detected field.
This can be achieved in a scheme for a cavity-optomechanical system driven on both the red and the blue sideband~\cite{clerk2008back}, or by modulating the optomechanical coupling strength~\cite{clerk2008back}.
As we discussed in Sec.~\ref{sec:QOMS:Hamiltonian}, the coupling between the cavity mode $\hat a$ and the mechanical oscillator $\hat b$ in the frame rotating with the cavity frequency $\omega_\mathrm{c}$ is given by $\hat H_I \approx - \hbar g_0 ( \alpha^* \delta \hat a  + \alpha \delta \hat a^\dag )( \hat b^\dag + \hat b)$. A drive can be modeled by $\hat H_\mathrm{d} = \alpha(t) \hat a + \alpha^*(t)\hat a^\dagger$ with $\alpha(t)$ the complex drive amplitude.
For a drive on the red sideband at $\omega_\mathrm{c}-\omega_\mathrm{m}$, with $\omega_\mathrm{m}$ the mechanical frequency, we have $\alpha(t)\propto e^{\mathrm{i}(\omega_\mathrm{c}-\omega_\mathrm{m})t}$ giving rise to the interaction Hamiltonian $\hat H_I \propto \hat a^\dagger \hat b + \mathrm{h.c.}$ in the frame rotating with $\omega_\mathrm{c}$ for the photons and $\omega_\mathrm{m}$ for the mechanical modes and in which we neglected counter-rotating terms. Similarly, a drive on the blue sideband at $\omega_\mathrm{c}-\omega_\mathrm{m}$, $\alpha(t)\propto e^{\mathrm{i}(\omega_\mathrm{c}+\omega_\mathrm{m})t}$, gives rise to the interaction Hamiltonian $\hat H_I \propto \hat a^\dagger \hat b^\dagger + \mathrm{h.c.}$.
Combining both the drive on the red and the blue sideband, we obtain
\begin{align}
    H_I \propto  \hat X_\mathrm{c} \hat X_{\rm{m}}.
\end{align}
That is, the interaction Hamiltonian couples the quadrature $\hat X_\mathrm{c} = (\hat a^\dag + \hat a)/\sqrt{2}$ of the cavity to the  quadrature $\hat X_{\rm{m}}$ of the mechanical mode such that both $\hat X_{\rm{m}}$ and $\hat X_\mathrm{c}$ are constants of the motion since $[H_I, \hat X_{\rm{m}}] = [H_I, \hat X_\mathrm{c}] = 0$. Since the mechanical oscillator only couples to the $\hat X_{\rm{m}}$ quadrature of the cavity, information on its motion only propagates into the conjugate cavity quadrature $\hat P_c = i (\hat a^\dag - \hat a)/\sqrt{2}$.
This allows us to repeatedly (or continuously) measure $\hat P_\mathrm{c}$ and infer $\hat X_{\rm{m}}$ to arbitrary precision.
For a more detailed theoretical discussion of this scheme, we refer to~\cite{clerk2020optomechanics} and the reviews~\cite{braginsky1996quantum,clerk2010introduction}.
The exact conditional dynamics of an optomechanical system driven on both sidebands were analyzed in detail in~\cite{brunelli2019conditional}.

The scheme described above was experimentally implemented in a superconducting electromechanical device in~\cite{suh2014mechanically}, allowing the detection and reduction of back action and the detection of a single quadrature below the zero-point fluctuations in the optical domain in~\cite{shomroni2019optical}.
BAE was also realized in hybrid optomechanical systems of a macroscopic mechanical oscillator and a spin oscillator~\cite{moller2017quantum}. 
A more elaborate BAE scheme relies on constructing an effective oscillator out of two and measuring a collective variable~\cite{woolley2013two}, which was experimentally realized in a microwave circuit with two mechanical oscillators~\cite{ockeloenkorppi2016quantum} achieving a measurement precision below the zero-point fluctuations. Back-action-noise can also be canceled through the addition of an ensemble of cold atoms, which act as a negative-mass oscillator and allow for sensing beyond the SQL~\cite{motazedifard2016force}. Furthermore, in a scheme with four drives~\cite{mercier2021quantum}, back-action was evaded, and the entanglement between the two oscillators was demonstrated. 

Apart from improving the measurement precision, BAE schemes result in the squeezing of the mechanical modes. In a generic, parametrically driven mechanical oscillator, the attainable amount of squeezing is limited due to the onset of parametric instability. However, continuous BAE measurements allow us to overcome this 3dB limit for squeezing~\cite{lei2016quantum}. Beyond the schemes discussed above that rely on measuring a single quadrature, an alternative approach to cancel quantum noise and overcome the SQL of force sensing using coherent feed-forward quantum control was proposed in~\cite{tsang2010coherent}. Another proposal shows that mechanically driving the oscillator within the BAE scheme leads to a monotonic response of the cavity, which can be beneficial for sensing force gradients~\cite{arvidsson2024sensing}.

Yet another direction for resolving the energy levels of a mechanical oscillator in an electromechanical experiment~\cite{dellantonio2018quantum} is to use QND measurements to bring us a step closer to understanding how quantum jumps between phonon states work, which is challenging since the coupling to an environment makes it difficult to detect mechanical mode occupation.
The notion of QND variables was generalized to a quantum-mechanics-free subsystem~\cite{tsang2012evading}, i.e., subsystems in which all observables commute and their expectation values are governed by classical equations of motion.


\subsubsection{Additional weak-force detection schemes}

Theoretical proposals for force sensing with massive quantum systems generally take one of two approaches: they either show that the SQL can be circumvented through novel protocols, such as back-action evading measurements (see Sec.~\ref{sec:back:action:evasion}) or the addition of quantum resources, or they consider the fundamental sensitivity that the systems can achieve, often quantified by the classical and quantum Fisher information (see Sec.~\ref{sec:quantum:fisher:information}). Both approaches limit the precision of the measurement. 
Generally, the Fisher information quantifies the precision that can be achieved beyond the SQL.

Apart from back-action evasion, quantum resources such as squeezing and entanglement are required for beating the SQL~\cite{zhang2021distributed}. In optomechanical systems, squeezing of both optical and mechanical motion can be implemented in a number of ways (see Sec.~\ref{sec:squeezing} for an overview of experiments). For example, in a mirror-in-the-middle optomechanical setup which results in two coupled cavity modes~\cite{xu2014squeezing}, the SQL is surpassed due to the resulting squeezing of the light. 
The inclusion of a single- or two-mode parametric amplifier (PA)~\cite{mollow1967quantum} with either $\hat a^{\dag 2} + \hat a^2$ (single-mode) or $\hat a \hat b + \hat a^\dag \hat b^\dag$ (two-mode), results in sensing precision beyond the SQL~\cite{motazedifard2019force, zhao2020weak}. Further, in dissipative optomechanical systems, a PA can counteract the negative effects of mechanical damping, which allows us to go beyond the SQL~\cite{huang2017robust}. Advantages through squeezing can also be achieved by adding a nonlinear medium in the cavity~\cite{peano2015intracavity}. Squeezing has also famously been shown to improve the precision of LIGO~\cite{aasi2013enhanced, buikema2020sensitivity}. 

Beyond squeezing, entanglement plays a crucial role in sensing and is a key ingredient for achieving a sensitivity that scales with the Heisenberg limit~\cite{zhuang2018distributed}. 
Most importantly, by performing measurements with $N$ entangled sensors, we may go beyond the $1/\sqrt{N}$ scaling achieved with independent probes and possibly obtain a scaling with $1/N$. 
EPR-entangled states have, for example, been proposed for use in LIGO~\cite{ma2017proposal}. 
In~\cite{brady2022entanglement}, the use of an array of mechanical sensors connected by entangled light was proposed, with applications for dark-matter searches (see Sec.~\ref{sec:equivalence:principle}). 
However, it has been shown that sensors with multicarrier optical modes do not outperform their single-mode counterparts~\cite{branford2018fundamental}. 
It has been experimentally demonstrated that using two optically entangled mechanical membranes leads to a 40\% improvement in the shot-noise dominant regime~\cite{xia2023entanglement} and allows a scaling better than $1/\sqrt{N}$. 

Another method for improving the precision of quantum sensors involves noise mitigation and engineering the surrounding noise bath. A structured non-Markovian environment was found to amplify the susceptibility for weak-force sensing with an optomechanical sensor~\cite{zhang2017optomechanical}. More broadly, the use of quantum error correction techniques has been proposed for quantum metrology~\cite{dur2014improved, kessler2014quantum}, even to the extent that the Heisenberg limit can be achieved~\cite{zhou2018achieving}. 

Yet another proposal for high-precision sensing, albeit challenging, is through the use of macroscopic superpositions in the sense of a large mass being in a quantum superposition of two distinct spatial locations. One method for generating such superpositions, particularly effective for large masses, is to couple a spin with a mass through a magnetic field gradient (a Stern-Gerlach mechanism, as described in Sec.~\ref{sec:entanglement:mediated:gravity}) \cite{scala2013matter,wan2016free,bose2016matter,bose2017spin,marshman2020mesoscopic,marshman2021large,zhou2022catapulting,zhou2023mass,margalit2021realization}, which followed on from general ideas to couple ancillary systems such as a quantized electromagnetic mode in a cavity with a mechanical object \cite{bose1997preparation,mancini1997ponderomotive,bose1999scheme,marshall2003towards,armata2017quantum,qvarfort2018gravimetry} or other ancillary quantum systems (superconducting qubits, etc) \cite{bose2006qubit,bose2006entangling,johnsson2016macroscopic}. Such quantum superpositions can be used to detect weak forces to a precision linear in time (essentially Heisenberg scaling), as the accumulated relative phase between the superposed components grows linearly in time. Moreover, at the end of such quantum ancilla-induced interferometry, the phase can be sensed by just measuring the ancilla. Example applications in the gravitational context involve detection of accelerations to very high sensitivity \cite{johnsson2016macroscopic,qvarfort2018gravimetry,marshman2020mesoscopic}, gravity gradient noise \cite{torovs2021relative}, space debris \cite{wutorovs2023quantum}, as well as the possibility to detect gravitational waves with a meter-sized compact interferometer for nano-objects \cite{marshman2020mesoscopic} (applications also exist outside the gravitational domain, eg, to detect neutrinos \cite{kilian2023requirements}). 

The quantum Fisher information (QFI) (see Sec.~\ref{sec:quantum:fisher:information})  allows us to consider sensitivities beyond the SQL. In the linearized optomechanical regime, the QFI has been considered for squeezed state inputs~\cite{lee2022quantum}. In the nonlinear regime of optomechanics (see Sec.~\ref{sec:QOMS:Hamiltonian}), the QFI was computed for detecting a constant~\cite{qvarfort2018gravimetry, armata2017quantum}, as well time-dependent gravitational potentials including gravitational waves~\cite{qvarfort2021optimal}. The QFI was also computed for an opto-magnon-mechanical setup, where the optomechanical system senses small changes in the separation between two magnets~\cite{iakovleva2023zeptometer}.

For further reading on sensing with mechanical resonators, we refer to the following dedicated reviews on sensing, which cover levitated systems~\cite{rademacher2020quantum}, hybrid optomechanical-BEC systems~\cite{motazedifard2021ultraprecision}, and cavity optomechanics~\cite{liu2021progress,li2021cavity}.

\subsubsection{Weak-force detection with BECs} \label{sec:BEC:gravimetry}

The mass of a BEC is generally lower than that of a composite quantum resonator, which means that it generally couples more weakly to gravity (see Sec.~\ref{sec:State-of-the-art} for a comparison of experimental parameters). However, the fact that all atoms in a BEC are identical makes it possible to control it extremely well in the laboratory. As such, BECs have been explored for force sensing. Gravity sensing with BECs can be done using trapped atoms, or atoms in free fall. In the free fall case, the precision depends on the time of flight. The time of flight in atom interferometry can be increased by using Bragg diffraction and Bloch oscillations of a BEC to slow down the particles~\cite{abend2016atom}. In such schemes, interactions are undesirable because they reduce the coherence time of the interferometer~\cite{pereira2017trapped}. However, interactions can be used to prepare initial states that have higher sensitivities~\cite{szigeti2020high}. Nevertheless, 
spatial interferometers cannot be reduced in size without losing precision. An alternative that could resolve this limitation is trapped BECs. 
Interactions in a trapped BEC give rise to phonons. Phonon modes are sharp in frequency while the atoms are completely delocalized within the trapped potential. Recent work shows that interferometry in the frequency domain using phonon modes can be used to miniaturize detectors while retaining high precision~\cite{howl2023quantum}. In frequency interferometry, the precision is limited by the lifetime of the states, not by the size of the system.  
 Squeezed states of phonon modes can be used to measure the gravitational field and its gradient with high precision~\cite{bravo2020phononic,bravo2019quantum} since the frequency of the modes is affected by the gravitational field. Phonon modes can also be used to measure oscillating gravitational fields, such as the acceleration and gradient of an oscillating mass close to the BEC~\cite{raetzel2018dynamical}. 
Single phonon measurement precisions have been reached in BEC analog experiments~\cite{steinhauer2022confirmation}. The most relevant limiting factor is particle loss due to three-body recombination. 
The resonance of phonons modes to external gravitational fields has been proposed to detect high-frequency gravitational waves and searches for dark matter~\cite{sabin2014phonon,howl2023quantum}. A BEC trapped in a double well has been proposed in searches of dark energy~\cite{hartley2019quantum}, and a proposal to show that gravity degrades entanglement between two BEC in a space-based experiment was presented in~\cite{bruschi2014testing}. While there are only theoretical proposals, the center of mass oscillations of a BEC has been used to measure Casimir-Polder forces in the lab~\cite{harber2005measurement}. This work shows that BEC technology is useful in measuring very small forces.

\subsubsection{Deviations from the Newtonian potential} \label{sec:modified:gravity}
An open problem in modern physics is the discrepancy between the observation of a small cosmological constant and the predicted value from particle physics theory~\cite{padilla2015lectures}. Modified gravity theories (MGTs) provide a solution to this dilemma, in that some of them predict deviations from general relativity while simultaneously addressing the discrepancy with particle physics. For a review of MGTs, see~\cite{clifton2012modified}. 
To address the fact that no deviations from general relativity have thus far been observed, mechanisms are introduced to explain the absence of large deviations in tests that have thus far been performed. For example, one such proposal known as a chameleon mechanism~\cite{khoury2004chameleon,khoury2004chameleonb,brax2004detecting} resolves the discrepancy via the introduction of a screening mechanism that depends on the local mass density. Note, however, that they do not solve the cosmological constant problem. According to the proposal, deviations in regions with high density, such as the solar system, are suppressed. Instead, high vacuum and extremely sensitive laboratory tests are needed. For current bounds on chameleon theories, see the Figures in~\cite{burrage2018tests}. 

Most MGTs can be parameterized into the following Yukawa-like modification to Newton's potential:
\begin{align}
V(r) = - \frac{G M_S m_p}{r}\left( 1 + \alpha \, e^{- r/\lambda} \right),  \label{eq:Yukawa}
\end{align}
where $M_S$ is the source mass, $m_p$ is the probe mass (not to be confused with the Planck mass $M_P$), $\alpha$ is a dimensionless modification to the strength of the potential, and $\lambda$ is a length-scale beyond which the modification is exponentially suppressed. Current solar-system tests of Newton's laws have considerably constrained the free parameters of such modified theories; see, e.g., Figure 8 in~\cite{murata2015review}. The parameter regimes that remain to be excluded include small $|\alpha|$ and $\lambda$, which correspond to the detection of extremely weak forces at short range. 

The main avenue for searches for MGTs with massive quantum systems is via precision tests of gravity, which we covered in the previous section. Several experiments have already been performed with mechanical resonators in the classical regime to bound deviations from the Newtonian potential, including with cantilevers~\cite{chiaverini2003new}. A key advantage for mechanical resonators such as levitated systems is that they are relatively confined in space and, therefore, can be used to test extremely short-length scales. 
See~\cite{moore2021searching} for a review of searches for new physics with optically levitated sensors. 
Additional proposals have been put forward for tests with levitated optomechanical devices~\cite{blakemore2021search, chen2022constraining}, as well as specific tests of the chameleon mechanism~\cite{betz2022searching}. 
However, while the larger mass of mechanical resonators increases the strength of the Newtonian potential and thus the deviations, their larger volume brings with it additional challenges. For example,  in the context of the chameleon mechanism, the large radius of, for example, a levitated nanomechanical resonator was found to additionally screen the interaction~\cite{qvarfort2022constraining}. Another major challenge is the presence of Casimir forces, which increase for small distances. See~\cite{onofrio2006casimir} for a review of measurements of Casimir forces in the context of searching for deviations from the Newtonian potential.

\subsubsection{Tests of the equivalence principle and dark matter searches} \label{sec:equivalence:principle}
The equivalence principle (EP) states that all forms of matter couple to gravity in the same way. An additional formulation known as the weak equivalence principle (WEP) states that gravitational mass and inertial mass are the same. Violation of the EP can be indicative of modified theories of gravity~\cite{hui2009equivalence} or physics beyond the Standard Model~\cite{damour2012theoretical}.

As we pointed out in Sec.~\ref{sec:postulates}, there is a conflict between the EP and quantum mechanics.
Nevertheless, several ideas have been put forward for testing the EP with quantum systems. Perhaps the simplest is a classical test of the Eötvös ratio, which defines the correlation between inertial mass and gravitational mass and which serves as a test of the WEP. The Eötvös ratio is defined as 
\begin{align}
    \eta_{A - B} = 2 \frac{|a_A- a_B|}{|a_A + a_B|} = 2 \frac{ |(m_i/m_g)_A - (m_i/m_g)_B|}{|(m_i/m_g)_A  + (m_i /m_g)_B|}, 
\end{align}
where $a_A$ and $a_B$ are the accelerations of bodies $A$ and $B$, and where $m_i$ and $m_g$ are the inertial and gravitational mass, respectively. The advantage of using quantum systems mainly pertains to the increased precision that they offer as sensors. Most of the measurements of the Eötvös ratio with quantum systems have been carried out through atom-interferometry (see e.g.~\cite{schlippert2014quantum, duan2016test,  overstreet2018effective, albers2020quantum, asenbaum2020atom}). The best Eötvös ratios achieved to-date is that of the MICROSCOPE mission, at $(1.5 \pm 2.3\,  (\mathrm{stat}) \pm 1.5 \, (\mathrm{sys})) \times 10^{-15}$~\cite{touboul2022microscope}. See also Sec.~\ref{sec:tests:with:atoms} for an overview of state-of-the-art tests with cold atoms. 

In general, there appears to be no clear consensus in the community on how the EP should be formulated for quantum systems since there are often additional aspects that need to be taken into account.
One of the earliest works on this topic showed that for the simple case of a particle in an external gravitational field, the WEP does not apply to a quantum-mechanical description of the problem~\cite{greenberger1968role}. 
While the classical equations of motion can be made independent of mass, the same is not true for quantum mechanics since mass enters into the quantization rules. However, the opposite point of view has also been argued. Starting from linearized gravity perturbations as a massless, spin-two gauge field coupled to itself and to matter, the equivalence principle must hold for quantum systems for the theory to be consistent~\cite{davies1982quantum}. It has also been argued that, for quantum particles in free fall, their expectation values for position and momentum are consistent with the WEP~\cite{viola1997testing}.

The influence of internal degrees of freedom of quantum systems on the formulation of the EP has been raised in several works. Since the EP stipulates equivalence between mass and energy, it must take into account the internal (potentially superposed) energy states of a quantum system. Based on this, a quantum formulation of the EP has been proposed, which requires equivalence between the rest, inertial and gravitational internal energy operators~\cite{zych2018quantum}.  An experimental test based on this proposal was put forwards in~\cite{orlando2016test} using trapped spin$-\frac{1}{2}$ atoms and later performed using a Bragg atom interferometer~\cite{rosi2017quantum}. The test provided constraints on the off-diagonal elements of the mass operators and additional constraints of the Etvös ratio for the WEP. Similarly, the WEP can be explicitly considered for internal degrees of freedom. Two distinct formulations of the WEP were proposed in~\cite{anastopoulos2018equivalence}. The first states that the probability distribution of position for a free-falling particle is the same as the probability distribution of a free particle (up to a mass-independent shift of its mean).  The second states that any two particles with the same velocity wave function behave identically in free fall, irrespective of their masses. It has also been stipulated that a quantum version of the EP should be linked to a notion of causality~\cite{hardy2018construction} since it is always possible to transform to a quantum reference frame in which we have a definite causal structure in the local vicinity of any point. Here, the notion of a quantum reference frame refers to frameworks developed in~\cite{giacomini2019quantum,guerin2018observer} and related work~\cite{de2022quantum}. 

Tests of the EP can also aid the search for dark matter since some dark matter models (such as light scalar dark matter~\cite{hees2018violation}) predict deviations from the EP. 
 See~\cite{carney2021mechanical,kilian2024dark} for an overview of quantum sensing with mechanical resonators for the detection of dark matter. 
It has been proposed that mechanical oscillators with masses below or around one kilogram operating near the standard-quantum limit could be used to detect ultra-light dark matter candidates~\cite{carney2021ultralight}. In addition, an optical cavity with mirrors made of different materials could facilitate coupling channels for vector dark matter~\cite{manley2021searching}. 
Heavier dark-matter candidates can also be detected by scattering off a mechanical resonator. A large array of femtogram masses could potentially detect dark matter candidates around 10 keV, with the advantage that they also provide directional sensing~\cite{afek2022coherent}. In addition, a recent white paper focuses on detecting deviations in momentum kicks resulting from exotic decay processes~\cite{brodeur2023nuclear}. Here, the position of levitated spheres is carefully monitored using displacement sensing.

%
\subsection{Gravitational decoherence,  semi-classical models, self-energy and gravitationally-induced wavefunction collapse} \label{sec:grav:dec:section}
Gravity can impart a coherent signal on the quantum state, which can be detected using quantum metrology tools (see Sec.~\ref{sec:quantum:enhanced:sensing}). However, there are a number of theoretical proposals where the quantum state no longer follows a unitary evolution when interacting with gravity. Here we cover proposals ranging from decoherence arising from quantum and stochastic gravity to modifications to the Schrödinger equation.  We also refer to the following comprehensive reviews dedicated to these topics~\cite{ anastopoulos2022gravitational, bassi2017gravitational}. 

\subsubsection{Gravitational decoherence} \label{sec:gravitational:decoherence}
%
Decoherence is the process by which off-diagonal elements in the density matrix of a quantum system are gradually reduced to zero. There are many proposals for how an external gravitational field interacts with the quantum system to cause decoherence, as well as dissipation and thermalization, in the regime of gravity at low energies. 
Unlike other sources of decoherence, such as from fluctuating electromagnetic fields,  gravitational decoherence is universal, and its influence cannot be shielded. 

The common starting point for most gravitational decoherence proposals is the linearized metric
\begin{align}
    g_{\mu \nu}  = \eta_{\mu \nu} + h_{\mu \nu}, 
\end{align}
where $\eta_{\mu \nu}$ is the Minkowski background spacetime, and $h_{\mu \nu}$ denote the fluctuations. Fluctuations can emerge within the perturbative quantum theory of gravity (see Eq.~\eqref{eq:perturbative}), can be postulated in a fundamentally classical theory of gravity (see the discussion around  Eq.~\eqref{eq:grav:random}), or could also emerge as a consequence of a minimum length scale of the spacetime fabric~\cite{hossenfelder2013minimal}.  Regardless of its physical origin, the fluctuations of the gravitational field are expected to decohere a quantum system similarly as any other fluctuating field (see Sec.~\ref{sec:open:system:dynamics}). 

In most of the proposals, the structure of the dynamics of a massive quantum system moving along one axis is captured by the following master equation:  
\begin{equation} \label{eq:basic:master}
     \frac{\partial}{\partial t} \hat \rho  = - \frac{C}{2} [\hat{A}, [\hat{A}, \hat \rho]],
\end{equation}
where we have omitted the Hamiltonian terms for brevity, and $C$ ($\hat{A}$) is a constant prefactor (an operator) specific to the model (see Eq.~\eqref{eq:Lindblad:equation} with only the jump operator $\hat{L}_1=\sqrt{C}\hat{A}$). It is then easy to obtain the decoherence rate $\gamma$ in the eigenbasis of the operators $\hat{A}$. Suppose  $a_1$ and $a_2$ are two real-valued eigenvalues of the operator $\hat{A}$.  By applying $\langle a_L \vert$ ($\vert a_R\rangle$) from the left (right) on Eq.~\eqref{eq:basic:master}, and multiplying by $2$,  we readily find:
\begin{equation} \label{eq:basic:rate}
\gamma= {C \Delta a^2},    
\end{equation}
where we have defined $\Delta a= a_L- a_R$ (which can be interpreted as the superposition size). When $\hat{A}$ is not Hermitian, the analysis in Eqs.~\eqref{eq:basic:master} and~\eqref{eq:basic:rate} requires generalizations (see details in the referenced works below).

In~\cite{anastopoulos2013master,blencowe2013effective} gravitational waves (forming an environmental bath) were considered as a source for the fluctuations $h_{\mu \nu}$. It was found that a free particle should decohere with the operators in Eq.~\eqref{eq:basic:master} given by kinetic energy $\hat{A}=\hat{p}^2/2m$, where $\hat{p}$ is the momentum operator and $m$ the particle mass. The characteristic decoherence rate is given by 
\begin{equation}
    \gamma= \frac{9}{32\pi \tau_P } \frac{T_P}{\Theta} \left(\frac{\Delta E}{E_P}\right)^2,
\end{equation}
where $\tau_P$ ($T_P$) is the Planck time (Planck temperature), $\Delta E$ ($E_P$) is the difference in kinetic energy {of a zero
momentum and a finite momentum state} (Planck energy), and $\Theta$ is a free parameter of the model. In earlier works, a general non-Markovian master equation for the interaction between $N$ gravitating quantum particle was derived~\cite{anastopoulos1996quantum}, a complementary analysis was given in~\cite{oniga2016quantum}, and a generalization for photons is discussed in~\cite{lagouvardos2021gravitational}.

Decoherence due to the emission of gravitational waves was studied in~\cite{suzuki2015environmental}.
A relation between decoherence and the classical limits in terms of the quadrupole radiation formula and backreaction dissipation was discussed in~\cite{oniga2017quantum}.  The analysis from~\cite{torovs2023loss} recovered the classical results for a linear quadrupole and showed that only systems with quadrupoles would decohere, while a free particle would not decohere. For the simplest case of a harmonically trapped particle (which has a linear quadrupole), it was found that the decoherence operator is $\hat{A}=\hat{b}^2$ (with $\hat{b}$ the mode operator of the harmonic oscillator). The associated decoherence rate for number states $\vert n \rangle$ is given by
\begin{equation}\label{formula1}
    \gamma = \frac{64 \omega}{15} \left(\frac{E}{E_P}\right)^2 (\langle\hat{n}^2\rangle- \langle\hat{n}\rangle) ,
\end{equation}
where $\omega$ is the frequency of the harmonic trap,  $E=\hbar \omega$ ($E_P$) is the difference between the energy levels (Planck energy), and $\hat{n}$ denotes the number operator.
Starting from Einstein's equivalence principle, it was found that the emission of quantized gravitational waves can only happen via the transition $n\rightarrow n-2$, which is prohibited for the states $\vert 1\rangle$ and $\vert 0\rangle$.  The number states $\vert 0\rangle$,  $\vert 1\rangle$, or any superposition of these states, are thus protected from decoherence via quantized gravitational waves resulting in a vanishing decoherence rate in Eq.~\eqref{formula1}. For a discussion about the bremsstrahlung effects, see~\cite{weinberg1965infrared}. 

The decoherence effect induced by gravitons in the context of gravitational wave detectors has been discussed in~\cite{parikh2020noise,parikh2021quantum}, while in~\cite{kanno2021noise} the analysis considered also matter-wave interferometry. The approximate decoherence rate obtained for an interferometer consisting of two paths (i.e., left and right paths) is given by:
\begin{equation}
    \gamma= 10\, \Omega_m \left(  \frac{m v}{m_p c}\right)^{\frac{1}{2}},
\end{equation}
where $m$ ($m_p$) is the mass of the system (Planck mass),  $2v$ is the relative speed between the states following the left and right paths of the interferometer,  and $\Omega_m$ is a high-frequency cut-off (which is inversely proportional to the superposition size, $\Delta x$, i.e., $\Omega_m\sim c/\Delta x$).

A class of stochastic models can be obtained by considering the non–relativistic limit of classical field equations and assuming stochastic fluctuations of the metric. Starting from the Klein–Gordon equation, it was found that the decoherence operator is $\hat{A}=\hat{p}^2/2m$, where $\hat{p}$ is the momentum operator, and $m$ the particle mass~\cite{Breuer2009Metric} . The decoherence rate is given by 
\begin{equation}
    \gamma= \frac{\tau_c}{\hbar^2} \Delta E^2,
\end{equation}
where $\tau_c$ is a free parameter characterizing the correlation time of the stochastic bath, and $\Delta E$ is the difference in kinetic energy. A generalized analysis using the Foldy-Wouthuysen method capturing higher order corrections has been performed in~\cite{asprea2021gravitational}.

Another proposal for decoherence is related to composite quantum particles such that individual parts of the systems follow different geodesics. In~\cite{pikovski2015universal}, it was shown that gravity entangles the internal and center-of-mass degrees of freedom, which in term decoheres the center-of-mass degrees of freedom of a system, e.g., of a crystal. An equation of the form in Eq.~\eqref{eq:basic:master} was obtained with the decoherence operator $\hat{A}=\hat{x}$, where $\hat{x}$ is the center-of-mass position operator. The decoherence rate was given as:
\begin{equation}
\gamma= \frac{\sqrt{N} g k_b T \Delta x}{\sqrt{2} \hbar c^2}, 
\end{equation}
where $N$ is the number of degrees of freedom in the crystal, $g$ is the Earth's gravitational acceleration, $k_b$ is the Boltzmann constant,  $T$ is the temperature, and $\Delta x$ is the spatial superposition size. The proposal has been the source of much discussion in the community (see~\cite{bassi2017gravitational} and~\cite{pikovski2017time} for a summary of the discussions).

Experimental signatures on matter-wave interferometers have been analyzed in~\cite{lamine2006decoherence, asprea2021gravitational,wang2006quantum}.
While many of these proposals suggest different mechanisms behind the gravitationally induced decoherence, many result in similar reductions of the density matrix elements. As a result, there are tests that search for gravitational decoherence regardless of its origin. 
Some of the first proposals for testing gravitational collapse and decoherence with optomechanical systems were those in~\cite{bose1999scheme} and~\cite{marshall2003towards}. In this protocol, a single optical mode is passed through a beam-splitter and into an interferometer, wherein one of the arms interacts with a mechanical resonator according to the optomechanical Hamiltonian in Eq.~\eqref{eq:optomechanical:hamiltonian}. 
The protocol has been analyzed with the addition of mechanical position-damping noise~\cite{bassi2005towards,adler2005towards} (see Sec.~\ref{sec:master:equations}). Additionally, the protocol has been examined in the high-temperature limit~\cite{bernad2006quest}. It was further developed in~\cite{kleckner2008creating}, where a number of practical aspects were taken into account. To date, such an experiment has not yet been performed. It has been pointed out that the phases picked up through the optomechanical interaction can partially be reproduced through classical dynamics~\cite{armata2016quantum}. 

Should gravity cause decoherence, it is possible that the mechanism itself relies on a modification of quantum mechanics. If the experiments are still modeled using standard quantum mechanics, it can become difficult to distinguish the decoherence effects accurately. Such a scenario has been considered in Ref~\cite{pfister2016universal}, where a general information-theoretic measure of decoherence was proposed. 
While many approaches consider continuous variable systems, there are also results for qubit states~\cite{kok2003gravitational}.

\subsubsection{Nonlinear modifications} \label{sec:sch-new}

A key cornerstone of quantum mechanics is that the Schr\"{o}dinger equation is a linear equation for the state of the system $\vert\psi\rangle$, and that the expectation value of an observable $\hat{O}$ is a bilinear function of the state, e.g., $\langle  \psi  \vert \hat{O} \vert \psi\rangle$.  We can, however, devise a modification of Quantum mechanics where the dynamics become nonlinear in the state $\vert\psi\rangle$ or construct expectation values with a non-bilinear dependency on the state~\cite{weinberg1989testing,weinberg1989precision}.  Such modifications are motivated by the measurement problem (see Sec.~\ref{sec:postulates}) as well as emerge from elementary considerations about semi-classical gravity. Semi-classical gravity is viewed by some as an effective theory, i.e., an approximation to a fundamental quantum theory of gravity. Within this approximation, one can investigate the back-reaction of matter on the gravitational field, generalizing the results of quantum field theory in curved spacetime (see Sec.~\ref{sec:QFTCS}), as well as model classical stochastic fluctuations of the gravitational field~\cite{hu2008stochastic}. An alternative viewpoint advocated by others is that semi-classical gravity is not a mere approximation but the fundamental theory where gravity remains classical whilst matter is quantized~\cite{kibble1981is}.

A conceptual start for such a nonlinear modification of quantum mechanics is given in the semi-classical Einstein equations~\cite{moller1962theories,rosenfeld1963quantization,ruffini1969systems}:
\begin{equation}
G_{\mu\nu}=\frac{8\pi G}{c^4}\langle \hat{T}_{\mu\nu}\rangle, \label{3C1}
\end{equation}
where on the left-hand side, we have the classical Einstein tensor $G_{\mu\nu}$, and on the right-hand side, the expectation value of the quantum stress-energy tensor $\hat{T}_{\mu\nu}$ taken with respect to the state of the quantum matter. The coupling in Eq.~\eqref{3C1} is arguably the simplest way to couple a classical gravitational field to quantized matter, but more importantly, it is the expected theory when the matter is in well-localized states. In such a case, matter can still be approximately described using a classical stress-energy tensor $T_{\mu\nu}$ such that $  T_{\mu\nu}\approx \langle \hat{T}_{\mu\nu}\rangle$, but beyond this regime, e.g., when we have spatial superpositions,  there is no consensus about its validity as we discuss below. 

In the non-relativistic regime, the gravitational field acting on 
by a particle in the state $\vert\psi\rangle$ thus depends on the value of $\langle \psi\vert \hat{T}_{\mu\nu} \vert\psi \rangle$.
Hence we expect the associated nonlinear Schr\"{o}dinger equation 
to have a cubic dependency on the state of the matter system, i.e.,  
\begin{equation}
i\hbar\frac{\mathrm{d}}{\mathrm{d}t} \vert \psi\rangle \propto G \langle \psi \vert \hat{T}_{\mu\nu} \vert\psi \rangle\vert \psi\rangle.    \label{SN11}
\end{equation}
Such an equation has been proposed in~\cite{diosi1984gravitation,penrose1996gravity,penrose1998quantum}, and has become known as the Schr\"{o}dinger-Newton equation.  A derivation from first principles of Eq.~\eqref{SN11} is however still a subject of debate~\cite{christian1997exactly,adler2007comments,anastopoulos2014newton,anastopoulos2014problems,giulini2012schrodinger}.

Such a hybrid quantum matter-classical gravity model has its appeal in conceptual simplicity of Eq.~\eqref{3C1}, with testable predictions differing from those arising from the framework of perturbative quantum gravity (see Sec.~\ref{sec:QLG:and:PQG}). However, unlike the latter, which is a fully consistent relativistic theory, deterministic nonlinear modifications of the Schr\"{o}dinger equation, such as the Schr\"{o}dinger-Newton equation, are at odds with the requirement of no-faster than light signaling to make them, at least conceptually, unsatisfactory~\cite{gisin1989stochastic,polchinski1991weinberg}. 

Nonetheless, to date, no laboratory experiment has been able to rule out the Schr\"{o}dinger-Newton equation. Furthermore, it has been suggested that it might be possible to resolve the issue of superluminal signaling if one takes into consideration the measurement problem with a suitable prescription of the wave-function collapse~\cite{bahrami2014schrodinger,Bera2015Stochastic}, and hence the predictions of the Schr\"{o}dinger-Newton equation might still remain valid in specific domains (see Sec.~\ref{sec:grav:collapse} for a discussion of possible modifications). For a discussion within the context of Generalized Probabilistic Theories (GPT) and its relation to classical state space, see~\cite{Mielnik1974Generalized,Mielnik1980Mobility,galley2022no}.
We here provide below a summary of the current experimental endeavors to test the Schr\"{o}dinger-Newton equation. 

Starting from the semi-classical Einstein equations in  Eq.~\eqref{3C1} we obtain in the non-relativistic limit the one-particle Schrödinger-Newton equation~\cite{diosi1984gravitation}:
\begin{equation}
    i\hbar \frac{d}{dt}\ket{\psi_t} =  \frac{\hat{p}^2}{2m} \ket{\psi} -  G m^2 \int \mathrm{d} \mathbf{s} \frac{\vert \psi(t, \mathbf{s}) \vert^2}{|\hat{\mathbf{r}} - \mathbf{s}|} \ket{\psi}, \label{SN1}
\end{equation}
where  $m$ is the mass of the particle, $\hat{r}$ ($\hat{p}$) is the position (momentum) operator, and we have introduced the wavefunction $\psi(t,\mathbf{s})=\langle \mathbf{s} \vert\psi_t\rangle$. The generalization to the $N$-particle case can be obtained from Eq.~\eqref{SN1} by replacing the source of the gravitational field with $\vert \psi(t,\mathbf{s}_1,..,\mathbf{s}_N)\vert^2$:
\begin{equation}
    i\hbar \frac{d}{dt}\ket{\psi_t} =  -  G m^2 \sum_{j,k} \int \prod_l \mathrm{d} \mathbf{s}_l \frac{\vert\psi(t,\mathbf{s}_1,..,\mathbf{s}_N)\vert^2}{|\hat{\mathbf{r}}_j - \mathbf{s}_k|} \ket{\psi}, \label{SN2}
\end{equation}
where we have omitted the kinetic terms for brevity, and $j,k=1,..N$.

Eqs.~\eqref{SN1} and ~\eqref{SN2} form the starting point for a number of experimental proposals. While some analytical results can be obtained~\cite{tod1999analytical} in most situations, one has to resort to numerical simulations to make quantitative predictions using the nonlinear  Schrödinger-Newton equation, similarly as in the case of the formally similar Gross-Pitaevskii equation~\cite{gross1961structure,pitaevskii1961vortex}. The key prediction of the Schr\"{o}dinger-Newton equation is the modification of the free spreading of the wavefunction as the last term in Eq.~\eqref{SN1}, with its Newtonian-like 1/r dependency, can be viewed as a self-gravity term which tends to localize the system in space. There are a number of papers investigating the free-spreading in space with the required parameter regime for experimental tests~\cite{colin2016crucial,giulini2011gravitationally,carlip2008is,moroz1998spherically,bahrami2014schrodinger} as well as proposals to test secondary effects in harmonic traps such as squeezing~\cite{yang2013macroscopic} and energy shifts~\cite{grossardt2016optomechanical}. 

Additional dependencies on the state of the system $\vert\psi\rangle$ can also be introduced to other parts of the quantum formalism~\cite{sorkin1994quantum}. Motivated by considerations about general covariance, it has been argued that all quantities in physical theories must be dynamical~\cite{norton1993reports} suggesting corrections to the Born rule~\cite{berglund2022gravitizing}. Cubic corrections to the Born rule $\mathcal{O}(\vert\psi\rangle^3)$, i.e., triple interference phenomena,  have been theoretically discussed in the context of the Talbot interferometer~\cite{berglund2023triple}.
The class of nonlinear modifications introduced in ~\cite{weinberg1989testing,weinberg1989precision} have also been recently analyzed in the context of gravitationally induced entanglement~\cite{Spaventa2023On}.

%
\subsubsection{Nonlinear and stochastic modifications} \label{sec:grav:collapse}
%
The quest of unifying quantum mechanics and gravity into a single theory and the measurement problem from quantum foundations appear to be two distinct problems at first (see Sec.\ref{sec:postulates}). However, using an elementary analysis, it was shown that there appears to be a deep conflict between the superposition principle in quantum mechanics and the equivalence principle of general relativity~\cite{penrose1986gravity}. Such a result could be viewed as another hint for the necessity to modify gravity i.e., constructing a quantum theory of gravity. However, any theory where the superposition principle remains valid, would not resolve the tension between quantum and classical physics, thereby leaving unanswered the measurement problem. Another option is that the conflict instead signifies the need to also modify quantum mechanics to accommodate notions of gravity, i.e., gravitization of quantum mechanics~\cite{penrose1986gravity,penrose1996gravity,penrose1998quantum,penrose2014gravitization}. Such a theoretical program, whilst still in its tentative state, suggests that it might be possible to consistently couple classical and quantum systems (in this context, gravity and matter, respectively) as well as solve the measurement problem simultaneously. We, however, remark that other programs for the emergence of classicality, fully compatible with (unmodified) quantum mechanics, are also considered in the literature~\cite{giulini2000decoherence}. 

The measurement problem of quantum mechanics, still to this day subject of controversy, has its roots in the two prescriptions for the evolution of quantum systems: on the one hand, the Schr\"{o}dinger equation is deterministic and linear, while, on the other hand, the wave-function collapse postulate induces a stochastic and nonlinear evolution of the state. It was shown that it is possible to combine the two types of prescriptions into a single dynamical law, thus avoiding the dichotomy, with quantum dynamics the limit for microscopic systems and classical dynamics the limit for macroscopic systems~\cite{ghirardi1986unified}. 

The structure of such modifications forms the basis for the family of spontaneous wave-function collapse models~\cite{bassi2003dynamical, bassi2013models}, with the basic form given by the following stochastic differential equation~\cite{ghirardi1990markov}
\begin{equation}
    \frac{d }{ dt} \vert \psi_t\rangle=\sqrt{\lambda} (\hat{A}-\langle  \hat{A} \rangle) \frac{d W_t}{dt} \vert \psi_t \rangle -\frac{\lambda }{2} (\hat{A}-\langle \hat{A} \rangle)^2 \vert \psi_t\rangle, \label{scm}
\end{equation}
where $\vert \psi_t\rangle$ is the state-vector, $\hat{A}$ is the operator, $\langle \psi_t \vert \hat{A}\vert  \psi_t \rangle$ is the expectation value, $dW_t$ is the Wiener increment, and $\lambda$ is the coupling rate. The models based on Eq.~\eqref{scm} make a series of predictions that are expected to be tested with the next generation of experiments, e.g., loss of interferometric visibility, anomalous heating of free systems, and X-ray emission~\cite{bassi2013models,carlesso2022present}. In contrast to the case of deterministic nonlinear modifications (see Sec.~\ref{sec:sch-new}), the stochastic nature of the evolution in  Eq.~\eqref{scm} conspires with the nonlinear terms to avoid the possibility of superluminal signaling, making such models conceptually more appealing, albeit a relativistic extension of such models is still an open problem~\cite{bedingham2014matter}. Collapse models have already been tested experimentally, including by mechanical systems in a non-interferometric way~\cite{vinante2016upper, helou2017lisa, vinante2020testing, forstner2020nanomechanical}, which provide the strongest experimental bound as of today, see~\cite{carlesso2022present} for a review on testing collapse models. 

As such, the connection between dynamical collapse models of the form in Eq.~\eqref{scm} and gravity remains tentative to date, with a derivation from first principles still an open question~\cite{bahrami2014role}. Nonetheless, using an elementary analysis, considering a spatial superposition, it has been argued that the system should decohere within a time given by $\tau=\frac{\hbar}{E_g}$, where $E_g$ is the gravitational self-energy of the difference between the mass distributions of the two states in superposition~\cite{penrose1986gravity,penrose1996gravity,penrose1998quantum,penrose2014gravitization}. For example, for a spherical mass distribution we have the following formula~\cite{penrose2014gravitization}:
\begin{equation}
    E_g= 
    \begin{cases}
      \frac{G m^2}{R} (2 \lambda^2 -\frac{3}{2}\lambda^3 +\frac{1}{5}\lambda^5) & \lambda \leq 1 \\
      \frac{G m^2}{R} (\frac{6}{5}-\frac{1}{2\lambda})  & 1 \leq \lambda  
    \end{cases}, \label{self-energy}
\end{equation}
where $m$ is the total mass, $R$ is the particle radius, $\lambda=\Delta x/(2R)$, and $\Delta x$ is the superposition size.

Although self-gravity has been extensively investigated in general relativity~\cite{lynden-bell1961stellar}, writing a fully consistent relativistic spontaneous collapse model remains an open problem. In the non-relativistic limit, one can nonetheless construct a wavefunction collapse model inspired by Newtonian gravity~\cite{diosi1987universal,diosi1989models}, which recovers the prediction for Penrose's decoherence time $\tau$~\cite{diosi2005intrinsic,diosi2007notes}. A drawback of the proposed model is, however, that it requires a short-distance cut-off to avoid the divergence for point-like microscopically mass distributions~\cite{ghirardi1990continuous}. We briefly discuss experimental bounds on the short-distance cut-off, usually labeled as $R_0$, at the end of this section.

There have been a series of investigations aiming to derive the collapse of the wave function from an underlying mechanism related to random fluctuations of spacetime:
\begin{equation} \label{eq:grav:random}
    g_{\mu\nu}=\bar{g}_{\mu\nu}+\delta g_{\mu\nu},
\end{equation}
where $g_{\mu\nu}$ is the spacetime metric, $\bar{g}_{\mu\nu}$ denotes a fixed background, and $\delta g_{\mu\nu}$ denotes the stochastic fluctuations. One of the earliest such attempts posited that the wave-function collapse could be induced by real-valued fluctuations of the space-time metric related to the Planck scale~\cite{karolyhazy1966gravitation}. However, the obtained model is still compatible with the superposition principle, and it seems to be at odds with the predicted X-ray emission from charged particles~\cite{diosi1993calculation}. An alternative idea with complex-valued fluctuations of the space-time metric was also proposed~\cite{adler2004quantum}, with a possible model of the basic form of Eq.~\eqref{scm} constructed in~\cite{gasbarri2017gravity}. 

A collapse-like dynamics of the form in Eq.~\eqref{scm} also appears in the context of quantum measurement and control~\cite{wiseman2009quantum}, which has been exploited to construct models of semi-classical gravity. Specifically, by continuously measuring the system, one has access to the signal $I(t)$ given by Eq.~\eqref{eq:measurementRecord}.
The signal $I(t)$ is used in experiments to gather information about the state of the system as well as to control the evolution of the system at future times, i.e., by creating a feedback loop. However, in this context, the system is not measured by an actual experimentalist or physical measurement apparatus, but it is instead postulated that such dynamics, resembling continuous measurements, is a fundamental law of nature~\cite{diosi2018how}. Such an approach has been used in the Kafri-Taylor-Milburn model~\cite{kafri2014classical} to construct a semi-classical (linearized) Newtonian interaction. We recall that the quantum interaction, arising in standard quantum mechanics, is given by
\begin{equation}
    -\frac{G m_1 m_2}{\vert (d +\hat{x}_1)-\hat{x}_2 \vert } \approx \frac{2 G m_1 m_2}{d^3} \hat{x}_1\hat{x}_2,  \label{standard}
\end{equation}  
where $d$ is the mean distance between the two masses $m_1$, $m_2$, and $\hat{x}_1$,$\hat{x}_2$ are the position operators, respectively (see Eq.~\eqref{eq:newt_pot_exp}). In~\cite{kafri2014classical}, using a formalism reminiscent of quantum measurement and control outlined above, one instead finds in place of Eq.~\eqref{standard} a modified potential: 
\begin{equation}
     \frac{G m_1 m_2}{d^3} (\hat{x}_1\langle \hat{x}_2\rangle+\langle\hat{x}_1\rangle \hat{x}_2).
\end{equation}
A related approach has also been considered in~\cite{tilloy2016sourcing}, where the matter density of the system $\rho$  is continuously monitored, producing the signal given by
\begin{equation}
    \rho_t= \langle \rho_t \rangle + \delta \rho_t,
\end{equation}
where $\langle \rho_t \rangle$  is the expectation value taken with respect to the state of the system. In~\cite{tilloy2016sourcing} it has been shown that when $\rho_t$ is the source of Newtonian potential $\phi$ in the Poisson equation, i.e. $\nabla^2 \phi=4\pi G \rho_t$, one is able to recover the standard quantum Newtonian interaction among particles in Eq.~\eqref{standard}, as well as the terms appearing in the Diosi model discussed above~\cite{diosi1987universal,diosi1989models}. 

Another approach that is related to the Diosi model is given by hybrid quantum-classical models~\cite{oppenheim2018post}. In such models, working in the ADM formalism~\cite{arnowitt1959dynamical} (see e.g.,~\cite{poisson2004relativist} for an introduction), the gravitational field is described by a probability density in (classical) phase space $\rho(z)$, where we attach to each point $z$ in phase-space a distinct density matrix $\hat{\sigma}(z)$. As a result, the total state of the system (comprising gravity and matter) is described by the classical-quantum state given by:
\begin{equation}
    \hat{\rho}_{\text{cq}}= \int d\ \rho(z) \vert z \rangle \langle z\vert \otimes \hat{\sigma}(z).
\end{equation}
We can construct a master equation for such a dynamics; by tracing away the state of gravitational field  $\rho(z)$ we obtains a master equation for the matter state (while tracing away the matter system one obtains a Fokker-Planck like equation for the gravitational field providing correction to general relativity). In the former case, we can, by suitably restricting the general form of the initial equation, recover the master equation arising from Eq.~\eqref{scm} such as the one given by the Diosi model~\cite{oppenheim2022gravitationally}.  

Testing spontaneous collapse models falls into two broad categories: interferometric and non-interferometric tests. In the former, the signature is a loss of interferometric visibility~\cite{torovs2018bounds}, where the current record is provided by experiments with macromolecules~\cite{fein2019quantum}, while the latter is a broad class of all other experiments~\cite{carlesso2022present}. To test the idea put forward by Penrose, one has to resort to direct tests of the superposition principle in interferometric tests (as a mathematical model has not yet been constructed). One way to test this idea is to use Bose-Einstein condensates~\cite{fuentes2018quantum,howl2019exploring}. On the other hand, the model put forward by Diosi, precisely formulated as discussed above, can also be tested with indirect non-interferometric tests. The model depends on a single free parameter $R_0$, which can be interpreted as the localization length-scale, which has been constrained using X-ray emission to values $R_0 < 5\times 10^{-10}$m~\cite{donadi2021underground,arnquist2022search}. 

\subsection{Entanglement mediated by gravity} \label{sec:entanglement:mediated:gravity}
It is not yet known whether gravity is fundamentally a quantum force. In this section, we will present arguments in favor of the view that detecting entanglement induced by gravitational interaction between two masses could finally help to settle this issue. We will also review proposals of  how to generate such entanglement. 
The realistic possibility of such an experiment has captured attention only recently. One reason for this raised interest was a proposal that showed that masses as small as micron-sized crystals, which can be isolated by levitation, can be fruitfully combined with spin qubit and quantum gate technologies developed for quantum computation to generate and witness gravitational entanglement \cite{bose2016matter,bose2017spin}. Simultaneously, a rationale based on a known result from quantum information theory, along with some basic assumptions, was also presented, which justified why such entanglement would evidence the quantum nature of gravity \cite{bose2016matter,bose2017spin,marshman2020locality} (see also \cite{marletto2017gravitationally} for alternative rationale). An important fact, noticed for the first time in the above references, is that one has to prepare and coherently maintain quantum superpositions having large spatial spreads for the entanglement growth to be realistically measurable. This, of course, makes the experiments challenging. We will review this class of schemes, as well as the recent rapid growth of literature which has been triggered by the above works. These dwell on the need to identify all the necessities needed for a practical realization, alternative proposals, as well as other ways to justify the conclusions about the quantum features of gravity from such experiments.    

In a historic verbal debate with other researchers on whether gravity is quantum, Feynman~\cite{dewitt2011role} advocated a thought experiment involving one mass in a spatial quantum superposition displacing a second mass due to its gravity (see Fig.~\ref{fig:Feynman-exp} of Sec. ~\ref{sec:controlling:massive:systems}). Although he stressed using quantum amplitudes to describe the setup, he did not explain {\em why} such an experiment will demonstrate the quantum character of gravity, nor did he clarify {\em what} to measure in order to make such conclusions. Also, crucially, he did not realize that the second mass also has to be in a highly delocalized state (spatial quantum superposition of comparable delocalization as the first mass) for the experiment to have a realistic duration.  Subsequently, a version of Feynman's setup was suggested by supposing that the large mass in a spatial superposition was a Bose-Einstein condensate~\cite{lindner2005testing}.
Such a superposition demands {\em all} atoms in one well or another; it is not an easy state to generate. It was then suggested that the interference pattern of a mesoscopic mass in a momentum state (again, not easy to prepare) be measured, scattering gravitationally from this superposition. However, a single particle measurement does not suffice to reveal entanglement without extra assumptions or procedures. Moreover, it is not justified exactly why such an interference pattern would constitute a test for quantum-natured gravity. In \cite{schmole2016micromechanical}, after presenting measurements of gravity for the smallest masses (milligram scale) to date, it was remarked that reaching a quantum coherent regime for such masses can generate entanglement gravitationally. However, an estimate of the required regime and states was not the subject of that work. On the side of logic, \cite{kafri2013noise} {\em defined} a classical force as one which cannot generate entanglement, and found that such a classical force leads to excess noise. They then suggested a design using tethered torsional oscillators to detect this excess noise, which is challenging, and requires $10^3$ s per run -- if that noise is not found, then one can conclude a quantum (coherent) coupling. Thus, the emphasis in the experimental proposal was to rule out a specific classical interaction rather than verify gravitationally generated entanglement. Similarly, \cite{krisnanda2017revealing} showed that if an inaccessible system entangled two quantum probes, then it would display nonclassical correlations with the probes in the form of quantum discord. It was then suggested that this could potentially detect the nonclassicality of gravity without going into the formulations of any scheme. Thus, an explicit scheme with calculated numbers, which proposed to exploit contemporary developments in levitation and quantum technologies, along with a simple rationale from quantum information theory, was necessary \cite{bose2016matter,bose2017spin} to inspire confidence in the testability of the quantum character of gravity in the laboratory.

\subsubsection{Gravitationally interacting interferometers based protocol} \label{sec:schematic:protocol}
\begin{figure}[t]
   \centering
   \includegraphics[width=.5\textwidth]{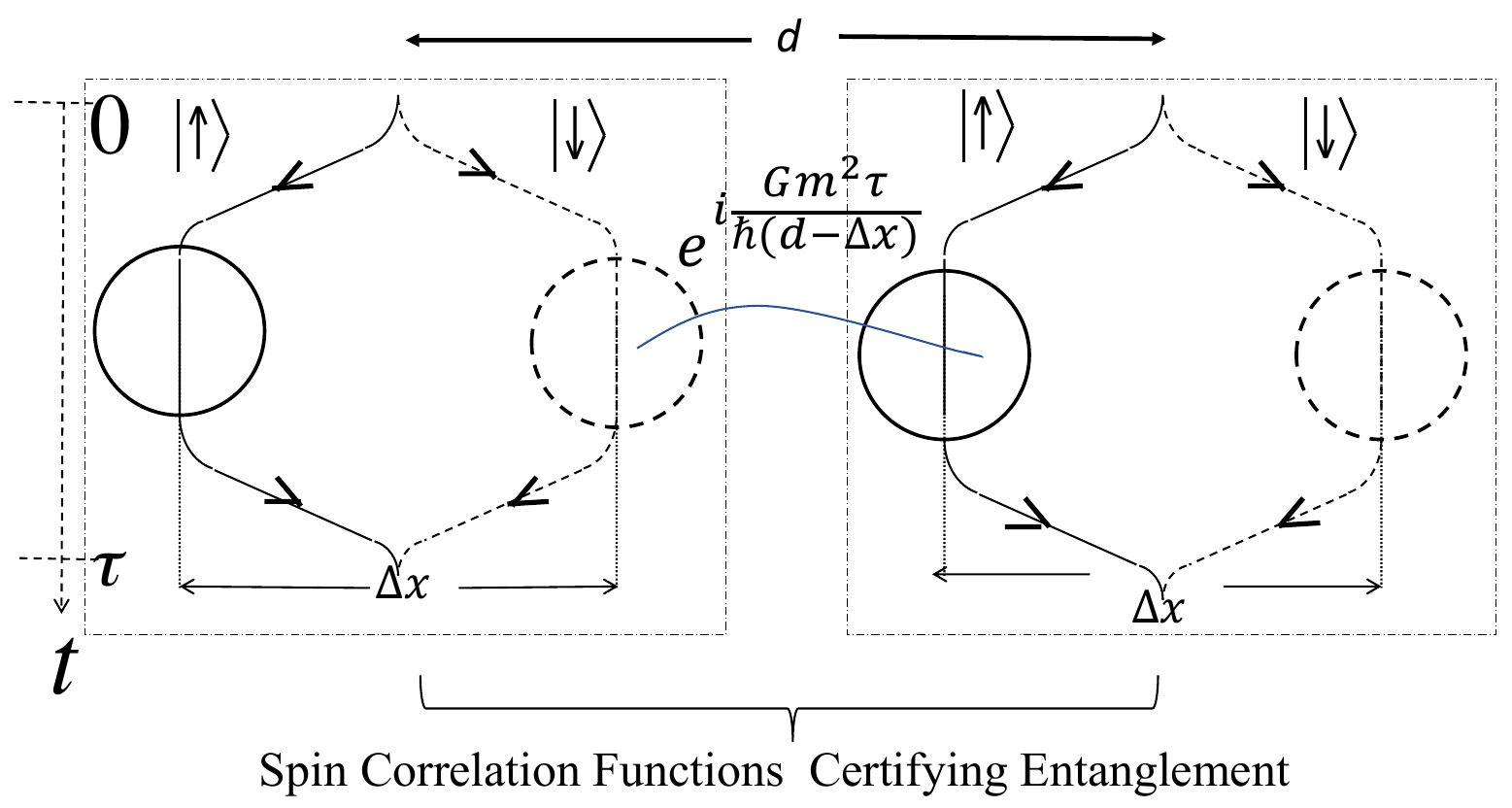}
   \caption{\textbf{Mechanism of gravitationally generated entanglement.}
   Mechanism of entanglement of two masses through the phase evolution due to gravitational interaction.  The phase evolution due to only the most prominent interaction, between the $|R\downarrow\rangle_1$ and $|L\uparrow\rangle_2$ is explicitly shown. The figure is a modified version of~\cite{bose2017spin}.
   }
   \label{Gravent}
\end{figure}

We outline here the protocol presented in~\cite{bose2016matter,bose2017spin}. 
Contemporary in publication (2017) is also~\cite{marletto2017gravitationally}, although, as that deals only schematically with the same idea without outlining explicit schemes,  we present the below in accordance to~\cite{bose2017spin}.  We consider two masses,  labeled by $j=1,2$, each with a spin embedded in it. A particularly relevant experimental example for these masses would be a diamond nano-crystal hosting a Nitrogen-Vacancy (NV) center, which is a highly coherent spin-1 system used in the area of quantum computation~\cite{bargill2013solid,wood2022long,hensen2015loophole,march2023long}. However, any other crystal with an embedded spin with a long coherence time would suffice, and generically, we require only two spin states, which we label as $|\uparrow\rangle$ and $|\downarrow\rangle$. The two masses, labeled as $j=1,2$, are each created in a quantum superposition of well-separated Gaussian states $|L\uparrow\rangle_j$ and $|R\downarrow\rangle_j$ by means of the Stern-Gerlach effect.  So,  we imagine that ideally, a spin embedded in each mass $j$ is placed in a quantum superposition of spin states,: $\frac{1}{\sqrt{2}}(|\uparrow\rangle_j+|\downarrow\rangle_j)$, where after a spin-dependent force (as in Stern-Gerlach) is applied to the masses so that they move from their initial central positions given by Gaussians $|C\rangle_j$ to evolve to 
\begin{equation}
|\psi\rangle_{j} = \frac{1}{\sqrt{2}}(|L\downarrow\rangle_j+|R\uparrow\rangle_j).
\end{equation}
This is shown in the upper half of Fig.~\ref{Gravent} as the point at which the trajectories achieve their maximal splitting $\Delta x$.
After achieving a certain maximal splitting $\Delta x$, the Stern-Gerlach force (spin-dependent splitting process) is stopped, for example, by either switching off the magnetic field or mapping the electronic spins to nuclear spins (this is shown as shoulders in the interferometers of Fig.~\ref{Gravent}), and the masses are allowed to translate in parallel next to each other.  We are going to assume (just for simplicity of presentation, although this assumption may be hard to fulfill) that the superposition is created so fast that the phase accumulation due to gravitational interaction during this time is negligible.  During the parallel motion {\em after} the creation of the superposition, the four configurations $LL,LR,RL$, and $RR$ (where the former refers to mass 1,  and the latter refers to mass 2) have different energies due to their Newtonian interaction, and thus their quantum phase evolutions happen at different frequencies
\begin{align}
& \omega_{RL} \sim  \frac{G m_1 m_2}{\hbar (d-\Delta x)}, \qquad\qquad\qquad\omega_{LR} \sim  \frac{G m_1 m_2}{\hbar (d+\Delta x)}, \nonumber\\ 
&\qquad\qquad\qquad \omega_{LL}=\omega_{RR} \sim \frac{G m_1 m_2}{\hbar d}.
\label{phase}
\end{align}
For simplicity, as well as for appreciating the maximal efficiency of the process,  we consider the situation when the superposition splitting is much larger than the distance of the closest approach of the masses, i.e.,  $\Delta x >> d-\Delta x$.  In that case, we can simplify to a situation where only $\omega_{RL}$ is prominent, while the other frequencies are negligible (taken to be zero with respect to $\omega_{RL}$).  Then the evolution of the state at a time $\tau$ is 
\begin{align} \label{evolved}
 |\Psi(t=\tau)\rangle_{12} &=|L\downarrow\rangle_1\frac{1}{\sqrt{2}}( |L\downarrow\rangle_2+|R\uparrow\rangle_2)\\
 &+|R\uparrow\rangle_1\frac{1}{\sqrt{2}}(e^{-i\omega_{RL}\tau} |L\downarrow\rangle_2 + |R\uparrow\rangle_2)\} .\nonumber
\end{align}
It is by inspection of the state, we conclude that for {\em any} value of $\omega_{RL}\tau \neq 2k\pi$,  where $k=$ integer, the state is entangled as it {\em cannot} be factorized into a product state of the two qubits as $( |L\downarrow\rangle_2+|R\uparrow\rangle_2) \neq (e^{-i\omega_{RL}\tau} |L\downarrow\rangle_2 + |R\uparrow\rangle_2)$.  In fact, for $\omega_{RL}\tau \sim \pi$, the state is a maximally entangled state of two qubits (the qubit states being defined by two orthogonal states $|L\downarrow\rangle$ and $|R\uparrow\rangle$. 

To compute the highest possible value of the frequency $\omega_{RL}$ one needs to identify the minimum value of $d-\Delta x$.  This, in turn, depends on the range within which you can bring the two masses without electromagnetic interactions swamping gravity.  It is possible, in principle, to make masses neutral (by shining UV radiation on the masses or the enclosure).  We also assume that it is possible to make the masses free of internal charge multipoles (how to achieve this is not yet fully solved for realistic nano and micro-crystals).  Suppose the electrostatic interactions between the masses are fully eliminated; there is still the Casimir interaction.  The ratio of the Casimir to the Gravitational interaction is given by
\begin{align}
 \frac{U_{\text{Casimir}}}{U_{\text{Gravity}}}\sim \frac{23}{4\pi} (\frac{3}{4\pi})^2\left(\frac{\epsilon-1}{\epsilon+2}\right)^2\frac{m_p^2}{\rho^2 (d-\Delta x)^6},
\end{align}
where $m_p$ is the Planck mass and $\rho$ is the density of the masses.  As we have to use some material to get the dielectric constant and density properties, we choose a diamond, which is a good candidate for the experiment, as it can host an embedded spin as an NV center defect, as stated before. If we want gravity to dominate by a factor of 10 over the Casimir interaction,  we get the minimum $d-\Delta x \sim 157 $\,$\mu$m~\cite{vandekamp2020quantum,schut2023micron}.   Putting this value in 
Eq.~\eqref{phase}, we get,  for micron-sized objects (radius $\sim 1 $\,$\mu$m,  mass $\sim 10^{-14}$\,kg)
\begin{align}\omega_{RL} \sim  \frac{G m_1 m_2}{\hbar (d-\Delta x)} \sim 0.4 {\text Hz}.
\label{phasevalue}
\end{align}
Now, how are we going to detect the entanglement generated as above? 
It is here that the rest of the interferometer shown in Fig.~\ref{Gravent} is important.  The paths are now recombined once again using Stern-Gerlach forces so that $|L\downarrow\rangle_j \rightarrow |C\downarrow\rangle_j$ and $|R\uparrow\rangle_j \rightarrow |C\uparrow\rangle_j$.  Then the state of the two embedded spins becomes
\begin{align} \label{spins}
 |\downarrow\rangle_1\frac{|\downarrow\rangle_2+|\uparrow\rangle_2}{\sqrt{2}}+|\uparrow\rangle_1\frac{e^{-i\omega_{RL}\tau} |\downarrow\rangle_2 + |\uparrow\rangle_2}{\sqrt{2}}.
\end{align}
The entanglement of these spins can be {\em verified} by measuring spin-spin correlation functions and combining them to construct an entanglement witness.  A good entanglement witness in this context (which works for smaller time evolution durations in comparison to the witness in~\cite{bose2017spin}) is~\cite{chevalier2020witnessing,guff2022optimal}
\begin{align} 
{\cal W}=\mathbb{1}-\sigma_x^1 \sigma_x^2 -\sigma_y^1 \sigma_z^2 -\sigma_z^1\sigma_y^2.
\end{align}
If,  after measuring the correlations,  the expectation value $\langle W\rangle <0$, then the state of the two spins is entangled.  As the only interaction was gravitational,  verifying the entanglement of these spins is equivalent to verifying the gravitationally generated entanglement. 

 Now,  in a real experiment,  it is possible that $\Delta x$ is achieved slowly so that a significant contribution to gravitational entanglement happens even during the growth of the superposition. Thus, according to the protocol of entanglement generation, $\tau$ should be an effective time that correctly captures the total entanglement growth rate during the evolution of the size of the superposition. Moreover, it is also possible that the ideal case of $\Delta x\gg d-\Delta x$ is not easily achievable. In fact, it is perhaps more likely, at least in the earliest experiments, that $\Delta x =\chi d$, where fraction $0<\chi<1$. In this general case, the entanglement developed, as well as the entanglement witness, depends only on a total phase, which one may call the ``entangling phase", defined as 
 $\phi_{\mathrm{ent}}=(\omega_{LR}-\omega_{LL})\tau+(\omega_{RL}-\omega_{LL})\tau$. For the configuration of interferometers given in Fig.~\ref{Gravent} we have, for small enough values of the entangling phase $\langle W\rangle \sim -\phi_{\text{ent}}$~\cite{chevalier2020witnessing}, 
 with
 \begin{equation}
     \phi_{\text{ent}}=\frac{G m_1 m_2\tau}{\hbar d}\frac{2\chi^2}{1-\chi^2},
     \label{phient}
 \end{equation}
 which, for $\chi \ll 1$ becomes
\begin{equation} \label{ent:phase:2}
    \phi_\text{ent} = \frac{2 G m_1 m_2 (\Delta x)^2 \tau}{\hbar d^3}. 
\end{equation}
From Eq.~\eqref{phient}, it becomes clear that the fraction $\chi$ should be chosen to be as close as possible to unity for a higher magnitude of the entanglement witness, enabling a lower number of measurements to determine it. However, in the regime of $\chi \ll 1$, from Eq.~\eqref{ent:phase:2}, one observes that we can, in principle, either choose a light mass $m$ and a large superposition size $\Delta x$ (the case discussed Sec.~\ref{sec:entanglement:mediated:gravity}) or alternatively, a heavy mass and a small superposition size without affecting the accumulated entangling phase. For example, with a mass of $m=1\,\text{kg}$, a superposition size of $\Delta x=10^{-14}\,\text{m}$, and an inter-particle separation $d=7\,\text{cm}$ (commensurate with the dimensions of such an object taking a standard density for a nanocrystal, say, that of diamond, of $3.5\times 10^3$\, kg/m$^{-3}$), and a time $\tau=1\,\text{s}$, a $\phi_\text{ent}\sim 0.2$ is obtained. However, such masses are usually tethered, which offers extra decoherence channels rather than being levitated. How to achieve, by squeezing and free expansion, a {spinless} measurement of a two-qubit entanglement witness (in terms of spatial qubits) has also been shown~\cite{yi2021massive,yi2022spatial}. 
 
\subsubsection{Alternative protocols} \label{sec:noise:and:mods}

Instead of an interferometric scheme, we can also consider nearby harmonic oscillators with mechanical frequency $\omega$ that are interacting gravitationally \cite{qvarfort2020mesoscopic,krisnanda2020observable}. We can obtain a figure of merit for the generated entanglement from Eq.~\eqref{ent:phase:2} by setting the delocalization to be the zero-point motion $\Delta x=\sqrt{\hbar/(2 m \omega)}$ (see Eq.~\eqref{eq:zpm}) and the interaction time to be $t=1/\omega$:
\begin{equation}
    \eta = \frac{2 G m }{ \omega^2 d^3}, \label{ent:phase:3}
\end{equation}
where we have defined  $\eta\equiv \phi_\text{ent}$ following the notation from~\cite{krisnanda2020observable}. Choosing the separation $d$ of the masses to be about $1.5$ times their radius, the above expression becomes solely dependent on their densities and $\omega$~\cite{krisnanda2020observable}. In order to achieve considerable entanglement, we again require $\eta \sim 1$. For that, for example, even with the densest material (Osmium), one has to accomplish the entire protocol with each mass in a $\omega \sim$ mHz trap over an interaction time of $10^3$ s, over which it will be very difficult to retain quantum coherence. Thus to achieve considerable entanglement, one has to use quantum states far more spatially spread (essentially similar in spread to the superpositions mentioned earlier) \cite{krisnanda2020observable,weiss2021large,cosco2021enhanced}, or non-Gaussian resources (i.e., we need to prepare non-Gaussian initial states) or with non-negligible nonlinear couplings (i.e., cubic or higher order terms in the position operators beyond the expansion in Eq.~\eqref{standard}) as discussed in~\cite{qvarfort2020mesoscopic}.  The generated entanglement can be read out using two optomechanical setups separately monitoring each of the two masses, i.e., each mass is a mirror that is coupled to an optical field that can then be measured (see Secs.~\ref{sec:optomechanical:Hamiltonian} and~\ref{sec:control}). Specific optomechanical configurations to measure the gravitationally-induced entanglement have been considered using the single-photon nonlinear regime in a quantum Cavendish experiment~\cite{balushi2018optomechanical,matsumura2020gravity},  and in the linear regime with the cavity driven by a coherent laser field~\cite{miao2020quantum}. Furthermore, in~\cite{datta2021signatures} an optomechanical scheme for measuring the differential motion of the two mirrors is given, and it is argued that detecting gravitationally-induced squeezing of the differential motion should be experimentally more accessible than detecting quantum entanglement between the two masses, and consequently, the conclusions that can be drawn would be different.

We can also consider the experimental situation with unequal masses~\cite{bose2017spin}. In place of $m^2$ and $\Delta x^2$ in Eq.~\eqref{ent:phase:2} we have   $m_1 m_2$ and $\Delta x_1 \Delta x_2$, respectively, where $\Delta x_j$ is the superposition size of the mass $m_j$ ($j=1,2$). To generate substantial entanglement, we again require $\phi_\text{ent}\sim1$. One scheme in this direction uses extremely different masses, an atom (in an atom interferometer), and a massive oscillator coupled to it gravitationally \cite{carney2021using}. Here, treating the atom as a spatial qubit gives a cyclic decoherence-recoherence dynamics of the two-mass system, which is robust to the thermal state of the oscillator, an interesting property that also underpins some previous optomechanical \cite{bose1997preparation,mancini1997ponderomotive,bose1999scheme,marshall2003towards,armata2017quantum,qvarfort2018gravimetry} and qubit-oscillator \cite{bose2006qubit,scala2013matter} schemes. The decoherence-recoherence of the spatial qubit is then suggested as evidence of entanglement, although this is reliant on the assumption of the high purity of the joint qubit-oscillator state. Another variant of this is a case with two qubits \cite{pedernales2022enhancing}: a nanoscale mass as a spatial qubit, which is gravitationally coupled to a massive mediating oscillator, which is, in turn, coupled electromagnetically to another qubit. By measuring entanglement between the two qubits, the quantumness of the gravitational interaction can be tested. While in a manner similar to the above, robustness to the thermal state of the intermediary is present (also reminiscent of geometric phase gates in ion traps \cite{molmer1999multiparticle,solano1999deterministic,milburn2000ion}), one still has the challenge of preparing a sufficiently spread spatial quantum superposition state of a nanoscale mass. Other works have also considered experimental configurations with modified geometries and multidimensional systems~\cite{tilly2021qudits}. Specific schemes with three or four~\cite{schut2022improving},  as well as an array of particles~\cite{miki2021entanglement,ghosal2023distribution} have been analyzed. 

It has also been noted that the relativistic regime is required to probe the spin nature of the gravitational interaction~\cite{bjerrumbohr2015bending,scadron2006advanced,carney2022newton,biswas2023gravitational}. In~\cite{bose2022mechanism} the leading order post-Newtonian terms in an experimental situation with harmonic oscillators have been considered, and in~\cite{aimet2022gravity} photons from two separate interferometers are let to entangle.

  Notably, a hybrid optomechanical scheme testing the quantum counterpart of the light bending by gravity was given in~\cite{biswas2023gravitational}. The scheme consists of a harmonic oscillator of mass $m$ placed at the origin and of a circular path of radius $r$ for
the optical field confined to a half-ring cavity (i.e., the mechanical oscillator and the cavity photons are conceptually replacing the two massive interferometers shown in Fig.~\ref{Gravent}). The interaction between the trapped mass and the photons is purely gravitational, and it reduces to the form of the cavity optomechanical interaction in Eq.~\eqref{eq:optomechanical:hamiltonian} with the coupling $g_0$, in this case, arising from the quantum light-bending interaction.  

A figure of merit, similar to the ones in Eqs.~\eqref{ent:phase:2} and~\eqref{ent:phase:3}, can also be constructed for these latter cases, albeit the details can depend on the specific experimental configuration and on the interaction. To get a tentative idea about the order of magnitude of the entanglement phase with photons, we can use the relation  $m  =\hbar \omega/c^2$. For example, replacing one of the masses in Eq.~\eqref{ent:phase:2} with $m  =\hbar n \omega/c^2$ (where $\omega$ is the frequency of the optical field), we find a figure of merit for entanglement between a photon and a massive particle:
\begin{equation}
    \phi_\text{ent} \sim \frac{2 G m  \omega \Delta x^2 \tau}{c^2 d^3}, \label{ent:phase:4}
\end{equation}
which can be used to gauge the order of magnitude of the entanglement phase. Taking the ratio of Eqs.~\eqref{ent:phase:4} and  Eq.~\eqref{ent:phase:2}, we thus see that the effects are suppressed by $\hbar \omega/ ( m c^2)$, and as such, achieving an entanglement phase of order unity requires a very large photon number to enhance the effect.

There are also proposals for testing gravitationally-induced self-interactions of matter~\cite{anastopoulos2014newton, anastopoulos2013master,anastopoulos2020quantum}. While in the previous scheme, one was interested in the interaction between two distinct systems $\propto T_{\mu\nu}^{(A)} T_{\alpha\beta}^{(B)}$ one can also consider self-interaction terms $\propto T_{\mu\nu}^{(A)} T_{\alpha\beta}^{(A)}$ and $\propto T_{\mu\nu}^{(B)} T_{\alpha\beta}^{(B)}$, where $T_{\mu\nu}^{(A)}$ and $T_{\mu\nu}^{(B)}$ denote the stress-energy tensors corresponding to systems $A$ and $B$. We can readily see how such terms emerge in the Newtonian limit. The interaction with quantum matter from Eq.~\eqref{eq:grav_lagrang} reduces to:
\begin{equation} \label{eq:density:source}
    \hat{H}_\text{int} = \frac{1}{2} \int d\bm{r} \, \hat{\rho} (\bm{r}) \hat{\phi} (\bm{r}),
\end{equation}
where  $\hat{H}_\text{int}=-\int \hat{\mathcal{L}}_\text{int}(\bm{r}) d\bm{r}$ is the interaction Hamiltonian, $\hat{\rho}$ is the matter density, and  $\hat{\phi}$ is the Newtonian potential. The matter-density $\hat{\rho}$ is also a source for $\hat \phi$ with the solution given by the familiar Newtonian potential. From Eq.~\eqref{eq:density:source} we thus find:
\begin{equation} \label{eq:density:source2}
    \mathcal{L}_\text{int} = -\frac{G}{2} \int  d\bm{r}\int d\bm{r}' \,  \frac{\hat{\rho} (\bm{r}) \hat{\rho} (\bm{r}') }{\vert \bm{r} -\bm{r}'\vert}.
\end{equation}
Eq.~\eqref{eq:density:source2} is suggestive for a figure of merit based on the entanglement phase from Eq.~\eqref{ent:phase:2}. Since we have only one system we set $\Delta x \sim\ d$ to find:
\begin{equation} \label{ent:phase:7}
    \phi_\text{ent} \sim \frac{G m^2 \tau}{\hbar d}, 
\end{equation}
where $d$ is to be interpreted as a characteristic length scale of the problem (e.g., the wavefunction spread). A scheme with BECs was investigated, where  $\hat{\rho} \propto \hat{a}^\dagger\hat{a}$ (with $\hat{a}$ the mode of a BEC). We thus find from Eq.~\eqref{eq:density:source} a Kerr nonlinearity which induces non-Gaussianity~\cite{howl2021non,haine2021searching}. Recently a scheme has also considered using the self-interaction of photons~\cite{mehdi2023signatures}. In place of Eq.~\eqref{ent:phase:7} we can use $m  =n \hbar \omega/c^2$ (where $\omega$ is the frequency of the optical field) to find the figure of merit $\phi_\text{ent} \sim 2 G \hbar \omega^2 \tau/ (c^2 d)$ which can be again enhanced by considering a large number of photons. It should be noted that gravitationally mediated entanglement experiments will also be able to test various variants of gravitational theories  \cite{marshman2020locality,elahi2023probing,vinckers2023quantum,chakraborty2023distinguishing}. Several foundational questions involving the nonclassical behavior of gravity can also be probed with similar setups \cite{kent2021testing,etezad2023paradox}, including the nonclassical behavior of gravity under a measurement \cite{hanif2023testing}.

\subsubsection{Major challenges}

Proposals for testing gravitationally-induced quantum phenomena discussed in this section face a series of experimental challenges specific to the experimental implementation. 

One major difficulty is the achievement of a large superposition for the interferometry-based schemes (for schemes using Gaussian wavepackets, it translates to obtaining a very large delocalization of the wavepacket \cite{weiss2021large,cosco2021enhanced}, which is a problem of similar nature). A large mass requires a strong force to create a quantum superposition of components separated by $\Delta x$.  Some of the early protocols of Stern-Gerlach-based creation of superpositions~\cite{scala2013matter,wan2016free} have been found to have limitations of the achievable growth rate of $\Delta x$~\cite{pedernales2020motional,marshman2021large}.  Some solutions have been investigated~\cite{zhou2022catapulting,zhou2023mass}, and this splitting rate is still a work in progress. 
 
If the electromagnetic interactions between the masses can be screened~\cite{schmole2016micromechanical,schut2023relaxation,schut2023micron}, then the masses can be brought closer ($d$ decreased), and consequently, the requirement of $\Delta x$ can be alleviated. For example, the most optimistic results known to us in this context of using {\em both} screening and trapping~\cite{schut2023micron}. For a screening material of $1$\,$\mu$m thickness, $d\sim 11$\,$\mu$m, masses $m_1 \sim m_2 \sim 10^{-14}$\,kg, then, for $\Delta x \sim 0.65$\,$\mu$m and a $\tau \sim 1$\,s, $\phi_{
\text{ent}}\sim 0.01$ is obtained, which requires $\sim 10^4$ repeats of the experiment.

The other important obstacle is, of course, maintaining coherence. In short, in presence of decoherence at a rate $\Gamma$ the witness becomes $\langle {\cal W}\rangle \sim \Gamma \tau -\phi_{\text{ent}}$~\cite{chevalier2020witnessing,guff2022optimal,schut2023micron}.
Thus, in order to have a negative expectation value of the witness, one has to keep the growth rate of the entangling phase above the decoherence rate.
Here we provide some general considerations about noise and decoherence for the figure of merit given in Eq.~\eqref{ent:phase:2} for concreteness. The requirements on the force noise spectra $S_{FF}$ can be estimated from the decoherence rate $\Gamma$:
\begin{equation}
    \Gamma =\frac{S_{FF}(\omega_\text{exp}) \Delta x^2}{\hbar^2},
\end{equation}
where $\omega_\text{exp}=1/\tau$ is the characteristic frequency of the experiment. We require $\tau < \Gamma ^{-1}$ to have sufficiently long coherence times~\cite{bose2017spin}. Specific noise and decoherence sources have been considered, as well as methods for its mitigation~\cite{vandekamp2020quantum,pedernales2020motional,rijavec2021decoherence,weiss2021large,torovs2021relative,wu2023quantum,gunnink2022gravitational,fragolino2023decoherence,yi2022spatial}. In the above, it is perhaps important to emphasize that primary sources are the collisions with background gas, and blackbody radiation emission \cite{romeroisart2011quantum,vandekamp2020quantum}. There could also be electromagnetic noise of various forms \cite{fragolino2023decoherence}. Most interestingly, gravitational and inertial noise also plays an important role in the decoherence of large quantum superpositions, and some ways of mitigation have been worked out \cite{torovs2021relative}. Very importantly, various systematic noises affect Stern-Gerlach interferometry, when that is the mechanism of creating large spatial superpositions, such as due to phonons \cite{henkel2022internal,henkel2023universal} and rotations \cite{japha2022role}. In this context, the coherence of spins is also important, and achieving dynamical decoupling has also been considered~\cite{wood2022spin}.

\subsubsection{Implications} \label{sec:implications}

Considering one observes the entanglement between two masses due to their gravitational interaction, what can we conclude from that? Essentially, it verifies one prediction of a fully quantum counterpart of Einstein's equations
\begin{equation}
\hat{G}_{\mu\nu}=\frac{8\pi G}{c^4} \hat{T}_{\mu\nu}\label{3Q1}.
\end{equation}
In this sense, it verifies a prediction common to all approaches in which gravity is treated as a quantum field, eg, \cite{dienes1997string,rovelli2008loop,donoghue1994general}.
Alternately, it falsifies {\em all} hybrid theories of quantum sources in a state $|\psi\rangle_{\mathrm{Source}}$ leading to classical gravity $G_{\mu\nu}$ (while Eq.~\eqref{3C1} is a special case of that, there could, of course, be much more general stochastic theories of the above hybrid nature~\cite{oppenheim2018post,kafri2014classical,diosi1998coupling,galley2023any}). We present our arguments below in favor of the view that observation of entanglement is {\em inconsistent} with gravity being a classical field/curvature even when defined in the above, very general, sense. We should, at once, state that such a conclusion is possible if one makes (i) an appropriate (very standard) definition of a classical field, and (ii) a minimal assumption. 

Let us first define what a classical field is. Namely, it is an entity with a probability distribution over fixed values (numbers) at every point in space-time. Thus, retaining the symbol usually used for the gravitational metric, we would define a classical gravitational field as an entity defined by probabilities $P^{(j)}$ and corresponding metrics $g_{\mu,\nu}^{(j)}$:
\begin{equation}
  \{ P^{(j)}, g_{\mu,\nu}^{(j)}(\vec{r},t) \},
\end{equation}
where $\mu,\nu=0,..,4$, and $\vec{r},t$ are spacetime points. This definition is broader than just having a unique metric $g_{\mu,\nu}(\vec{r},t)$ defined everywhere in spacetime as we are allowing for probabilities. The allowance for probabilities makes it possible for gravity to be a {\em statistical} field, while still being classical. {\em Quantum} is more stringent, as it necessitates {\em quantum superpositions} of different configurations $g_{\mu,\nu}(\vec{r},t)$. As long as we {\em disallow superpositions}, then even with fluctuations (probabilities), a field is classical. Now comes the assumption. This is namely the assumption that two masses outside each other's {\em support} (by support, we mean their positions, or, if quantum, their wavefunctions, or if a second quantized matter field, then the localized mode which they occupy) can only interact with their local field and not {\em directly} with each other. This makes the field a {\em mediator}. Within the domain of non-relativistic experiments that would be feasible in the foreseeable future, we cannot {\em prove} the necessity of the mediator, and we appeal to what is known from the rest of physics, namely that there is no action at a distance in our known domain of physics.  

Under assumptions (i) and (ii), the operations that can happen between the masses due to their interactions with their local gravitational field are {\em Local Operations and Classical Communications (LOCC)}, which {\em cannot} create entanglement. Thus it follows very simply that if entanglement is observed between the masses due to their gravitational interaction, then either gravity is {\em not} a classical field as per the definition (i) (i.e., it is {\em nonclassical}) or the assumption of a mediator (ii) is violated. This was the justification presented in~\cite{bose2017spin}. Within this setting of exchange of a mediator between the masses, only a highly quantum mediator, namely a {\em virtual} (off-shell) particle (a quantum superposition of all energies) is necessary for the continuously coherent generation of entanglement, as has been shown through a fully relativistic treatment in~\cite{marshman2020locality} (for a treatment that also shows the retardation in the growth of entanglement, see \cite{christodoulou2023locally}). Alternately, it has been shown that the presence of entanglement also necessitates an {\em operator valued interaction} between masses, which is not possible with a classical mediator~\cite{bose2022mechanism}.   

 Another way to interpret the results of the experiment is that it evidences a quantum superposition of geometries that one of the masses produces, on which the other mass evolves~\cite{christodoulou2019possibility}. If the quantum superposition of geometries (i.e., quantum-natured gravity) is disallowed, then no superposition develops. The conditions for justification of a non-classical nature of gravity within the framework of generalized probability theories have been presented~\cite{galley2022no}. Within the effective quantum field theory description of gravity, it has also been argued that once the Newtonian interaction enables entanglement, then the other degrees of freedom have to be quantized for consistency \cite{belenchia2018quantum,carney2022newton,danielson2022gravitationally}, along with an expression for a quantum state of gravity associated to a mass in a quantum superposition \cite{chen2023quantum}. Indeed, if the condition of mediator, providing the L part of the LOCC, is not imposed, one can still draw interesting conclusions from the generation of entanglement, as discussed in~\cite{fragkos2022inference}, while the modifications needed to draw conclusions about quantum natured gravity even without the assumptions have been stated, after a relativistic treatment, in \cite{martin2023gravity}. It is worthwhile to also note here that LOCC is more restrictive than just disallowing entanglement generation \cite{lami2023testing}.

\subsection{Other tests of gravity} \label{sec:other:tests}
There are a number of ways in which gravity can affect quantum systems beyond the topics of precision gravimetry, decoherence, and entanglement. Here we detail additional tests of gravity and related effects.

\subsubsection{Tests of the generalized uncertainty principle} \label{sec:GUP}

Many quantum gravity theories predict the existence of a finite and minimum length scale at least as small as the Planck length $l_P = \sqrt{\hbar G/c^3} \approx 1.6 \times 10^{-35}$\,m~\cite{garay1995quantum,hossenfelder2013minimal}. The emergence of a finite length implies a generalized uncertainty principle (GUP) because the fundamental position uncertainty can no longer be reduced to zero. The GUP widely considered reads
\begin{equation} \label{eq:GUP}
   \Delta  x \Delta p = i \hbar \left( 1 + \beta_0 \left(\frac{ l_P \Delta p}{\hbar} \right)^2 \right), 
\end{equation}
 where $\Delta x$ and $\Delta p$ denote the uncertainties in the operators, and where $\beta$ is a dimensionless constant that indicates the strength of the modification. The minimal length arises because the uncertainty in $\Delta x$ can no longer be made infinitesimally small. 
 Associated with the GUP is also the modified commutator relation
 \begin{align} \label{eq:GUP:commutator}
     [\hat x, \hat p] =i \hbar  \left( 1 + \beta_0  \left( \frac{l_P \hat p }{\hbar} \right)^2 \right). 
 \end{align}
 Bounding the parameter $\beta$ in Eq.~\eqref{eq:GUP} through experiments also bounds new physics below the length scale $\sqrt{\beta} l_P$~\cite{das2008universality}. 
 
The existence of a finite length scale and GUP was first put forward in string theory~\cite{veneziano1986stringy, amati1989can}, but were later also derived using general mode-independent properties of quantum gravity theories. 
For example, a generalized Gedanken experiment for the measurement of the area of the apparent horizon of a black hole in quantum gravity leads to the emergence of a GUP~\cite{maggiore1993generalized}. There also exists an algebra that gives rise to the modified commutator relation in Eq.~\eqref{eq:GUP}, just like the operator $\hat x$ and $\hat p$ satisfies $[\hat x, \hat p] = i \hbar$~\cite{maggiore1993algebraic}.  Model-independent arguments for the measurement of micro-black holes allow us to arrive at a GUP~\cite{scardigli1999generalized}.  
The influence of minimal length scales on quantum states has been widely considered. 
 There are quantum-mechanical implications of a GUP and finite length, which were analyzed in~\cite{kempf1995hilbert}, including the localization of wavefunctions in space and the effects on harmonic oscillators. Harmonic oscillators with minimal length scales were also considered in~\cite{chang2002exact}, where the effects on electrons trapped in magnetic fields were also considered, as well in~\cite{lewis2011position}. In addition, an equivalent formulation of the GUP but with a maximum observable uncertainty in the momentum, rather than a minimum uncertainty in the position, has been formulated~\cite{petruzziello2021generalized}. 

A number of proposals for laboratory experiments to test GUPs with massive quantum systems have been put forward. 
By using pulsed optomechanics (see Sec.~\ref{sec:optomechanical:Hamiltonian}), it was shown that the effects of a GUP should create changes in the trajectories in phase space traced out by a massive system in~\cite{pikovski2012probing}. The scheme was later extended in~\cite{kumar2018quantum}. Along similar lines, mechanical oscillators near the Planck mass ($m_P \approx  22$\,ng) were analyzed, where the modified dynamics were directly compared with the unmodified~\cite{bawaj2015probing}. A further proposal considered a pendulum, where continuous rf measurements of the frequency of an electromechanical oscillator can help to further bound $\beta_0$~\cite{bushev2019testing}. The radiation-pressure noise can also contain information about the GUP, as proposed in~\cite{girdhar2020testing}.
The sensitivity to the modified commutator relation was shown to improve in the vicinity of exceptional points~\cite{cui2021detecting}, and quadratic corrections were shown to affect the noise spectrum of an optomechanical system~\cite{sen2022probing}. A recent result is the evaluation of improved constraints on minimum length models by using a low-loss phonon cavity~\cite{campbell2023improved}. 

It should be noted that additional considerations suggest that the observed effects from a GUP scale with $N^{-a}$, where $N$ is the number of particles of the composite system and $a$ is a parameter to be determined~\cite{kumar2020quantum}. The strength of this scaling is unclear but should be taken into account in experiments. It has therefore been proposed that bounds on GUPs can also be obtained through the use of atoms~\cite{chatterjee2021violation}. In another work, it was pointed out that different modifications of the canonical commutator yield the same commutator relation in Eq.~\eqref{eq:GUP:commutator}~\cite{bishop2020modified}, which necessitates the need for caution when interpreting experimental results. 

The current leading bound for $\beta_0$ appearing in Eq.~\eqref{eq:GUP} is $\beta_0< 5.2 \times 10^6$, which was calculated using pendulum measurements~\cite{bushev2019testing}.
Including the dependency on the number of constituent particles as discussed above would, however, change the bound for $\beta_0$ as well as its physical interpretation within the considered GUP model~\cite{kumar2018quantum}.
Other bounds on $\beta_0$ have been derived from astronomy~\cite{scardigli2015gravitational} as well as from gravitational-waves~\cite{das2021bounds}. See~\cite{scardigli2015gravitational} for a comparison between bounds obtained from different experiments available at the time.

%
\subsubsection{Tests of the gravitational Aharonov-Bohm effect} \label{sec:grav:aharonov} 

The Aharonov-Bohm effect was originally introduced for electrons in a constant magnetic field, which picks up a phase depending on the (spatially dependent) vector potential that can be measured in an interferometer.
Fundamentally, the phase difference stems from an action difference between the interferometer arms which would not be accessible classically.
Similarly, a gravitational field can induce such action differences (even in the absence of forces), giving rise to a scalar gravitational Aharonov-Bohm effect~\cite{audretsch1983neutron}.
Ultra-cold atoms were proposed as an experimental platform to detect this effect~\cite{hohensee2012force} due to their long coherence times. The gravitational Aharonov-Bohm effect was successfully detected with a light-pulse $^{87}\text{Rb}$ atom interferometer and a kilogram-scale source mass~\cite{overstreet2022observation} allowing to directly probe space-time curvature~\cite{roura2022quantum}. Further proposals include tests using quantum systems in free-fall \cite{chiao2023gravitational}, as well as of the vector gravitational Aharonov-Bohm effect \cite{chiao2014gravitational}.

%
\subsubsection{Tests of gravity through resonances and phonon excitations} \label{sec:phonon:test} 

The first attempts to detect gravitational waves involved resonant mass antennae ~\cite{misner1973gravitation}. Weber searched for gravitational waves using resonances with phononic modes of aluminum bars now known as Weber bars \cite{weber1969evidence,ferrari1982search}. In the experiments Weber's bars reached temperatures of a few Kelvin, not cold enough to suppress noise sufficiently nor to reach the quantum regime. Proposals to use a kilogram-scale superfluid $^4$Helium resonator promise better sensitivities by reaching mKelvin temperatures \cite{lorenzo2014superfluid,singh2017detecting}. 

A quantum version of a resonant antenna has been proposed using resonances of phonon modes in BECs that can reach nanoKelvin or picoKelvin temperatures~\cite{sabin2014phonon,kohlrus2017quantum}. At these temperatures, it is possible to prepare highly sensitive quantum states that could be used to detect long-lived high-frequency gravitational waves (between $10^3-10^7$Hz). The quantum thermodynamical properties of resonances in relativistic quantum fields were studied in \cite{bruschi2020thermodynamics}, showing that the BEC phonon resonance antenna is a quantum thermal machine capable of extracting energy from the gravitational wave. Resonant effects with BEC phonons can be used to detect gravitational accelerations and gradients in the Newtonian approximation~\cite{raetzel2018dynamical}, including the search for modifications of Newtonian dynamics (known as MOND)~\cite{fernandez2023testing} and also to measure relativistic corrections~\cite{ahmadi2014relativistic,lock2017dynamical,howl2023quantum}. 

 Recent proposals consider searching for signatures of quantum gravity using multi-atomic states of cold atoms~\cite{haine2021searching} (see also Sec.~\ref{sec:entanglement:mediated:gravity}), BEC phonons~\cite{howl2021non} and massive quantum acoustic resonators~\cite{tobar2023detecting,tobar2024detecting}. These proposals aim at detecting single gravitons through their direct interaction with matter, rather than through decoherence. A proposal to detect quantum gravity using Weber bars cooled down to sub-miliKelvin temperatures \cite{aguiar2011past}, suggests searching for quantum gravity signatures by testing modifications of the energy moment uncertainty principle \cite{Bhattacharyya2020generalized}. Signals detected using these new detector concepts could be correlated with independent classical detections of gravitational waves by laser interferometry to ascertain their origin. There are also new proposals to use laser interferometers to search for signatures of the quantum nature of gravitational waves~\cite{parikh2020noise,parikh2021quantum,parikh2021signatures} and looking for geontropic vacuum fluctuations from quantum gravity~\cite{verlinde2021observational,bub2023quantum}.

Detecting gravitational waves and gravitons using table-top experiments that make use of the high sensitivity of quantum technologies promise to open a new area in the study of fundamental physics.

%
\subsubsection{Tests of quantum field theory in curved spacetime and analog gravity} \label{sec:QFT:CS} 
Quantum field theory in curved spacetime predicts that spacetime dynamics produces entangled excitations in quantum fields~\cite{ball2006entanglement,fuentes2010entanglement} and that the presence of horizons gives rise to decoherence~\cite{fuentesschuller2005alice, adesso2009correlation,alsing2012observer}, see Sec.~\ref{sec:QFT:CS}. Underpinning these effects is parametric amplification, where particles are created out of the quantum vacuum by moving boundary conditions or horizons. In this sense, there is a deep connection between the dynamical Casimir effect where by changing the length of a cavity, the vacuum state of the electromagnetic field changes producing entangled particles~\cite{moore1970quantum,fulling1976radiation, bruschi2012voyage}; parametric down-conversion where a medium change produces entangled photons~\cite{kwiat1995new}; and effects of quantum field theory such  
as Hawking radiation~\cite{hawking1974black} and the creation of particles by the expansion of the universe~\cite{polarski1996semiclassicality,birrell1982quantum}, among other interesting effects. This connection is made evident through the mathematical formalism of both quantum optics and quantum field theory in curved spacetime, where Bogoliubov transformations produce mode-mixing and two and single-mode squeezing of modes. In~\cite{friis2013relativistic}, the formalism of continuous variable quantum information is applied to quantum field theory in curved spacetime to compute entanglement in relativistic settings. Quantum field theory has been demonstrated numerous times in the flat case, however, they key predictions of the theory in the presence of gravity are currently out of experimental reach. Systems such as black holes are not accessible to experimentation, and most predicted effects are too small. Consider, for example, the dynamical Casimir effect, where producing excitations via oscillating mirrors requires velocities close to the speed of light. Oscillating a microwave mirror at a frequency of 2 GHz with a
displacement of 1 nm produces velocities of only $v\approx10^{-7}c$. At these velocities, approximately one photon is produced per day. However, moving the mirror at these speeds requires an input of mechanical power of 100 MW, and at the same time, a temperature of $\approx 20$ mK is needed to ensure that the field is in the vacuum state. For this reason, it has become fashionable to simulate effects in the lab. Photon creation by a moving boundary condition was demonstrated using a superconducting circuit where the electromagnetic flux going through a SQUID produced a boundary condition moving at a third of the speed of light~\cite{wilson2011observation}.

Analog experiments help test consistencies within a given mathematical model. An alternative that involves using a massive system in a small-scale lab is to test the key predictions of the theory using Bose-Einstein condensates (BECs). Atomic interactions in a BEC produce phonons, which are a massless quantum field that obey a Klein-Gordon equation in an effective curved metric~\cite{visser2010acoustic,fagnocchi2010relativistic,sabin2014phonon,hartley2018quantum},
\begin{align} 
\square \psi = \frac{1}{\sqrt{-G}}\partial_\mu \left(\sqrt{-g}\,G^{\mu \nu} \partial_\nu \psi \right)\,,
\end{align}
where $G := \text{det}(\boldsymbol{G})$ is the effective metric given by
\begin{align} 
G^{\mu \nu}=\rho\frac{c}{c_s}\left(g_{\mu\nu}+\left(1-\frac{c^2_s}{c^2}\right)\frac{u_\mu u_\nu}{|u_\mu| |u_\nu|}\right). 
\end{align}
Here $g_{\mu\nu}$ is the spacetime metric, $\rho$ is the density of the condensate, $c$ the speeds of light and $c_s$ is defined by
\begin{equation}
	c_s^2 := \frac{c^2 c_0^2}{|u_\alpha u^\alpha| + c_0^2}\,,
\end{equation}
with $c_0^2:=\lambda\rho \hbar^2/2m^2$ and the four-vector $u_\mu$ is the flow associated with the phase of the wave function of the BEC bulk. The speed of sound in the BEC is $c_0$, and the density $\rho$ of the BEC may depend on space and time coordinates. By choosing or changing the density, the speed of sound, or the velocity field, it is possible to simulate certain spacetime metrics. 

One metric that can be simulated this way is that of a black hole, which makes it possible to test Hawking's prediction for black hole radiation~\cite{hawking1975particle, hawking1974black}. 
By starting from the expression for an irrotational fluid, $\vec{\nabla} \times \vec{v} = 0$, where $\vec{v}$ is the velocity of the fluid, the analog to the Schwarzschild metric is~\cite{unruh1981experimental}
\begin{equation}
ds^2 = \rho \left( ( c^2 - v^2) dr^2 - \frac{1}{1 - \frac{v^2}{c^2}} dt^2 - r^2 d\Omega^2 \right),
\end{equation}
where $\rho$ is the density of the fluid, $v$, as before, is the velocity of the fluid, and $c$ is the speed of the particles, also referred to as the speed of sound. 
The Hawking temperature is then given by 
\begin{equation}
T_H = \frac{\hbar}{4 \pi k_B c } \left[ \frac{d}{dx} \left( c^2 - v^2 \right) \right]. 
\end{equation}
The $T_H$ of a BEC should be about 1 nK~\cite{unruh1981experimental}. 
A BEC black hole emitting Hawking radiation has been both theoretically simulated~\cite{carusotto2008numerical} and experimentally realized\cite{lahav2010realization, steinhauer2014observation, steinhauer2022analogue,steinhauer2022confirmation}. Beyond BECs,  an analog of Hawking radiation can also be observed in quantum optics~\cite{philbin2008fiber} using the nonlinear Kerr effect that arises in certain dielectric media.

We note that computer and analog simulations on their own cannot falsify nor verify theories. As a result, current analog gravity experiments cannot be said to test general relativity directly. 
Fortunately, experimental settings are reaching scales where the key predictions of quantum fields in curved spacetime are becoming testable in the laboratory. Theoretical studies have shown that actual changes of the spacetime metric $g_{\mu\nu}$ can, in principle, produce observable effects on phonon states~\cite{sabin2014phonon}. This effect is at the heart of the proposals to detect high frequency persistent gravitational waves~\cite{sabin2014phonon}, search for dark matter and dark energy~\cite{howl2023quantum}, opening the possibility of testing the predictions of quantum field theory in curved spacetime in the lab. 
Some aspects of the interplay of quantum fields and general relativity could be tested in space-based experiments where photon entanglement has been distributed across thousands of kilometers (for a review, see~\cite{sidhu2021advances}). At these scales, relativity kicks in since the proper time on Earth is different from the proper time on a satellite. Theoretical studies have shown that spacetime curvature affects the propagation of light wavepackets on Earth, affecting quantum communications~\cite{bruschi2014spacetime}. This effect goes beyond the gravitational phase shift predicted by special relativity. The curvature of the spacetime around the Earth can flatten traveling wavepackets, as well as decohere quantum state,s and these effects can be used to measure spacetime parameters~\cite{kohlrus2017quantum,kohlrus2019quantum,bruschi2014quantum}.

\section{Experimental pathways towards tests of gravity} \label{sec:experiments}

In the previous section, we reviewed proposals for testing the overlap between quantum mechanics and gravity with massive quantum systems. Here we review experimental advances toward the regime where the dynamics of quantum systems are affected by gravity. 

\begin{figure*}
    \centering
    \includegraphics[width=\textwidth]{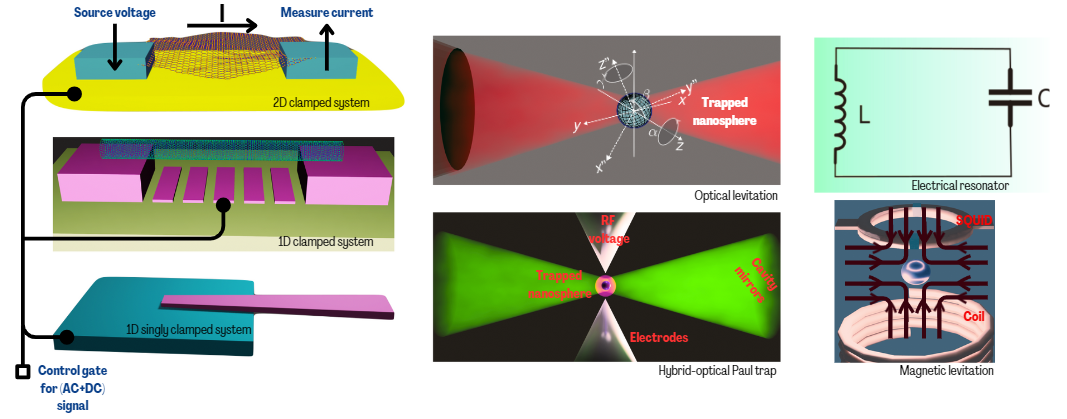}
    \caption{\textbf{Illustration of state-of-the-art mechanical systems.} This figure encapsulates an array of resonators utilized in experimental efforts aimed at detecting the interplay between quantum mechanics and gravity. The leftmost column showcases mechanical resonators: a 2D clamped resonator (Graphene), a 1D clamped resonator (A suspended carbon nanotube), and a 1D singly clamped beam resonator, illustrating the diversity in mechanical systems. The central column depicts optical levitation systems: a standalone optical levitation and a hybrid Paul-optical levitation system, demonstrating the integration of optical techniques. The rightmost column presents an electrical resonator and a magnetic levitation system, representing the incorporation of electromagnetic methodologies. Together, these systems exemplify the wide range of experimental apparatuses employed in the quest to uncover the interplay between quantum mechanics and gravity.
    }
    \label{fig:MechanicalSystem}
\end{figure*}

\begin{figure*}
    \centering
    \includegraphics[width=\textwidth]{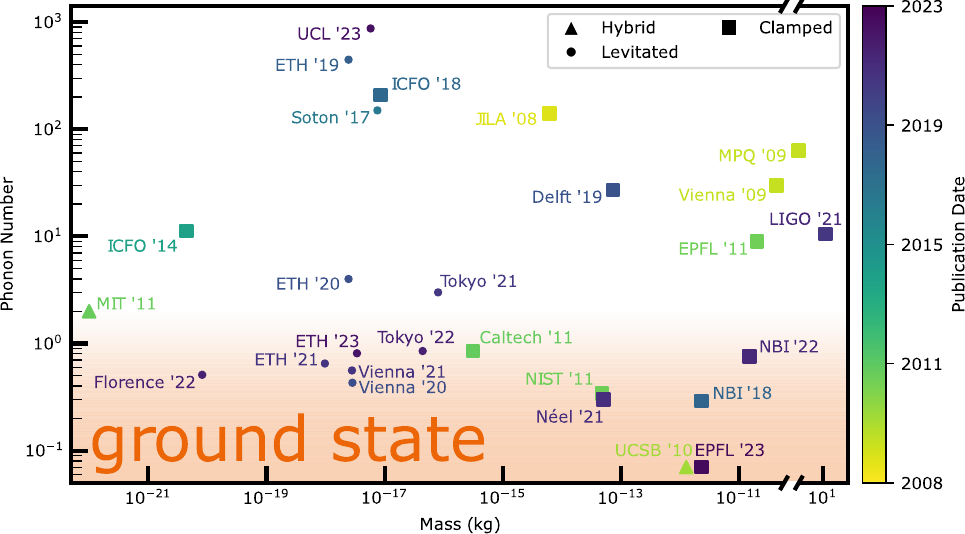}
    \caption{\textbf{Masses and phonons of massive quantum systems}. The plot shows the masses of systems controlled in the laboratory plotted against the mechanical occupation number (phonons). The colors indicate the year of publication, while the symbols indicate the type of system. Where phonon occupancy numbers were not explicitly stated in the publication, we have estimated them using $N = k_B T/(\hbar \omega_m)$, where $N$ is the number of thermal phonons, $k_B$ is Boltzmann's constant, $T$ is the temperature, and $\omega_m$ is the reported mechanical frequency. We classify systems as hybrid whenever a qubit or similar system is used to interface with the mechanical element. The data points correspond to the following references:  
    JILA '08:~\cite{teufel2008dynamical}, Vienna '09:~\cite{groblacher2009demonstration}, MPQ '09:~\cite{schliesser2009resolved}, UCSB '10:~\cite{oconnell2010quantum}, EPFL '11:~\cite{riviere2011optomechanical}, NIST '11:~\cite{teufel2011sideband}, Caltech '11:~\cite{chan2011laser}, NBI '18:~\cite{rossi2018measurement}, Delft '19:~\cite{guo2019feedback}, Néel '21:~\cite{cattiaux2021macroscopic}, NBI '22:~\cite{seis2022ground}, EPFL '23:~\cite{youssefi2023squeezed}, ICFO '18:~\cite{debonis2018ultrasensitive}, ICFO '14:~\cite{moser2014nanotube}, MIT '11:~\cite{schleier2011optomechanical}, Soton '17:~\cite{vovrosh2017parametric}, ETH '19:~\cite{windey2019cavity}, ETH '20:~\cite{tebbenjohanns2020motional}, Vienna '20:~\cite{delic2020cooling}, Tokyo '21:~\cite{kamba2021recoil}, ETH '21:~\cite{tebbenjohanns2021quantum}, Vienna '21:~\cite{magrini2021real}, Tokyo '22:~\cite{kamba2022optical}, Florence '22:~\cite{ranfagni2022two}, ETH '23:~\cite{piotrowski2023simultaneous}, UCL '23:~\cite{pontin2023simultaneous}, LIGO '21:~\cite{whittle2021approaching-1}.    }
    \label{fig:phonons:vs:mass}
\end{figure*}

To showcase the diversity of systems available, we provide a graphical overview of different platforms in Fig.~\ref{fig:MechanicalSystem}. To further demonstrate advances that have been made in terms of controlling massive systems in the laboratory, we plot of masses vs. phonon numbers achieved for mechanical oscillator in Fig.~\ref{fig:phonons:vs:mass}. Here, the symbols represent the type of system, and the colors indicate the date of publication. For comparison, the largest Bose-Einstein condensates (which are not included in the plot) that have been created thus far contain around $10^{10}$ atoms \cite{vanderstam2007large}, which have a total mass of $4\times 10^{-16}$\,kg in the case of sodium. 
We start by reviewing the state-of-the-art experimental tests of gravity today (Sec.~\ref{sec:State-of-the-art}), then we provide an overview of key methods for controlling massive quantum systems in the laboratory, including preparing squeezed states, spatial superpositions, and entangled states (Sec.~\ref{sec:controlling:massive:systems}).

%
%
\subsection{State-of-the-art of experimental tests of gravitation with massive systems}
\label{sec:State-of-the-art}
\begin{figure}
    \centering
    \includegraphics[width=\columnwidth]{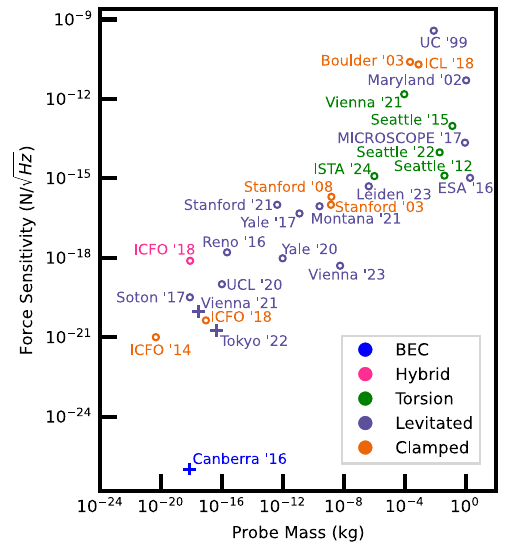}
    \caption{\textbf{Force sensitivities achieved with massive probe systems.} The colors correspond to  BECs (blue), hybrid systems (pink), torsion balances (green), levitated systems (purple), and clamped systems (orange). Plus signs indicate that the system contains less than one thermal phonon. We classify systems as hybrid whenever a qubit or similar system is used to interface with the mechanical element.  The data points correspond to the following references:  
   Canberra '16:~\cite{hardman2016simultaneous}, Boulder '03:~\cite{long2003upper}, Stanford '03:~\cite{chiaverini2003new}, Stanford '08:~\cite{geraci2008improved}, ICL '18:~\cite{pike2018broad}, ICFO '18:~\cite{debonis2018ultrasensitive}, ICFO '14:~\cite{moser2014nanotube}, ICFO '18:~\cite{tavernarakis2018optomechanics}, UC '99:~\cite{goodkind1999superconducting}, Maryland '02:~\cite{moody2002three}, Reno '16:~\cite{ranjit2016zeptonewton}, Soton '17:~\cite{hempston2017force}, Yale '17:~\cite{monteiro2017optical}, Yale '20:~\cite{monteiro2020force}, UCL '20:~\cite{pontin2020ultranarrow}, Montana '21:~\cite{lewandowski2021high}, Vienna '21:~\cite{magrini2021real}, Stanford '21:~\cite{blakemore2021search}, Tokyo '22:~\cite{kamba2022optical}, Leiden '23:~\cite{fuchs2023magnetic}, Vienna '23:~\cite{hofer2023high}, ESA '16:~\cite{armano2016sub}, MICROSCOPE '17:~\cite{touboul2017microscope}, Seattle '12:~\cite{wagner2012torsion}, Seattle '15:~\cite{terrano2015short}, Vienna '21:~\cite{westphal2021measurement}, Seattle '22:~\cite{shaw2022torsion}, ISTA '24:~\cite{agafonova2024laser}.}
    \label{fig:sensitivity:vs:mass}
\end{figure}

Here, we review experiments that have made headway toward testing aspects of gravity, such as precision force sensing. In Fig.~\ref{fig:sensitivity:vs:mass}, we plot the force sensitivities that have been achieved to date against the masses of the probe systems. We note that it is not always clear whether the values reported can be compared directly, as we do here. In some cases, such as for~\cite{hofer2023high}, the values plotted are based on predictions for the ideal experiment. We refer the interested reader to further consult the works in questions, which are cited in the caption of Fig.~\ref{fig:sensitivity:vs:mass}.

\subsubsection{Tests with atoms and molecules} \label{sec:tests:with:atoms}

Since its early demonstrations and pioneering work of Kasevich and Chu~\cite{kasevich1991atomic}, atom interferometry employing laser-cooled cold atoms~\cite{cronin2009optics} has been established as a precision technique for sensing~\cite{peters1999measurement, peters2001high,tino2021testing}, with the realization of sensitive gravimeters, gravity gradiometers, and gyroscopes. 
In addition to measuring the gravitational acceleration due to the earth $g$ with part-per-billion precision, atom interferometers have also been used to measure the Newton constant $G_N$~\cite{sorrentino2010sensitive,lamporesi2007source,rosi2014precision} at the 150 parts-per-million level and are promising for improving tests of the gravitational inverse square law at laboratory scales~\cite{tino2021testing}. Several theories, as described in Sec.~\ref{sec:modified:gravity} predict modifications of the Newtonian inverse square law with a Yukawa type deviation below the mm length scale, such as that described in Eq.~\eqref{eq:Yukawa}. As one particularly well-suited class of modified gravity theories, interferometry with cold atoms is ideal to study Chameleon forces~\cite{sabulsky2019experiment,jaffe2017testing} due to the screening effect present in larger scale test masses.  
Atom interferometry has also been used to test the Einstein Equivalence principle~\cite{schlippert2014quantum} at the part-per-trillion level~\cite{asenbaum2020atom} and has been proposed as a technique for precision tests of general relativity~\cite{dimopoulos2007testing}.  
In addition, atom interferometry has been identified as a promising method to search for gravitational waves in the mid-band ($\sim1$ Hz)~\cite{canuel2018exploring, abe2021matter,badurina2020aion}, between the sensitivity bands of the ground-based interferometer detectors~\cite{abbott2016observation} and LISA~\cite{seoane2013gravitational,amaroseoane2017laser}.  
Recent work has enabled the detection of the gravitational analog of the Aharonov-Bohm effect in precision atom interferometry~\cite{overstreet2022observation, overstreet2022inference}.

The sensitivity of atom interferometers as gravimeters are limited by the interrogation time $T$, which for free-fall interferometers scales as $\delta_{\phi}=k_{\rm{eff}}g T^2$, limiting earth-based experiments to times of order 1 second for a 10-meter drop path. Here $g$ is the Earth's gravitational acceleration and $k_{\rm{eff}}$ is the effective wave-vector of the momentum transfer in the beamsplitter pulse of a light-pulse atom interferometer. Large momentum transfer beamsplitters~\cite{rudolph2020large,siemss2022large} are a pathway for improved sensitivity when limited by interrogation time constraints. Recent work has demonstrated a momentum transfer of $102 \hbar k$ in $^{87}$Rb~\cite{chiow2011large}, $112 \hbar k$ in $^{174}$Yb~\cite{plotkinswing2018three}, and $141 \hbar$k in $^{88}$Sr~\cite{rudolph2020large}. 
Space-based approaches may permit significantly longer interrogation times, and alternatively, atom interferometry with atoms trapped in a lattice can extend interrogation times up to 20 seconds~\cite{xu2019probing}, with recent work recently surpassing a minute~\cite{panda2023probing}.

Atom interferometry has also been performed with ultra-cold atoms cooled to quantum degeneracy, both with bosonic (see for example~\cite{zoest2010bose,muentinga2013interferometry, Kovachy2015Quantum}) and fermionic~\cite{roati2004atom} atomic species.  
Bose-Einstein condensates (BECs) of ultracold atoms are a versatile platform that can be used for a variety of precision quantum sensing applications, where their slow wave packet expansion and coherence play an important role. Interferometry with Bose-Einstein condensed atoms has been proposed as a method to search for short-range deviations from Newtonian gravity~\cite{dimopoulos2003probing}. 
 Bloch oscillations of bosonic Sr atoms have been considered as a method to test short-range gravitational forces~\cite{ferrari2006long}. The largest condensates of ultracold atoms have realized atom numbers as large as 20-120 million~\cite{streed2006large,vanderstam2007large}, with reports of atom interferometry with 5 million atoms~\cite{hardman2016simultaneous}. Atom chip-based atomic interferometry experiments with Bose-Einstein condensates have been performed using as many as $4 \times 10^5$ atoms~\cite{jo2007long}. Key advantages when compared with mechanical oscillators are the environmental decoupling and quantum coherence as well as a mature toolbox for quantum state preparation and measurement. BECs and atom interferometers, in general, can, in principle, achieve atom shot-noise limited sensitivity, with a phase resolution scaling as $\delta_\phi \sim 1/\sqrt{N}$. Interferometry with phase resolution at the Heisenberg limit $\delta_\phi \sim 1/N$ is also possible with the aid of squeezed and highly entangled states~\cite{szigeti2021improving}.  A challenge has to do with the fragility of such highly entangled states due to environmental perturbations such as background gas collisions. 

Another modality of sensing with BECs involves observing their center of mass oscillations or collective modes.  For example, the center of mass oscillation of BECs has been used to measure Casimir-Polder surfaces between the condensate and a nearby surface~\cite{harber2005measurement}.  The dynamical response of the phonons of BECs has also been predicted to be very sensitive to acceleration due to the gravitational attraction of nearby masses, with sensitivities to oscillating masses at the hundred mg scale at millimeter separations~\cite{raetzel2018dynamical}. 
Experiments employing a BEC in a double-well are useful for a variety of fundamental physics tests and could have some advantages when compared to methods using solids~\cite{howl2019exploring}. BECs can be cooled down to picoKelvin regime lowering some sources of noise. Another advantage is that atoms are free to tunnel between wells and states such as two-mode squeezed states can be prepared involving atom superpositions between the two wells~\cite{esteve2008squeezing}. However, particularly challenging is to prepare Schrödinger cat or NOON states. These states are very sensitive to decoherence. The most limiting source of noise is three-body recombination~\cite{tolra2004observation}.

Also, atom interferometry with atoms trapped in an optical lattice has been suggested as a possible route towards observing quantum entanglement induced from the gravitational interaction with a mechanical oscillator~\cite{carney2021using}, albeit with additional assumptions~\cite{hosten2022constraints,streltsov2022significance,ma2022limits}.

\begin{figure*}
    \centering
    \includegraphics[width=1\textwidth]{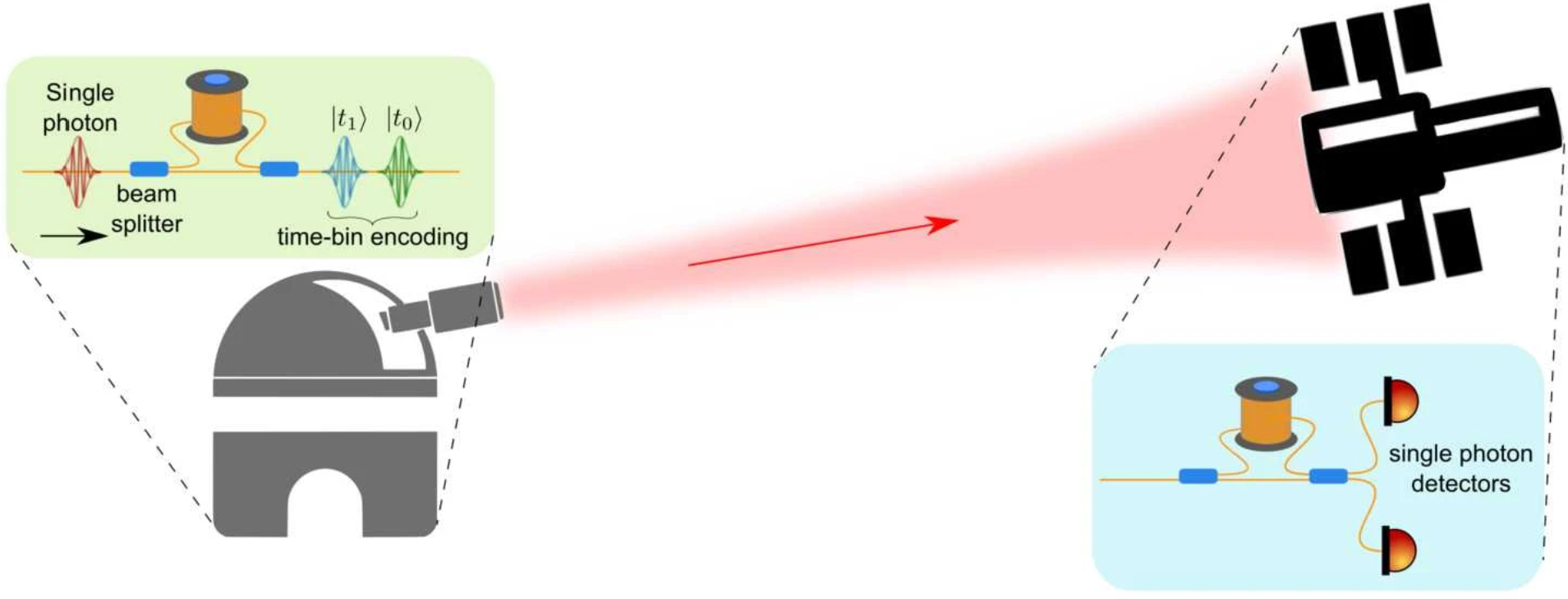}
    \caption{\textbf{Simplified diagram of the optical COW experiment in space.} A single photon is split into two paths using an unbalanced Mach-Zehnder interferometer (MZI). This photon is then transmitted to a satellite with an identical MZI. The interference observed at the satellite shows the phase shift caused by gravity. Adapted from \cite{mohageg2022deep}.
    }
    \label{fig:cow-exp}
\end{figure*}

\begin{figure*}
    \centering
    \includegraphics[width=1\textwidth]{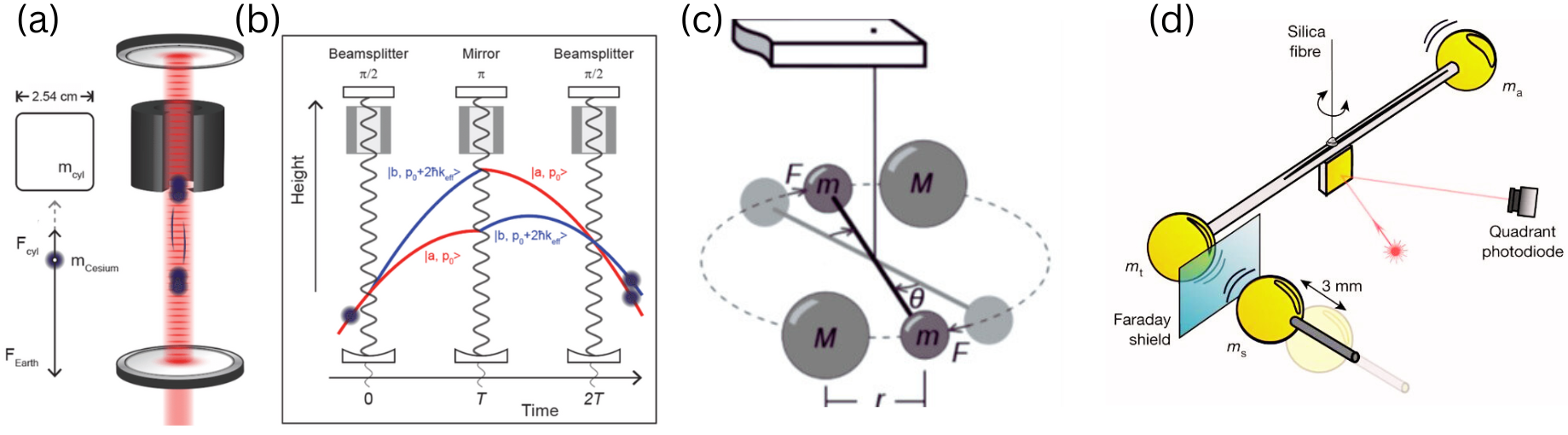}
    \caption{\textbf{Experimental setup and interferometry measurement.} 
    (a) Setup for measuring Cesium atoms' acceleration towards a 0.19 kg tungsten cylinder in ultrahigh vacuum. The cylinder is 2.54 cm in height and diameter, with a 0.5 cm axial through-hole and slot width. Differential measurements are used to study interactions with the mass. 
    (b) A Mach–Zehnder interferometer using Raman transitions in an optical cavity. Three laser pulses manipulate Cesium atoms, splitting, reflecting, and recombining them to measure phase differences, providing an ensemble-averaged acceleration over 110 ms across approximately 100,000 atoms. Adapted from \cite{jaffe2017testing}.
    (c) Schematic of a Cavendish torsion balance with dumbbell masses (M), external forces (F), and resulting rotation angle ($\theta$). 
    (d) Torsion pendulum as a gravitational acceleration transducer, containing two gold spheres (1 mm radius, 40 mm apart) on a glass capillary. One sphere serves as a test mass (90.7 mg), the other as a counterbalance (91.5 mg). A silica fiber with a diameter of 4 $\mu$m supports the pendulum with a 3.6 mHz torsional resonance. The torsion angle is detected optically. The source mass (92.1 mg) is harmonically moved 3 mm at 12.7 mHz to enhance the gravitational signal. Electrostatic interference is reduced using Faraday shield and discharging techniques. Adapted from \cite{westphal2021measurement}.
    }
    \label{fig:mat-wav-inter}
\end{figure*}

Finally, matter-wave interferometers for complex molecules formed of many atoms are also promising for tests of gravity as they benefit from the increase in mass in superposition~\cite{fein2019quantum}. One example is shown in Fig.~\ref{fig:mat-wav-inter} (a) and (b), where interferometry measurements assess the acceleration of Cesium atoms toward a tungsten cylindrical mass in an ultrahigh vacuum environment. This setup employs a Mach-Zehnder interferometer with Raman transitions to measure phase differences and determine acceleration, adapted from \cite{jaffe2017testing}. Additionally, the duration of free fall in Earth's gravity ultimately limits the mass of molecules in interferometry, while it is more difficult to pick up a COW-like phase in the typically near-field regime of operation of these large-mass interferometers (see Sec.~\ref{sec:Newtonian:Schrödinger} for a discussion of Collela-Overhauser-Werner (COW) phases on matter waves). Fig.~\ref{fig:cow-exp} illustrates the optical COW experiment conducted in space, demonstrating how a single photon split by an unbalanced Mach-Zehnder interferometer and transmitted to a satellite reveals gravity-induced phase shifts through observed interference~\cite{mohageg2022deep}. More details about molecule interferometry experiments are given in Sec.~\ref{Sec-exp-superposition}.

\subsubsection{Tests with neutrons}
Neutrons have proven to be a powerful probe for testing our understanding of gravity. The equivalence principle, which states that all objects fall at the same rate in a gravitational field regardless of their composition or structure, can be scrutinized using neutron matter-wave interferometry~\cite{rauch1975verification,colella1975observation,greenberger1983neutron} (see also Sec.~\ref{sec:Newtonian:Schrödinger}). In these experiments, a neutron beam is split into two paths. These two paths then interfere upon recombination, a process that allows for precise determination of the relative gravitational potential experienced by the neutrons in the two paths. The equivalence principle is tested by measuring the phase shift in the interference pattern. Such an approach has made it possible to verify the equivalence principle to a high degree of precision~\cite{lammerzahl1996equivalence,lammerzahl1998minimal}. One of the main advantages of neutrons in these experiments over atoms is that neutrons are uncharged, meaning they are free from the electromagnetic forces that can influence the movement of atoms. This allows the gravitational interactions to be studied with less interference from other forces, leading to a higher degree of precision in the results.

Later, with the advancement in the field, ultra-cold neutrons (UCN) emerged as a robust tool for testing gravitational theories. The UCNs originated from the insights of Enrico Fermi~\cite{fermi1936motion} who recognized the potential of slow neutrons to interact coherently while scattering, creating an effective interaction potential for neutrons passing through matter. This led to the concept of storing neutrons with very low kinetic energies, initially predicted by~\cite{zeldovich1959storage} and first realized experimentally by groups in Dubna~\cite{lushikov1969observation} and Munich~\cite{steyerl1969measurements}. A significant breakthrough was made by~\cite{nesvizhevsky2002quantum}, who observed quantized states of matter under the influence of gravity using UCNs. Their work has further opened up possibilities for probing fundamental physics, such as the equivalence principle~\cite{nesvizhevsky2002quantum}. More generally, the advancement in this field has led to UCNs becoming a robust tool for testing gravitational theories~\cite{steyerl1977quasielastic,ivanov2021quantum}. UCNs cooled nearly to absolute zero can be stored for extended periods, enabling precise measurements of the gravitational behavior of neutrons. In recent years, there have been advances in the production of UCNs, with~\cite{zimmer2011superthermal} reporting a world-best UCN density available for users, achieved with a new source based on the conversion of cold neutrons in superfluid helium. Experiments with UCNs aim to measure the gravitational free-fall of neutrons with high accuracy, offering a platform to test general relativity and other theories like modified Newtonian dynamics (MOND)~\cite{famaey2012modified}. Ultracold neutron (UCN) spectroscopy has been instrumental in constraining various theories and phenomena, including dark energy, chameleon fields,~\cite{jenke2014gravity} and new short-range forces~\cite{kamiya2015constraints}. 
In a recent experiment~\cite{haddock2018search} a pulsed neutron beam was deployed to probe Newton's law of universal gravitation on subnanometer scales. The results set a stringent upper bound on the magnitude of potential unaccounted-for forces, enhancing the foundation upon which we apprehend gravity.
Moreover, promising theoretical outlooks are unveiling new paths of exploration, including measurements of the gravitational redshift of neutrons. This involves observing the change in energy of a neutron due to a change in gravitational potential, establishing a promising technique for testing general relativity~\cite{roura2022quantum}. Advancements are also foreseen in the precision of measuring the electric dipole moment of neutrons, with potential assistance from quantum sensors based on weak-value amplification~\cite{pendlebury2015revised, altarev1986new, knee2013quantum}.
Other projects, such as the qBOUNCE and the GRANIT collaborations, aim to expand the understanding of gravity at short distances by examining gravitationally bound quantum states of ultra-cold neutrons~\cite{jenke2019testing,jenke2009q,kreuz2009method}. Studies such as these could impose stringent constraints on hypothetical fields and forces, further refining our understanding of gravity and providing insights that might push beyond the boundaries of currently accepted theories. For a more detailed review of gravity measurements by neutrons, we recommend~\cite{pokotilovski2018experiments}.

\subsubsection{Tests with torsion balances and clamped mechanical systems}
Sensitive torsion balances are a powerful and proven method for studying exotic short-range gravity~\cite{kapner2007tests,lee2020new}, equivalence-principle violation involving ordinary and dark~\cite{wagner2012torsion,shaw2022torsion} matter, and novel spin-dependent interactions~\cite{terrano2015short} as well as measuring the Newton constant~\cite{gundlach2000measurement}. They remain one of the most promising paths forward for these studies as their sensitivity continues to increase and the understanding of background noise and systematic errors from patch charges and other surface forces improves.

Current tests are often limited by environmental vibrations that can “kick” the pendulum, exciting it's fundamental and spurious (swing, bounce, and wobble) modes~\cite{wagner2012torsion}. This is particularly in short-range tests where patch charges couple to the spurious modes producing noise that dominates at small separations and limits the minimum attainable separation~\cite{lee2020new}. Time-varying environmental gravity gradients limit equivalence-principle tests. Both of these technical limiting factors could be addressed by the development of a suitable low-vibration underground facility. 

Torsion balance experiments (shown in Fig.~\ref{fig:mat-wav-inter} (c) and (d)) have typically employed relatively large source masses, well beyond the scale envisioned for achieving a quantum superposition. Work towards employing sub-mm-scale source masses and similarly miniaturized torsion pendula is underway~\cite{westphal2021measurement}. In this work, thus far the smallest source mass that has been used for a gravitational measurement is approximately the mm scale. While far from the scale where macroscopic quantum superpositions have been imagined in interference experiments, this work represents a step in this direction to bridge this gap. These experiments also tend to operate at low frequencies and are limited by the same environmental perturbations and thus could benefit from similar future low-noise facilities.

At even smaller length scales, microcantilevers~\cite{geraci2008improved,chiaverini2003new} and microfabricated torsion oscillators~\cite{long2003upper} have been used to obtain bounds on Yukawa type deviations of the Newton inverse square law at distances ranging from a few microns to tens of microns.  Cutting-edge nanofabrication technology is making it possible to routinely design advanced 2D and 1D clamped resonators with massive quality factors, for instance, suspended silicon nitride membranes and carbon nanotube resonators. 

Clamped mechanical systems interfaced with superconducting qubits have emerged as a fertile ground for probing the interplay between quantum mechanics and macroscopic objects. Early groundbreaking experiments demonstrated the feasibility of reaching the quantum ground state of mechanical resonators using superconducting circuits~\cite{oconnell2010quantum, teufel2011sideband}. Furthermore, laser cooling techniques have been adapted to cool nanomechanical oscillators into their quantum ground state~\cite{chan2011laser}. Recent work in \cite{Liu2020laser} reported laser cooling a clamped oscillator to an average occupation number as low as $\langle n \rangle =0.09$ phonons. 

Building upon this lineage of research,~\cite{youssefi2022squeezed} introduced a hybrid quantum system consisting of a superconducting circuit seamlessly integrated with a micromechanical oscillator. Achieving a thermal decoherence rate of 20.5\,Hz and a dephasing rate of 0.09\,Hz, they enabled the free evolution of a squeezed mechanical state over milliseconds. 
We anticipate that such advances will enable exploration of elusive phenomena that arise from the interplay between quantum mechanics and general relativity. 

\begin{figure*}
    \centering
    \includegraphics[width=0.70\textwidth]{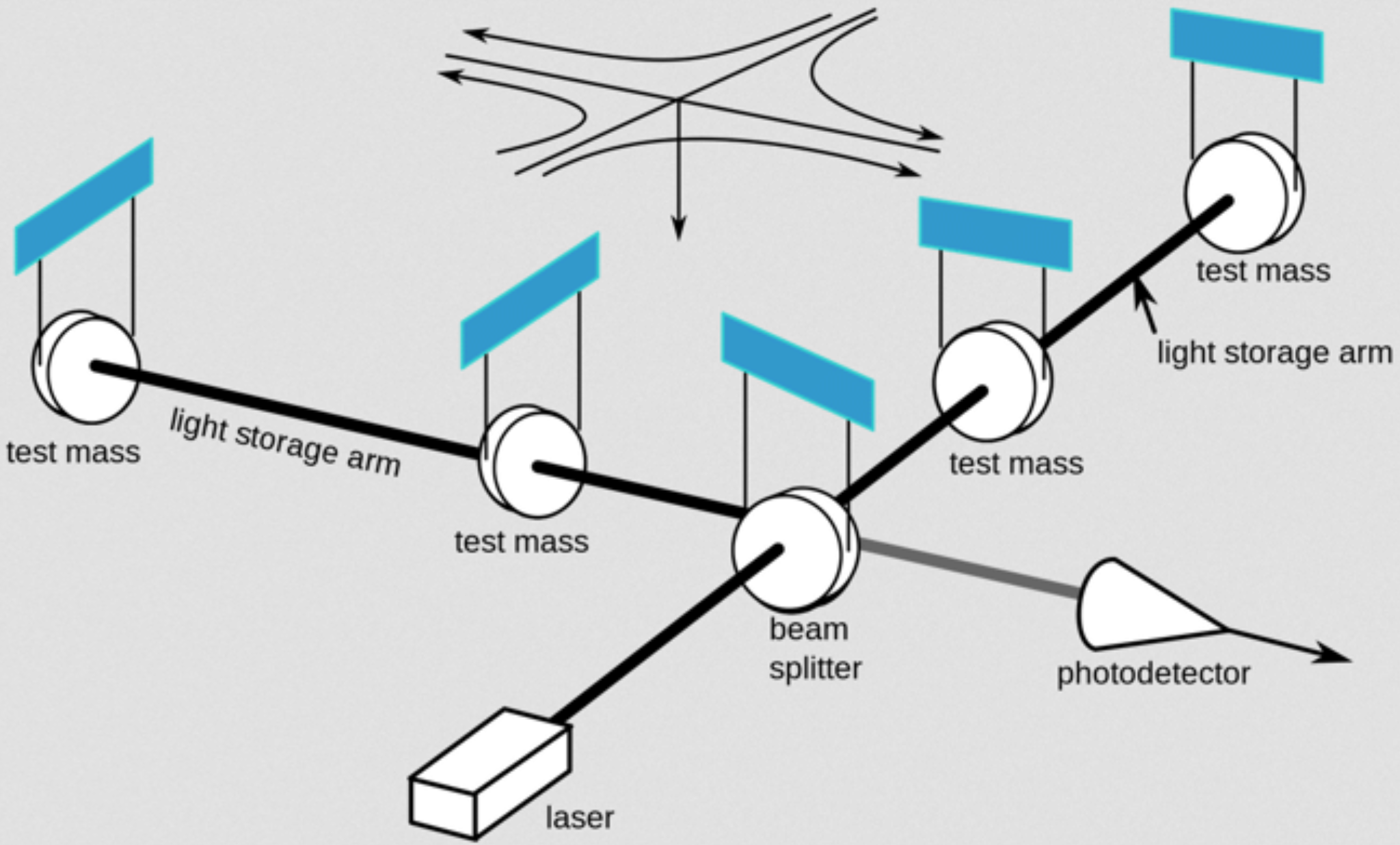}
    \caption{\textbf{LIGO laser interferometer setup.} The interference of light in the two arms leads to the detection of extremely small displacements. Adapted from \cite{scharpf2017simulation}.
    }
    \label{fig:LIGO}
\end{figure*}

Furthermore, LIGO-style experiments have also contributed significantly to the field. The Laser Interferometer Gravitational-Wave Observatory (LIGO)~\cite{abbott2016first}, as shown in Fig.~\ref{fig:LIGO}, has provided direct evidence for the existence of black holes and opened up a new avenue for exploring the nature of gravity. 

In conclusion, tests with clamped systems, including torsion balances, have proven effective in studying gravity and fundamental physics. Challenges such as environmental vibrations and technical limitations have motivated the development of low-vibration facilities and miniaturization efforts. Nanofabrication techniques have enabled advanced resonators, while hybrid quantum systems offer new avenues for investigating quantum physics and dark matter. Additionally, LIGO-style experiments have made groundbreaking contributions to our understanding of gravitational waves. Overall, these advances hold substantial potential to advance our understanding of gravitational phenomena. 

\subsubsection{Tests with levitated mechanical systems}
\label{sec:tests:levitated}
Levitated mechanical systems offer a platform to investigate the interplay between quantum mechanics and gravity in the low-energy non-relativistic regime. Since levitated systems are much isolated from their environment the center of mass motion can be very close to an ideal harmonic oscillator persisting at large Q-factors~\cite{geraci2010short}. Such isolation together with the, in principle, quantum-limited detection of the position and therefore motion of the mechanical system by light (or direct electrical or magnetic interactions) make them exceptional for testing quantum effects in gravity~\cite{caves1980measurement, aspelmeyer2014cavity}. 

Recently, the study of levitated mechanical systems has gone beyond the usual limits set by the gravitational law that governs how objects attract each other~\cite{arndt2014testing}. In~\cite{moore2021searching, priel2022dipole} the finer points of gravity-related phenomena were explored by probing two-particle interactions with levitated particles beyond established force laws based on precise force and acceleration measurements.

In some more detail, levitated systems provide a platform to generate and coherently control quantum effects in their motion via ground-state cooling, measurement-based schemes, and more~\cite{aspelmeyer2014cavity} and at the same time, come with sufficient mass for directly testing gravity effects on experimentally accessible time and magnitude scales. Quantum experiments with gravitating particles of the Planck mass $(m_{pl}= \sqrt{\hbar c/G})$~\cite{aspelmeyer2022when, ulbricht2021testing} become feasible. In addition, the levitated system has a full set of only six mechanical modes of translation ($x,y,z$) and rotation ($\alpha, \beta, \gamma$) which are developed to be used as quantum probes of gravity in the linear and nonlinear regimes~\cite{bateman2014field,gosling2024levitodynamic,kilian2024dark}. Recent experiments achieved the simultaneous cooling of all those modes~\cite{pontin2023simultaneous, kamba2023nanoscale}, which opens the door for quantum state preparation~\cite{delic2020cooling}. Rotational states provide a unique setting with their intrinsic nonlinearities for quantum experiments~\cite{stickler2018probing, schrinski2022interferometric}.

Besides the prospects of using levitated mechanical systems for force~\cite{winstone2018direct} and inertial sensing~\cite{teufel2011sideband}, direct gravity probes are emerging. Gravity affects levitation directly but is negligible in small-mass particle optical levitation~\cite{ashkin1971optical,rademacher2022measurement}, while relevant in larger mass Paul ions~\cite{paul1990electromagnetic}, In Meissner-superconducting and diamagnetic traps,  $g$ influences the trapping position and has a clear effect~\cite{cano2008meissner}.
Static gravity was measured with levitated optomechanics by turning off the trap~\cite{frimmer2017controlling}. In addition, a Meissner levitated magnet has been used in a two-mass gravity detection experiment and has measured gravity at the level of attonewton gravity~\cite{fuchs2023magnetic}.

In optical levitation, nano and microparticles trapped in an ultra-high vacuum can be cooled down to their ground state of center-of-mass motion through radiation pressure forces exerted by optical cavities~\cite{libbrecht2004toward, romero2010toward, barker2010cavity, chang2010cavity}. This technique has been explored for over a decade to test short-range gravity forces~\cite{geraci2010short}. By employing optically levitated systems where microspheres are trapped and cooled in a vacuum, it has been possible to probe and measure gravitational effects with unprecedented precision at the micrometer scale~\cite{ranjit2016zeptonewton}.
Nanoparticles with a cooled center of mass temperature can also serve as a source for matter-wave interferometry experiments~\cite{bateman2014field}, which could be used for measuring gravitational acceleration and probing gravity at the micron length scale~\cite{geraci2015sensing}.

Recent experiments~\cite{timberlake2021probing} employed levitation via the Meissner effect, where two magnets suspended in a levitated state perturb each other's motion to measure the gravitational attraction between them. These experiments demonstrated the practicality of measuring gravitational acceleration for small masses, showcasing the potential for future improvements in experimental setups. Additionally, the gravitational constant (G) can be estimated from such measurements. Coupling to superconducting LCs in cryogenic environments. Paul ion trapping provides a stable trap for tuneable $e/m$ ratios~\cite{paul1990electromagnetic}.  The close technological heritage from atomic Paul trapping makes available a set of the center of mass motion state preparation protocols and tools for the manipulation, cooling, and control of charged nano- and micro-particles motion via electro-dynamical ion levitation~\cite{leibfried2003quantum, schneider2010optical}.

The LISA Pathfinder (LPF) mission, designed to detect gravitational waves in space, utilized electrostatic detection of freely falling masses on the level of kilogramm~\cite{armano2016sub}. LPF data, as well as those from earth-bound gravitational wave detectors, were employed to establish strong upper bounds on CSL and DP models and are space-based derivatives of ground-based optomechanical precision experiments, including gravitational wave detectors such as LIGO, VIRGO, GEO600, but also AURIGA~\cite{carlesso2016experimental}. LPF is similar to other space missions such as Gravity Probe B to test Lense-Thirring GR frame-dragging effects~\cite{everitt2011gravity}, and GRACE, GOCE the satellite gradiometry missions~\cite{tapley2004gravity,drinkwater2003goce}. The MICROSCOPE mission~\cite{touboul2017microscope} aimed to test the weak version of the equivalence principle, the basic principle of Einstein's theory of general relativity. Test masses made of different materials but with equal inertial masses were used in this experiment. By monitoring the motion of these masses over an extended period, MICROSCOPE sought to detect any deviations from the principle. The results of MICROSCOPE provided strong evidence in support of the principle, consolidating the predictions of general relativity.

In conclusion, levitated systems have emerged as a promising platform for investigating the interface between quantum mechanics and gravity. Recent advancements, such as Meissner effect-based levitation and space-based experiments like LISA Pathfinder and MICROSCOPE, have expanded our capabilities to study fundamental physics principles. The precise measurements achievable in levitated systems, coupled with the microgravity environment of space, contribute to our understanding of gravity, general relativity, and the fundamental laws of physics. To this end, the recent MAQRO proposal aims to explore levitated particle dynamics in space, which would open a pathway for matter-wave interference experiments with long interaction times, by not being subject to falling under Earth's gravity~\cite{kaltenbaek2022maqro,belenchia2022quantum}. Future advancements in levitated systems and their applications hold exciting prospects for furthering our knowledge of the quantum-gravity interface~\cite{rademacher2020quantum}.

\subsubsection{Approaches with hybrid systems} \label{sec-hybrid-grav}
Last but not least, there are also hybrid mechanical systems~\cite{rogers2014hybrid, treutlein2014hybrid, kolkowitz2012coherent} where systems other than optical fields/photons are coupled to mechanical motion. Hybrid mechanical systems are indeed very powerful experiments building on the well-studied Jayens-Cummings Hamiltonian physics of a two-level system coupled to a continuous variable - quantum harmonic oscillator - system, as discussed in Sec.\ref{sec:two:level:coupling}. Hybrid systems have been indeed at the forefront of demonstration of quantum states of very massive systems, while for relevance for probing gravity effects, one has to consider that many of the hybrid systems involve non-center-of-mass motional modes, such as vibration modes of membranes states~\cite{oconnell2010quantum} or acoustic modes in superconducting qubits~\cite{chu2017quantum}. Hybrid systems are arguably the most established mechanical quantum systems, and the quantum aspects of hybrid systems are discussed more in Sec.~ \ref{Sec-exp-superposition} as they are used for the generation of massive superpositions and as well as for demonstration of quantum entanglement between two large-mass systems. Typically, in hybrid systems, the mechanical modes are at high frequency (100 MHz to 10 GHz) which allows for cooling to the quantum ground state by cryogenics, usually in dilution-type refrigerators. Such ground state cooling gives direct access to the quantum regime and for quantum state preparation and coherent control schemes. An example where this has been done by continuous weak measurement (see for theory on measurement based control Sec.~\ref{sec:control}) based feedback techniques is the work by Siddiqi~\cite{siddiqi2021engineering}, which is built on a record-high detection efficiency of the measurement of more than 80\%. A more detailed account of the physics of hybrid systems is given in~\cite{aspelmeyer2014cavity}.

However, there are approaches for testing aspects of gravity by hybrid systems including for gravitational wave detection including coupling atoms to an optomechanical cavity which influences the atom-cavity interaction~\cite{camerer2011realization}, cooling the system by linking a superconducting qubit to a mechanical resonator for improved detection sensitivity~\cite{oconnell2010quantum} and integrating quantum dots with mechanical resonators or optical cavities for enhanced detection~\cite{bennett2010strong, yeo2014strain}. Also, the use of solid-state systems with mechanical resonators, coupled with optical cavities, proves promising for gravitational wave detection~\cite{arcizet2011single, kolkowitz2012coherent}. These various approaches leverage the unique properties of different components for higher sensitivity and precision in gravitational wave detection; for instance, the use of optomechanical systems that offer enhanced cooling by constructive quantum interference and suppressed heating by destructive interference, which is essential for precision control and quantum information processing~\cite{chen2015cooling}. Furthermore, modern hybrid systems allow for the exploration of the quantum-classical mechanics interface and demonstrate the potential for a paradigm shift from cryogenic to room temperature quantum experiments using hybrid nanoelectromechanical system (NEMS) resonators~\cite{tavernarakis2018optomechanics}. Such advancements reflect the rapidly evolving potential of hybrid systems in examining quantum physics at a macroscopic scale and as an avenue for quantum state generation in massive mechanical systems~\cite{liu2021gain, akram2015electromagnetically}. 

In conclusion, hybrid systems are expected to open new ways to inject quantum features into large-mass mechanical systems by coupling to qubit systems, and the first concrete steps have been taken already. Hybrid mechanical systems will play a key role to probe into gravity effects within the domain discussed in this review, because of the maturity of quantum controlling large-mass mechanical states.

%
\subsection{Controlling massive mechanical quantum systems in the laboratory} \label{sec:controlling:massive:systems}
%
Here we briefly review experimental achievements and theoretical proposals for large-mass mechanical systems in the quantum domain by sectioning into three classics of exemplary quantum states: squeezing, superposition, and entanglement. 

%

\subsubsection{Squeezing and swapping of mechanics} \label{sec:squeezing}
%
As we are interested in creating quantum states of mechanics, we here discuss the squeezing of mechanical degrees of freedom. We do not, however, discuss the application of squeezed light to mechanical oscillators as it was, for instance, used to advance the gravitational wave detectors VIRGO~\cite{schnabel2017squeezed} and eventually LIGO~\cite{collaboration2011gravitational, aasi2013enhanced}.

\textit{Squeezing in clamped optomechanics:} An early demonstration of directly squeezing the mechanical mode was by~\cite{rugar1991mechanical}, where control over the spring constant enabled to parametrically drive and thus amplify the mechanical motion of the oscillator. This approach allowed for noise suppression of -4.9 dB in one quadrature (see Sec.~\ref{sec:SQL}). Improvement in noise suppression has been theoretically proposed by using both detuned parametric driving and continuous weak measurement of the mechanical oscillator~\cite{szorkovszky2011mechanical}. Experimentally, noise suppression was demonstrated by weak measurement and achieved -6.2 dB in one quadrature~\cite{szorkovszky2011mechanical}.~\cite{pontin2014squeezing} use parametric feedback to stabilize one quadrature without affecting the other and achieve squeezing of -7.9 dB. The realization of quantum squeezing of a quadrature below the zero point fluctuations was achieved by~\cite{pirkkalainen2015squeezing, lecocq2015quantum, wollman2015quantum}. 

Further quantum state protocols, such as state transfer and swapping have also been shown as multi-mode optomechanics systems began to be explored. These naturally have a strong motivation as quantum information protocols but demonstrate powerful state control capabilities with a myriad of techniques. For example,~\cite{weaver2017coherent} demonstrates coherent state swapping between modes of two separate mechanical frequencies in the same cavity.

\textit{Squeezing in levitated optomechanics:} Adjacent to this, the novel capability to control the potential and thus the mechanical frequencies enabled squeezing via non-adiabatic pulses~\cite{rashid2016experimental}. A study of the scattered light revealed the signatures of squeezing on the scattered light generated by the mechanics oscillators~\cite{militaru2022ponderomotive}. Recently, squeezed states have been discussed for testing tiny effects such as those predicted by some form of quantum gravity~\cite{belenchia2016testing}. Squeezing is an operation to affect the mechanical state, it is one crucial operation in the universal toolbox of Gaussian state preparation, but is also discussed as a state preparation step for achieving non-Gaussian states such as quantum superposition in a levitated mechanical system~\cite{rieracampeny2023wigner}, and as well for generating quantum entanglement between two mechanical systems~\cite{cosco2021enhanced}.

In conclusion, squeezing is one option for generating out-of-equilibrium states of a continuously variable system. Squeezing of large-mass mechanical systems has been experimentally demonstrated. Squeezing generates highly sensitive states exhibiting a peculiar quantum signature~\cite{chowdhury2020quantum} and is used to control the effect of dynamical nonlinearities.

%
\subsubsection{Spatial superpositions of mechanical systems}\label{Sec-exp-superposition}
\begin{figure*}
    \centering
    \includegraphics[width=0.75\textwidth]{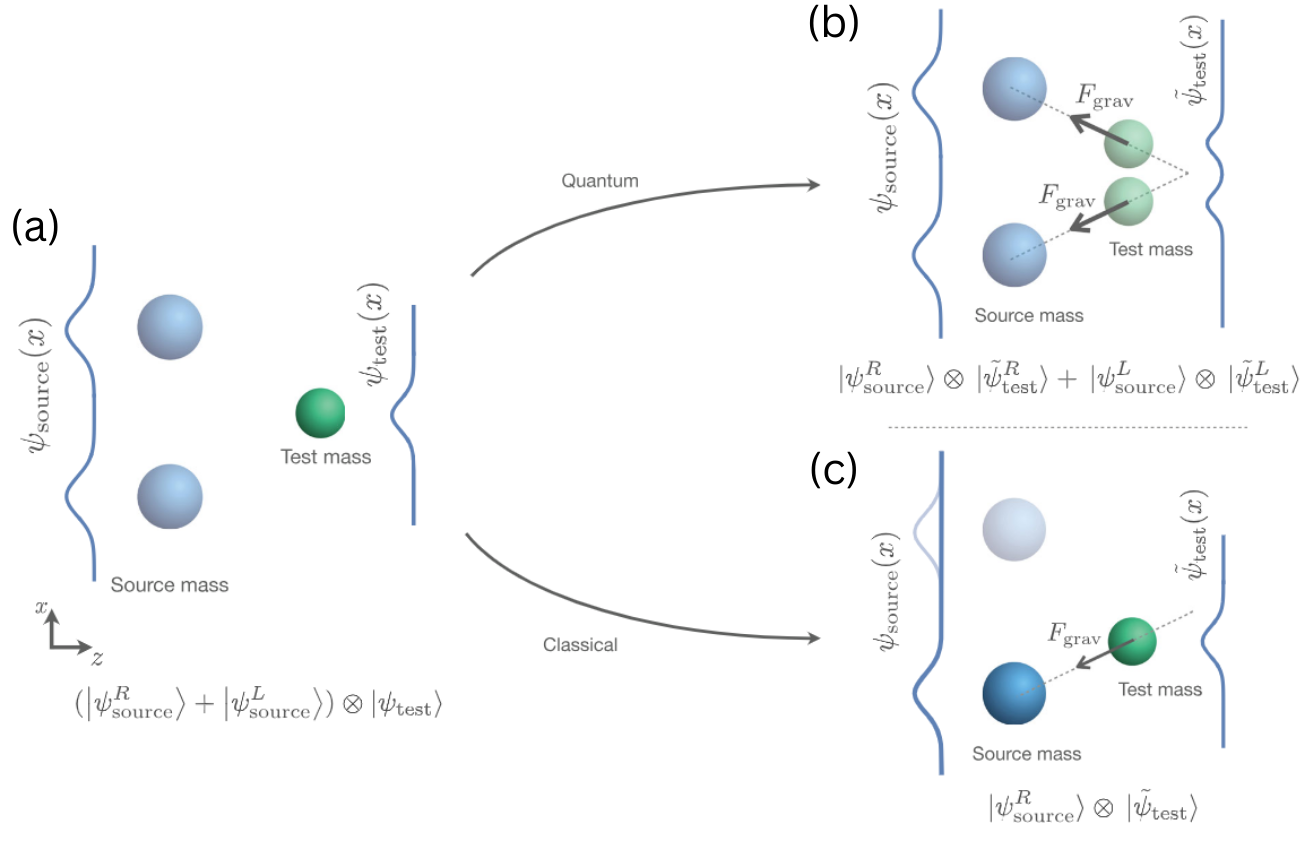}
    \caption{\textbf{Feynman's thought experiment.} (a) A gravitational field source (blue ball) is placed in a quantum superposition and interacts with a test mass (green ball). The system can develop into two scenarios: (b) If the field is in a quantum superposition, the test mass feels two different gravitational forces and also enters a superposition state. (c) If the field stays classical, the test mass feels a single gravitational force and moves accordingly. Adapted from \cite{pedernales2023origin}.
    }
    \label{fig:Feynman-exp}
\end{figure*}

%
The goal of this section is to review the state-of-the-art mechanical quantum systems and how they approach the regime for testing the overlap between quantum mechanics and gravity. One key aspect of this endeavor is the generation of the prototypical quantum state -- the spatial superposition state of sufficiently massive or macroscopic systems. An illustrative example is Feynman's thought experiment, shown in Fig.~\ref{fig:Feynman-exp}, where a gravitational field source in a quantum superposition interacts with a test mass, leading to different scenarios depending on whether the field remains in a superposition or collapses to a classical state~\cite{pedernales2023origin}. Theoretical proposals for achieving this or similar goals, such as ~\cite{bose2017spin,hanif2023testing} and others have been described in \ref{sec:entanglement:mediated:gravity}.

There are competing definitions of what "macroscopic" actually means and it strongly depends on what aspects of physics are to be tested. A single photon in a superposition or a pair of photons entangled over a thousand kilometers is arguably a large quantum system. while various measures of what quantum coherence at macroscopic scales actually means~\cite{leggett1980macroscopic, dur2002effective, bjork2004size, cavalcanti2006signatures, marquardt2008measuring, lee2011quantification}. Here, however, for the purpose of testing the overlap between quantum mechanics and gravity, it seems advisable to choose a macroscopicity measure which includes a set of three parameters about the quantum system and is able to compare a manifold of different physical systems in an objective way, such as the macro-measure based on matterwave superposition as put forward in~\cite{Nimmrichter2013Macroscopicity}. The measure $\mu$ is a function of the mass of the system in a spatial superposition, the spatial size of the superposition, and the time for the spatial superposition to exist. Using $\mu$, it becomes evident how wide beam splitting low-mass atomic fountains~\cite{Kovachy2015Quantum}, large-mass small zero-point motional optomechanical setups (LIGO)~\cite{whittle2021approaching-1} in continuously monitored low-phonon states and levitated mechanical systems compare and why levitated mechanics with mesoscopic masses look most promising to deliver the most macroscopic of superpositions. See Table~\ref{tab:macroscopicity} for a summary of values computed for the measure thus far. The current mass record in matterwave interferometry is for complex molecules at 28 kDa in~\cite{fein2019quantum}, which achieves $\mu > 14$.

\begin{table}[htbp]
    \begin{ruledtabular}
    \begin{tabular}{p{1.8cm} p{4.3cm} p{0.5cm} p{1cm}}
\multicolumn{2}{l}{\textbf{Experiment}}  & \textbf{Year} & $\mu$ \\ \hline
Mechanical resonators & Bulk acoustic waves~\cite{schrinski2023macroscopic, bild2023schrodinger}  & 2022 & 11.3 \\
 & Phononic crystal resonator~\cite{wollack2022quantum} & 2022 & $\sim9.0^*$ \\
 & Surface acoustic waves~\cite{satzinger2018quantum} & 2018 & $\sim8.6^*$ \\ \hline
Matter-wave interference & Molecule inteferometry~\cite{fein2019quantum} &  2019 & 14.0 \\
 & Atom interferometry~\cite{xu2019probing} & 2019 & 11.8 \\
& BEC interferometry~\cite{asenbaum2017phase} & 2017 & 12.4 
\end{tabular}%
\end{ruledtabular}
  \caption{Macroscopicity measure summarized in~\cite{schrinski2023macroscopic}. *estimated by the authors of~\cite{schrinski2023macroscopic}. }
\label{tab:macroscopicity}%
\end{table}%

Different ways to generate spatial superpositions have been proposed for mechanical systems, both clamped and levitated, and some have already been demonstrated experimentally. The technical challenge, if formulated in matterwave language, is to split the matter-wavefront coherently for a tiny de Broglie wavelength. This typically requires preparing, by cooling, some sort of coherent initial state, and a subsequent application of a coherence beamsplitter operation. For spatial superpositions, the prepared coherence length determines both the spatial resolution of the beamsplitter as well as the extend of the final superposition.  

Methods to realize beamsplitters inspired by established technology from matterwave interferometry with electrons, neutrons, atoms and complex molecules~\cite{cronin2009optics, juffmann2013experimental, arndt2014testing, hornberger2012colloquium, millen2020quantum}  the use of optical gratings~\cite{bateman2014field, geraci2015sensing}, by nonlinear interaction in a cavity~\cite{romeroisart2011optically,bose1997preparation,mancini1997ponderomotive,bose1999scheme} or levitated magneto-mechanical oscillators coupled to magnetic fields~\cite{romeroisart2012quantum} are in the mix of proposals, such as magnetic beamsplitter using ferromagnetic particles~\cite{rahman2019large}. 
For mechanical systems, ideas also include measurement-based multiple-pulsed schemes addressing the position and momentum-dependent continuous variables (feeding the cat to become fat)~\cite{vanner2011pulsed} as well as by continuous weak measurements protocols~\cite{rossi2018measurement} or advanced protocols from quantum metrology using dynamical model selection and classical and quantum hypothesis testing~\cite{schrinski2019macroscopicity, ralph2018dynamical, mcmillen2017quantum}, schemes which go even conceptually much beyond the classic scenario for generation of superposition states as well as in evidencing the appearance of non-classicalities, however, the same measure has to be applied to rank macroscopicity consistently.   

While all of the above addresses external ($x$ and $p$) degrees of freedom for generating superpositions, there are also promising ideas for addressing internal states such as isolated electron and nuclear spin states. If the coherence of such states can be extended to long enough times and coupled to $x$ and $p$ in a coherent fashion, then beautiful protocols for state preparation can be transferred from the rich toolbox of atomic two- and few-level physics. Such ideas have been put forward for harmonically bound systems~\cite{scala2013matter} as well as for free motion~\cite{wan2016free}. Again the massive and freely evolving quantum state keeps promise to become the most macroscopic quantum one. 

Superpositions of different energy states of the harmonic oscillator using strong coupling in cavity-QED-like (Jaynes-Cummings Hamiltonian, see Sec.~\ref{sec:two:level:coupling}) systems are a further option and have historically been the first demonstration for a quantum superposition of a seriously massive microwave-driven quantum system~\cite{teufel2011sideband}, for surface acoustic-wave phonons~\cite{satzinger2018quantum} and by using coupling to a superconducting qubit~\cite{wollack2022quantum}, while the spatial extend of the superposition is on the size of the amplitude of the zero-point motion in those systems and advanced techniques have to be applied to reach the defined macroscopic. The energy state superposition is mapped onto a motional or vibrational state~\cite{oconnell2010quantum, bild2023schrodinger, schrinski2023macroscopic}.   While large-mass hybrid mechanical systems can be prepared in quantum states~\cite{sletten2019resolving}, the macroscopicity of hybrid quantum superposition states has to be analyzed by a generalized measure as, for instance, the one shown in table~\ref{tab:macroscopicity}. \cite{bild2023schrodinger} have produced the so far largest spatial superposition state among hybrid systems for Planck mass acoustic modes. Hybrid mechanical systems are supreme quantum systems and their relevance for gravity measurements is discussed in Sec.\ref{sec-hybrid-grav}. 

While the above discussion is about linear motion, macroscopic superpositions can also be achieved by utilizing rotational mechanical degrees of freedom~\cite{stickler2021quantum, carlesso2017when}. These approaches are promising since the generated quantum state is potentially more protected from noise and decoherence. In addition, the technology for implementing angular superpositions is different from those needed for linear superpositions and could be easier to realize. 

For the realization of superposition experiments, aspects of decoherence such as by gas collisions and effects of black body photons, amongst others, have to be considered and understood~\cite{hackermuller2004decoherence, hornberger2003collisional, romeroisart2011quantum}. It is clear that decoherence puts severe constraints on any attempt to realize macroscopic quantum states, and each experiment has to focus on those aspects. Theoretical studies of decoherence effects have been carried out by using an open quantum system dynamical model on the level of master equations, see Sec.~\ref{sec:master:equations} for a more detailed summary. 

Another intriguing area of research focuses on the superposition of massive electromechanical resonators, which carries significant implications for exploring the effects of general relativity~\cite{gely2021superconducting}. To investigate these effects effectively, it is crucial to ensure that the coherence time of the superposition state exceeds the timescale associated with general relativity. In pursuit of this goal, one approach involves integrating clamped mechanical oscillators, such as silicon nitride membranes, with superconducting circuits. This integration allows for the preparation of these resonators in small cat/fock states, enabling experiments that probe the interplay between quantum mechanics and general relativity~\cite{liu2021gravitational,albrecht2014testing}.

Furthermore, the coherent coupling of mechanical vibrations in carbon nanotube resonators, controlled by the electronic spin of a nitrogen vacancy, has emerged as a fascinating avenue of research~\cite{qin2019proposal}. By cooling these resonators to their quantum ground state, it becomes possible for the mechanical phonons within them to exhibit both wave-like and particle-like behavior, effectively manifesting the essence of quantum superposition. This line of inquiry not only pushes the boundaries of quantum mechanics but also highlights the potential applications of electromechanical resonators in quantum information processing and quantum metrology.

Additionally, electromechanical systems offer opportunities for experiments involving Paul traps, which can shed light on the interplay between quantum mechanics and general relativity. In the work by~\cite{martinetz2020quantum}, a Paul trap is proposed to trap and cool a single charged nanoparticle to its quantum ground state. Through the controlled application of laser beams and the analysis of the nanoparticle's evolution, the researchers were able to demonstrate the sensitivity of the nanoparticle to gravity and place constraints on non-Newtonian gravitational interactions. Moreover, electrically levitated nanorotors, when coupled to a superconducting qubit, enable ultra-short timescale interference experiments by achieving a quantum superposition state through controlled rotational and translational motion. These developments represent exciting prospects for furthering our understanding of macroscopic quantum phenomena and their implications for our understanding of both quantum mechanics and gravity.

Based on careful considerations of thermal decoherence effects, a proposal for a superconducting and magnetic version of a superposition experiment has been made. The low temperature and extreme-high vacuum setting appear the most promising~\cite{pino2018chip}. The extreme settings where the quantum system is - as much as possible - decoupled from its environments can be analyzed in terms of the duration of free evolution (spreading) of the wavefunction and demands extremely small amplitudes (typically smaller than the spatial size of the de Broglie wavelength) of all guiding fields and also vibrations. This adds another serious technical demand to any experimental realization. 

As dictated by Schr\"odinger dynamics, the free evolution time grows proportional to the mass of the quantum system and is pushing realistic attempts to beat the existing mass record well beyond some 100 ms superposition lifetime. All decoherence effects and noises have to be controlled to be smaller than the evolution amplitudes during that same time. Ideally, the wavefunction is let alone for some seconds, as is at the core of proposals for macroscopic quantum superposition on a dedicated satellite in space~\cite{kaltenbaek2022maqro}. 

However, an alternative solution may come from boost or inflation operations, which accelerate the spread of the wavefunction significantly~\cite{romeroisart2017coherent}. The boost has to be coherent so it does not spatially resolve the position of the particle during the boost. Boosts have been demonstrated by Stern-Gerlach beam splitters for atoms on a chip~\cite{margalit2021realization}, and it seems possible to translate the same beam splitting technique to the much more massive NV-defect center diamond nanoparticles and other spin systems. 

Another severe experimental challenge is that the mechanical experiments, which are usually single-particle experiments, have to be repeated many times to achieve particle number statistics to show unique quantum features~\cite{neumeier2022fast}. All known measures of quantumness are statistical ensemble measures on the level of density matrix rather than the level of the wavefunction directly, as every single run of a quantum experiment or operation has a completely random outcome and cannot be predicted by quantum mechanics. As there is no coherent ensemble of massive particles, equivalent to, for instance,  an atomic BEC, each large mass single-particle experiment has to be repeated many times (say, at least 1000 times) under the exact same conditions. Still, many experimentalists have started taking on the challenge to work toward the first generation of a truly macroscopic quantum superposition, which will challenge our understanding of quantum mechanics as well as gravity and will hint at how the two important theories are connected fundamentally.   

In a recent study~\cite{romerosanchez2018quantum}, researchers explored the fascinating realm of ultra-strong coupling between a mechanical oscillator and an LC-resonator, achieved through magnetically induced electromotive force. This intriguing approach to coupling magneto-mechanical oscillators in the ultra-strong regime has opened up a wealth of possibilities, ranging from sensitive weak-force detection to advanced electro-mechanical state manipulation.

Furthermore, the use of levitated superconducting microrings offers a unique advantage over traditional microspheres by capitalizing on flux conservation within the ring structure~\cite{navau2021levitation}. This innovative approach provides a versatile platform for the design and optimization of magneto-mechanical ring oscillators. Notably, when it comes to generating quantum superpositions through ground state cooling, the separation between the peaks of the wave function must exceed the physical dimensions of the object. In this regard, ring geometries outperform traditional spherical counterparts, making magneto-mechanical ring resonators particularly suited for experiments investigating gravitational interactions. Recent breakthroughs have demonstrated precise control and levitation of high-Q superconducting microspheres~\cite{latorre2022superconducting, hofer2023high}. Appealing aspects of these systems include the mechanical frequencies on the order of 100Hz, the access to trap anharmonicities, and the scalability of mass of the levitated particle, which is of particular relevance for testing macroscopic quantum states and the role of gravity affecting quantum states. One viable approach involves utilizing a static magnetic trap formed by two coils configured in an anti-Helmholtz arrangement to achieve stable levitation, which was first analyzed in detail numerically for an on-chip configuration~\cite{latorre2020chip}.

In addition, nanomechanical oscillators can be effectively cooled to their ground state when levitated under an inhomogeneous field, a phenomenon facilitated by the Meissner effect~\cite{cirio2012quantum}. When these oscillators are inductively coupled to a flux qubit, a remarkable opportunity arises to create macroscopic entangled states within these sizable objects. This breakthrough holds significant promise for practical applications, especially in the precise measurement of force gradients, such as those encountered in the study of gravity~\cite{johnsson2016macroscopic}.

In conclusion, while quantum superpositions have been achieved in clamped and hybrid mechanical systems, we are still awaiting their experimental demonstration for levitated mechanics. There are plenteous ideas and approaches under active research and we expect the first levitated superposition within the next five years or so. 

\subsubsection{Entanglement in mechanical systems} \label{sec:mech:entanglement}
%
%
The quantum entanglement of massive objects has been a focal point in the intersection of quantum mechanics and gravity research. Understanding entanglement in macroscopic systems offers insights into the quantum-to-classical transition and theories related to wave-function collapse, as discussed in Section~\ref{sec:grav:dec:section} which delves into gravitational decoherence, semi-classical models, self-energy, and gravitationally-induced wavefunction collapse. Furthermore, the intricacies of entanglement mediated by gravity are elaborated upon in Section~\ref{sec:entanglement:mediated:gravity}. Various approaches have been explored to create and manipulate entanglement in massive quantum systems, aiming to unravel the mysteries of quantum mechanics and its connection to gravity. Several approaches have been proposed for entangling massive quantum systems, and some have been already demonstrated experimentally in clamped optomechanical systems in recent years. Quantum entanglement has not yet been demonstrated in levitated mechanical systems but is an active research objective.

In a clamped system,~\cite{eichler2014quantum} uses a Bose-Hubbard dimer, consisting of two bosonic modes with an onsite interaction strength, to experimentally demonstrate quantum-limited amplification and entanglement. The authors studied how this system responds in different parameter regimes, and by applying a coherent drive field, they were able to generate entangled photon pairs from vacuum input fields demonstrated through the measurement of cumulants. This study could prove useful for experimental studies related to non-equilibrium many particle physics in photonic systems as well as being applied towards massive resonators, which would be used for exploring wavefunction collapse theories at macroscopic scales due to gravitational interactions between them.~\cite{riedinger2018remote} demonstrated the generation of distributed entanglement between two nanomechanical phononic-crystal resonators by using a three-step protocol consisting of cryogenically cooling the two mechanical resonators, sending a weak pump pulse into a phase-stabilized interferometer, and creating a phonon. The joint state of the two mechanical systems was then entangled and the entanglement was verified by mapping the mechanical state onto an optical field in this measurement-based entanglement scheme. The simultaneous coupling of two nanomechanical resonators to a superconducting qubit in the strong dispersive regime was used to entangle the two nanomechanical devices~\cite{wollack2022quantum}.

Cavity optomechanical setups have also been demonstrated experimentally for generating entanglement between two massive mechanical oscillators~\cite{ockeloen2018stabilized} by a two-mode back-action evading measurement to verify entanglement in the cavity mode.~\cite{thomas2021entanglement} have experimentally achieved entanglement between a macroscopic mechanical oscillator and an atomic spin oscillator. This achievement was accomplished using a millimeter-sized dielectric membrane and an ensemble of roughly $10^9$ atoms within a magnetic field. The entanglement was confirmed by achieving an Einstein–Podolsky–Rosen variance below the separability limit. The process involved manipulating the light that passed through the two spatially separated systems, with the collective atomic spin serving as an effective negative-mass reference, thereby suppressing quantum back-action.

Further theoretical proposals for entanglement schemes making use of access to a discrete variable qubit quantum system and/or cavity modes include things like a coherent feedback loop~\cite{li2017enhanced} applied to two macroscopic mechanical resonators that were strongly coupled to a common optical mode.~\cite{asadian2016heralded} proposed using a sequence of pulses to periodically flip a qubit, synchronized with the resonator frequency. A conditional photon emission is then applied to the qubit to produce a single photon depending on the state of the qubit prior to the pulse to create a mechanically entangled coherent state, a Schr\"odinger cat state.~\cite{yi2021massive} discusses the use of large-scale spatial qubits to explore macroscopic nonclassicality and entanglement generated through the Casimir effect. As already mentioned in Sec.~\ref{sec:squeezing},~\cite{cosco2021enhanced} proposed a protocol for generating entanglement between two weakly interacting massive resonators, such as nanoparticles levitated by optical means or massive pendula tethered to a base. The protocol involves applying a continuously squeezed protocol to the two resonators and removing the squeezing quickly after generating the desired entanglement. They also proposed the reverse protocol to reduce the decay rate of the entanglement dramatically. 

Another promising approach and especially interesting for low-frequency mechanical oscillators, as proposed by~\cite{li2020stationary}, involves the preparation of entangled states between a massive membrane and a low-frequency LC (inductor-capacitor) resonator, and~\cite{li2021entangling} are proposing to use cavity magnomechanics for entangling vibrational modes of two massive ferromagnetic spheres.~\cite{xu2022optomechanical} focus on the entanglement dynamics of spatially separated local LC oscillators coupled to a long, partially metalized elastic strip through the optomechanical interaction which is proposed to be usable to observe quantum gravity-induced entanglement at low energies. Electromechanical systems have been considered as well,~\cite{khosla2018displacemon} introduced a novel approach involving the coupling of multiple electromechanical resonators to a common qubit. Such electromechanical systems also allow the verification of entanglement of mechanical oscillators via qubits ~\cite{bose2006entangling}. This coupling scheme results in the entanglement of these massive oscillators. Remarkably, this entanglement manifests itself in the form of quantum interference patterns observed in the displacement of the resonators.

 In conclusion, various protocols have been proposed, and some have already been demonstrated, with each offering unique insights into the interplay between quantum mechanics and gravity. These advancements not only contribute to our understanding of fundamental quantum principles but also hold the potential to validate or challenge extensions of the Schrödinger equation, furthering our grasp of the complex realm of quantum mechanics in gravitational contexts, including such exciting questions such as if gravity can be used to quantum entangle two particles, see Sec.~\ref{sec:entanglement:mediated:gravity}. We emphasize that large-mass mechanical systems are considered {\it the} system for experimental exploration at the quantum gravity interface.


%
\section{Outlook} \label{sec:discussion}

In this review article, we have outlined ideas and proposals for probing the interface between quantum mechanics and gravity with massive quantum systems. We emphasize that many unknowns still persist and that numerous questions remain completely open in the field. Gravity appears to behave differently in comparison with the other forces, in that it can be formulated as curvature and allows for an equivalence principle. A fully-fledged theory that consistently combines quantum mechanics and gravity might ultimately manifest in a completely unexpected manner. 

To conclude this review, we offer a few brief remarks on some of the outstanding questions raised throughout the article. Firstly, we have seen that even classical gravity is one of the least precisely tested forces, especially at small-length scales. New and better precision tests are needed to further constrain modified gravity theories and verify our current description of gravity. Quantum-enhanced sensors play an important role here. Equally intriguing are questions regarding the nature of gravity itself, and whether its interaction with quantum systems results in a decohering or entangling process. Here, tests with masses prepared in quantum superpositions could provide a key step toward establishing the correct theoretical description. Beyond non-relativistic quantum mechanics, the mathematical formalism of quantum field theory in curved spacetime enables the study of the interplay of quantum and relativistic effects at low energies. It has, however, not yet been demonstrated in the laboratory, and the formalism itself gives rise to additional tensions, such as the black-hole information paradox. 

All of the questions above warrant further study. Ultimately, to develop a theory that incorporates quantum and general relativistic effects in a consistent way, it is necessary to understand what principles are truly fundamental. Experiments and the development of quantum technologies play a key role here. In particular, increasing the masses and coherence times of quantum systems may allow for some of the proposals outlined in this article to be realized in the laboratory. The results could help guide future research, which thus far has mainly relied on mathematical and theoretical arguments. 

In summary, the prospect of using massive quantum systems to explore the interplay between quantum mechanics and gravity opens up a number of novel questions and exciting challenges. The field is ripe for exploration and potentially groundbreaking discoveries. We (the authors of this article) hope that this review may inspire many future discussions around these topics and look forward to learning about and partaking in potential discoveries in the future.

%
\section*{Acknowledgments}

We thank Angelo Bassi, Miles P.~Blencowe, Caslav Brukner, and Gary Steele for fruitful discussions. We thank Gerard Higgins and Onus Hosten for their help with compiling data for Figure \ref{fig:sensitivity:vs:mass}. We also thank Jie Li, Klaus Hornberger, Igor Pikovski, Tejinder Singh, Flaminia Giacomini, Michael Tobar, Anupam Mazumdar, Martin Plenio, Julian Pedernales, Jack Clarke, Nathan Inan, Witlef Wieczorek, Simon Haine, Eduardo Martin-Martinez, Magdalena Zych, and Liu Qui for helpful comments.
We acknowledge the EPSRC International Quantum Technologies Network EP/W02683X/1
Levitation Network for Advanced Quantum Technologies for supporting the writing of this article. 
This collaboration began as the UK Optomechanics Research Network (UniKORN) online seminar series (2020--2022). We thank all speakers, panel members, and audience participants who contributed to the seminars, which inspired us to write this article. 

IF thanks Eugene Jhong, John Moussouris, Jussi  Westergren and the Emmy Network for support and research funding and acknolwedges support from the Leverhulme Trust project {\it MONDMag} (RPG-2022-57). 
AG is supported in part by NSF grants PHY-2110524 and PHY-2111544, the Heising-Simons Foundation, the W.M. Keck Foundation, the John Templeton Foundation and ONR Grant N00014-18-1-2370. 
MR acknowledges funding from the EPSRC and STFC via Grant Nos. EP/N031105/1, EP/S000267/1, EP/W029626/1, EP/S021582/1 and ST/W006170/1.
SQ is funded in part by the Wallenberg Initiative on Networks and Quantum Information (WINQ) and in part by the Marie Skłodowska--Curie Action IF programme \textit{Nonlinear optomechanics for verification, utility, and sensing} (NOVUS) -- Grant-Number 101027183. Nordita is supported in part by NordForsk. 
MT would like to acknowledge funding by the Leverhulme Trust (RPG-2020-197).
HU acknowledges support from the QuantERA grant LEMAQUME, funded by the QuantERA II ERA-NET Cofund in Quantum Technologies implemented within the EU Horizon 2020 Programme, from the UK funding agency EPSRC grants (EP/W007444/1, EP/V035975/1, EP/V000624/1, EP/X009491/1), the Leverhulme Trust project {\it MONDMag} (RPG-2022-57), the EU Horizon 2020 FET-Open project {\it TeQ} (766900), the EU Horizon Europe EIC Pathfinder project {\it QuCoM} (10032223) and the European Space Agency for the ESA Payload Masters project {\it Op-To-Space}, as well as the UK Space Agency for the IBF project {\it A3S}.
CCW acknowledges funding received from the Winton Programme for the Physics of Sustainability, EPSRC (EP/R513180/1) and the European Union’s Horizon 2020 research and innovation programme under grant agreement no.~732894 (FET-Proactive HOT). SK acknowledges funding from the ERC via Grant Agreement No. 818751.

\bibliography{ref}

\end{document}